\documentclass[preprint, review,3p,times, 10pt, authoryear]{elsarticle}
\usepackage[most]{tcolorbox}
\usepackage{float}
\usepackage{cite}
\usepackage{amsmath,amssymb,amsfonts}
\usepackage{algorithmic}
\usepackage{graphicx}
\usepackage{textcomp}
\usepackage{comment}
\usepackage{footnote}
\usepackage{comment}
\usepackage{threeparttable}
\usepackage{romannum}
\usepackage{longtable}
\usepackage{lscape}
\usepackage{soul}
\usepackage{array, makecell}
\usepackage{setspace}
\usepackage{tikz}
\usepackage{pgfplots}
\pgfplotsset{width=10cm,compat=1.9}
\usepackage{longtable}
\usepackage{threeparttablex} 
\usepackage{booktabs}        

\usepackage[hyphens]{url}
\usepackage{hyperref}
\hypersetup{colorlinks=true,breaklinks=true,pdfauthor=author}
\usepackage{ragged2e}
\usepackage{blindtext}
\usepackage{verbatim}
\usepackage{sectsty}

\usepackage{subcaption}

\usepackage{etoolbox}
\usepackage{lmodern}

\usepackage{lipsum}
\usepackage{lineno}

\usepackage[rightcaption]{sidecap}
\graphicspath{ {./img/} }

\usepackage[noabbrev]{cleveref}
\newcounter{rownumbers}
\newcommand\rownumber{\stepcounter{rownumbers}\arabic{rownumbers}}
\usepackage{svg}
\usepackage{multirow}
\usepackage{rotating}
\usepackage{paracol}
\usepackage{multicol}

\usepackage[utf8]{inputenc}
\usepackage{colortbl}

\usepackage{listings}
\usepackage{xcolor}

\definecolor{codegreen}{rgb}{0,0.6,0}
\definecolor{codegray}{rgb}{0.5,0.5,0.5}
\definecolor{codepurple}{rgb}{0.58,0,0.82}
\definecolor{backcolour}{rgb}{0.95,0.95,0.92}

\lstdefinestyle{java}{
  commentstyle=\color{codegreen},
  keywordstyle=\color{magenta},
  numberstyle=\tiny\color{codegray},
  stringstyle=\color{codepurple},
  basicstyle=\linespread{0.8}\ttfamily\footnotesize,
  breakatwhitespace=false, 
  breaklines=true, 
  captionpos=b,  
  keepspaces=true, 
  numbers=none,  
  numbersep=0.5pt,  
  showspaces=false,  
  showstringspaces=false,
  showtabs=false,  
  tabsize=2
}
\definecolor{eclipseStrings}{RGB}{176,99,99}
\definecolor{eclipseKeywords}{RGB}{4,81,165}
\definecolor{eclipseBackground}{RGB}{191,191,191}
\colorlet{numb}{magenta!60!black}

\lstdefinelanguage{json}{
basicstyle=\linespread{0.9}\ttfamily\footnotesize,
commentstyle=\color{eclipseStrings}, 
stringstyle=\color{eclipseKeywords}, 
numbers=none,
numberstyle=\tiny\color{codegray},
numbers=none,
numberstyle=\scriptsize,
stepnumber=1,
numbersep=8pt,
showstringspaces=false,
breaklines=true,
string=[s]{"}{"},
comment=[l]{:\ "},
morecomment=[l]{:"},
literate=
*{0}{{{\color{numb}0}}}{1}
{1}{{{\color{numb}1}}}{1}
{2}{{{\color{numb}2}}}{1}
{3}{{{\color{numb}3}}}{1}
{4}{{{\color{numb}4}}}{1}
{5}{{{\color{numb}5}}}{1}
{6}{{{\color{numb}6}}}{1}
{7}{{{\color{numb}7}}}{1}
{8}{{{\color{numb}8}}}{1}
{9}{{{\color{numb}9}}}{1}
}
\usepackage{color}

\journal{arXiv}
\begin{document}
\begin{frontmatter}
\title{\textit{LEI}: Livestock Event Information Schema for Enabling Data Sharing}

\author[1,2,5]{Mahir Habib\corref{correspondingauthor}}\ead{mhabib@csu.edu.au}
\author[1,2,5]{Muhammad Ashad Kabir}\ead{akabir@csu.edu.au}
\author[1,2,5]{Lihong Zheng}\ead{lzheng@csu.edu.au}
\author[2,3,5]{Shawn McGrath}\ead{shmcgrath@csu.edu.au}

\affiliation[1]{organization={School of Computing, Mathematics and Engineering, Charles Sturt University}, city={Bathurst}, state={NSW}, postcode={2795}, country={Australia}}

\affiliation[2]{organization={Gulbali Institute for Agriculture, Water and Environment, Charles Sturt University}, city={Wagga Wagga}, state={NSW}, postcode={2678}, country={Australia}}

\affiliation[3]{organization={Fred Morley Centre, School of Animal and Veterinary Sciences, Charles Sturt University}, city={Wagga Wagga}, state={NSW}, postcode={2678}, country={Australia}}


\affiliation[5]{organization={Food Agility CRC Ltd}, city={Sydney}, state={NSW}, postcode={2000}, country={Australia}}

\cortext[correspondingauthor]{Corresponding author: Charles Sturt University, Panorama Ave, Bathurst, NSW 2795, Australia}%

\begin{abstract}

Data-driven advances have resulted in significant improvements in dairy production. However, the meat industry has lagged behind in adopting data-driven approaches, underscoring the crucial need for data standardisation to facilitate seamless data transmission to maximise productivity, save costs, and increase market access. To address this gap, we propose a novel data schema, Livestock Event Information (LEI) schema, designed to accurately and uniformly record livestock events. LEI complies with the International Committee for Animal Recording (ICAR) and Integrity System Company (ISC) schemas to deliver this data standardisation and enable data sharing between producers and consumers. To validate the superiority of LEI, we conducted a structural metrics analysis and a comprehensive case study. The analysis demonstrated that LEI outperforms the ICAR and ISC schemas in terms of design, while the case study confirmed its superior ability to capture livestock event information. Our findings lay the foundation for the implementation of the LEI schema, unlocking the potential for data-driven advances in livestock management. Moreover, LEI's versatility opens avenues for future expansion into other agricultural domains, encompassing poultry, fisheries, and crops. The adoption of LEI promises substantial benefits, including improved data accuracy, reduced costs, and increased productivity, heralding a new era of sustainability in the meat industry.


\end{abstract}




\end{frontmatter}

\pagenumbering{arabic} 

\section{Introduction}

More than half of Australian agricultural production is attributed to cattle, including milk and red meat. Australia prides itself on producing superior quality beef free from exotic diseases~\citep{Anon2022}. Meat \& Livestock Australia (MLA)\footnote{MLA is an organisation designated by the Australian Meat and Livestock Industry Act as the industry's marketing and research body.} launched data-driven livestock traceability tools through the National Livestock Identification System (NLIS)\footnote{NLIS is Australia's scheme for the identification and tracing of livestock and is crucial in protecting and enhancing Australia's reputation as a producer of quality beef and sheep meat~\citep{NSWGovernment2022NationalServices}.} in 1999~\citep{Bahlo2021LivestockAustralia}. The NLIS identifies livestock electronically and records all cattle movements in a central database~\citep{nlis2008}. This information on movement events has been used to meet market biosecurity standards and allow Australian producers\footnote{A sheep or cattle farmer~\citep{MeatLivestockAustralia2021GlossaryAustralia}.} to export livestock products around the world. Animal movements can facilitate the transmission of several potentially devastating diseases~\citep{Sellman2022}, including foot-and-mouth disease and bovine TB, between farms and countries~\citep{Fevre2006AnimalDiseases}. These diseases can have serious economic consequences, including trade restrictions on cattle and beef products, loss of productivity, and increased costs associated with disease prevention and control measures. Access to movement records is required to trace such movements, which is an important step in analysing and managing disease outbreaks~\citep{Iglesias2015}. The recording of these events relies on data standardisation~\citep{Hardin2022InternetProcessing}. There are opportunities to extend data recording to include a wider range of cattle events information and explore using a more decentralised architecture that can support further growth in the red meat industry~\citep{Bahlo2019TheReview}.

Data drives farmers' businesses throughout the agrifood supply chain~\citep{Whitacre2014}. However, when these data are recorded in disparate systems, it is not easy to work efficiently with buyers,\footnote{Buyers purchase stock at physical markets. A buyer can work for many different clients or only for one company, e.g. processor, feedlot, restocker, and backgrounder~\citep{MeatLivestockAustralia2021GlossaryAustralia}.} auditors,\footnote{Auditors act as a check of Australia’s red meat integrity system and conduct audits each year to ensure the management systems used by livestock producers are complying with rules and standards~\citep{IntegritySystemsCompany2021a}.} and regulators.\footnote{A regulator is a government agency or organisation responsible for regulating and overseeing the livestock industry~\citep{AustralianGovernment2021RegulatoryDepartment}, such as the Australian Pesticides and Veterinary Medicines Authority (APVMA) (website: \url{https://apvma.gov.au/}), and the Department of Agriculture, Fisheries and Forestry (DAFF) (website: \url{https://www.agriculture.gov.au/}). Also, the state and territory governments in Australia have responsibility for regulating certain aspects of the livestock industry within their respective jurisdictions, such as enforcing animal welfare laws and regulating the sale and transportation of livestock.} For example, producers collect much data on cattle management, health and growth, farming practices and business. Many participants in their value chain need these data to inform their work. Producers must fill out forms to demonstrate their ethical practices when selling cattle or their products. Most of the information being requested by quality assurance and compliance programs is universal, requiring the same data to be entered multiple times on different forms, taking valuable time and focus away from operations and increasing the chance of mistakes or non-compliance. It would be desirable to only need to enter data once and send it wherever it needs to go.
For livestock producers, this means being able to choose what information to share with whom and for how long without worrying about each form, leaving more time to focus on producing products, adding value, and getting the data to do the work instead of them.

The National Farmers’ Federation (NFF) has set a goal of increasing the farmgate output from \$80 billion to \$100 billion by 2030~\citep{nff}. To achieve this goal, Australian agriculture needs to invest more in data and agricultural technology, which can improve productivity, efficiency, and sustainability. However, a key challenge is the lack of standardisation of data across the industry.

Currently, there are different data standards for livestock events within a farm, the International Committee for Animal Recording (ICAR)\footnote{ICAR, an International Non-Governmental Organisation (INGO), provides an open secure network to share with, learn from, and interact with fellow members and related stakeholders in global animal production~\citep{ICAR2013}.} and Integrity System Company (ISC)\footnote{ISC One of the members of ICAR and entirely owned and operated by MLA.} have developed their own schemas for Animal Data Exchange (ADE). These schemas are based on the JSON format and are open-source. However, they have some limitations: 1) They do not capture all relevant events, such as castration and weaning. 2) They do not include some important information, such as farm ownership and event triggers. 3) They are designed for centralised data management control, which limits data sharing with third parties.

Furthermore, the challenges extend to data management approaches. While centralised data management control ensures data quality and security~\citep{T.2022}, it also leads to reduced data flexibility and diversity~\citep{Iftikhar2011}. On the other hand, open data frameworks enable more stakeholders to participate in data creation and use, which may increase data variety and value but can create challenges for data integration and interoperability~\citep{Aydin2022DesignDatab}.

However, accessing and using these data is not easy because they are often stored in different formats, systems, or platforms. This makes it hard to combine, analyse, or share them effectively. Additionally, emerging technologies such as microservices~\citep{Alshuqayran2016}, Internet of Things (IoT)~\citep{Farooq2019}, and the concept of Industry 4.0~\citep{Antonino2022} offer opportunities to create a more connected and intelligent industry, but also introduce new challenges due to the diversity of data formats. Furthermore, data sharing in the agri-food sector has several challenges, as stated by~\citet{Durrant2021}, including technical, social, and cultural challenges. Technical challenges include issues with data quality, transparency, privacy protection, interoperability, security, and standardisation. Social challenges include inequality, power dynamics, trust, incentives, governance, ethics, and legal issues. Cultural challenges include the lack of awareness, education, skills and collaboration among different stakeholders in the agri-food domain.

The purpose of this research is to develop a unified schema\footnote{A unified schema is a common data model that defines the structure, meaning, and relationships of data elements. It can facilitate data standardisation, integration, and interoperability across different sources and systems.} integrating data from different phases of the livestock lifecycle to support strategic decision making for the red meat industry. The lack of standardisation of data across industries, centralised data management control, and data integration and interoperability are all major challenges facing the red meat industry. These challenges make it difficult to meet changing and diverse market demands, and also limit the ability of the industry to innovate. To address these explicit challenges, this article makes the following novel and significant contributions to the field of livestock data standardisation and management:

\begin{itemize}
\item We proposed the Livestock Event Information (LEI) schema, a comprehensive and flexible schema that enables data sharing and organises data related to different types of livestock events. LEI standardises data on 34 cattle-related events, from calving to death, providing detailed information for better tracking, record keeping, and decision making in movements, deaths, insemination, and treatments. LEI is based on reliable ICAR and ISC data standards, but extends them with a layered structure that improves the scalability, accuracy, and efficiency of data collection. LEI also facilitates smooth data transmission among researchers, farmers, and industry professionals, allowing them to access and use data to make data-driven decisions and optimise their practices.

\item We presented a thorough discussion of the LEI, ICAR, and ISC attributes, highlighting the similarities and differences among them.

\item We compare the structure of LEI with that of ICAR and ISC using structural metrics, such as the number of attributes, number of levels, number of optional attributes, etc. We show that LEI has a more balanced and modular structure than ICAR and ISC, making it easier to maintain and extend.

\item We conducted a comprehensive case study to evaluate the effectiveness and applicability of the LEI schema compared to ICAR and ISC. Our study covers a range of scenarios and aims to provide valuable information for the Australian red meat industry. We used real-world data from cattle farms in Australia and measured LEI performance in terms of data quality, completeness, consistency, and usability.

\end{itemize}

The remainder of the paper is organised as follows:~\Cref{sec:background} provides a brief background on the Australian red meat industry and data on livestock events. After reviewing the existing data standards and the limitations of the ICAR and ISC schemas in~\Cref{sec:lit}, we present an overview and implementation of our proposed schema and its layers in~\Cref{sec:lei}. We also perform an implementation code for validation between the schema and the data.~\Cref{sec:evaluation} describes the structural metrics for schema evaluation that we use to compare the proposed schema with ICAR and ISC. A case study with real-world scenarios of the new schema is discussed in~\Cref{sec:casestudy}. Finally,~\Cref{sec:conclusion} concludes the paper and outlines future research directions.

\section{Background}\label{sec:background}
\begin{figure}[!ht]
\centering
\includegraphics[scale=0.9]{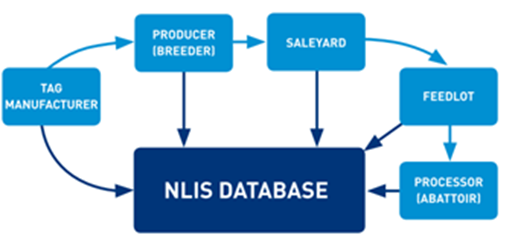}
\caption{NLIS central database~\citep{Victoria2009}}
\label{fig:nlis}
\end{figure}

The National Livestock Identification System (NLIS) is Australia's scheme for recognising and tracking livestock~\citep{nlis2016}. It is allowed by the federal and state governments and by the main producers, feedlots,\footnote{The feedlots are places where cattle are fed a high-protein grain-based diet to meet market specifications.} agents,\footnote{Agents sell and buy cattle on behalf of clients.} sale yards,\footnote{A sale yard is a physical marketplace where people can buy and sell livestock through an auction system.} and abattoir\footnote{An abattoir is a facility where cattle are butchered for human consumption.} businesses~\citep{Victoria2009}. In 1999, The NLIS was established to address the requirements of the European Union so it could allow livestock to be tracked rapidly, from the farm of calving to the abattoir, and for food safety issues. NLIS Cattle was enacted on July 1, 2004, and Sheep and goats were introduced on January 1, 2006. In 2006, recording livestock movements on a centralised database became compulsory~\citep{nlis2008}. As shown in~\Cref{fig:nlis}, tag manufacturers can access the NLIS central database to upload the NLIS device numbers issued to a Property Identification Code (PIC). Each device has a unique Radio Frequency Identification (RFID) number that is linked to the PIC of the source property~\citep{Trevarthen2007}.

The NLIS system benefits the red meat industry by identifying an animal's location at each stage of its life, decreasing the expense and societal impact of cattle disease, and maintaining access to the worldwide market. Keeping clients satisfied with the quality of their livestock and dairy products is another advantage of the NLIS. One of the main objectives of the NLIS system is to check the history of livestock and report missing or misplaced animal tags~\citep{Trevarthen2007}.

From a farmer's perspective, societal factors such as resistance to change to newer technology and lack of technical skills on farms are linked to a lack of data standards. They prefer to use technologies such as the mobile phone short message service (SMS)~\citep{Car2012} and the comma separated value files (CSV)~\citep{Chang2022}. Data standards make it easy for farmers to understand and accept new technology~\citep{Marshall2020}. However, data standardisation will increase the value of data while reducing exchange and reuse limitations. Furthermore, presenting unstructured data in a readable and structured format for humans and machines raises awareness and knowledge of how data providers\footnote{A data provider (also known as a data vendor or data supplier) is an organisation or corporation designated by farmers to provide them with data for their own use or to share with third parties.} acquire, use and share farm data. 

For herd management, producers need to record many data (for example, publisher ID, PIC, and RFID) periodically from the data provider's equipment, representatives, or other data-publishers.\footnote{Data publishers are the source of event information.} This is called event data. These event data are often shared among various data-consumers.\footnote{Data consumer (also known as the data subscriber) is the final destination for an information event.} The content of the event is known as event information.

The event does not need to be ``a change occurring in the status of the system" (that is, in our case, the birth or death of cattle changes the number of cattle on the farm, which impacts the production), as~\citet{Subirats-Coll2022} defines it. However, we can define the event as a record of the state of the system (that is, keeping a record of the weight of cattle or their status) in a certain period for analysis purposes (that is, intelligent processes that can be performed on that record for machine learning~\citep{Shahinfar2018}) or historical data. An event has a body and header to present event information. The description of the event is found in the event header, while the event details are found in the event body. Across event specifications, these components are constants.

Most of the time, producers have to access the information they require by logging into the portal via the data provider dashboard. Producers can repeat this procedure with most of the data providers they have appointed. Apart from this, some data providers can update the event information multiple times per day and send a daily email to the producers summarising the data collected by their machines. Also, on the data provider's website, farmers can view and download additional information. All this is an extra workload because producers must download all this data, refine it, and present it in paper or CSV files, as required per data-consumer, with the possibility of errors or missing data.

Significantly, most of these companies capture the data from producers' farms and provide it back to them. The diversity of companies negatively impacts the presentation of the data, since each company adheres to a different standard. On the other hand, farmers find it difficult to switch from one company to another due to different data standards. As a result, the onus is on the producers to re-standardise the data in a format that meets the needs of consumers and then share it with them.

\section{Related work}\label{sec:lit}
There are standards for transmitting agricultural data, such as those set by the International Organisation for Standardisation (ISO), which has over 1000 international standards specifically relevant to agriculture, with many more under development. These include standards for agricultural equipment such as tractors and moving equipment, irrigation, fertilisers, soil conditioners, feed machinery and supplies, animal feeding equipment, environmental sensors, and agricultural electronics~\citep{ios2017}.

These standards help ensure that data are exchanged accurately and seamlessly between agricultural service providers and farm management information systems. One such standard is ISO/TS 34700:2016, which applies to animals raised or kept to produce food or feed. However, it is essential to note that this standard does not apply to animals used for study and academic purposes, endangered animals and zoos, pet animals, homeless and wild animals, marine animals, slaughtering for public or animal health purposes guided by qualified administration, and humane killing traps for pest and fur animals~\citep{ISO2016}. Another standard in the agricultural industry is AgXML, which was developed to help facilitate electronic information communication throughout the agribusiness supply chain. This standard helps to ensure that information is exchanged accurately and quickly between various agricultural service providers~\citep{Santos2012}.

In addition, there is AgroXML~\citep{Schmitz2009}, a standard that facilitates data exchange for plant and crop production. This standard allows farm management information systems (FMIS) data to be submitted to external partners, such as agricultural service providers. AgroXML was developed by agricultural, machinery, and software architects in Germany and was initially applied only in Germany and the German language. However, since version 1.3, it has been converted to English. This standard is linked to ISOBUS, an ISO11783 standard protocol for information transfer between a tractor, implements (such as planters) and farm management applications running on personal computers~\citep{Nash2009}. There is also eDAPLOS, created by the Trade and Business Group of UN/CEFACT Europe. This initiative is based on previously created EDI messages for business dealings between consumers and farmers~\citep{Nash2009a}.
Furthermore, Scotland's government created the Scottish Beef Calf Scheme XML (SBCS) to help integrate with the government’s system. The government provides schemas to make integration easier, but it is important to note that the schemas do not cover several agricultural sectors and do not have expansion~\citep{karwowski2010ontologies}.
Lastly, mpXML (Meat and Poultry XML) is a prominent US standard for farm animal manufacturing. This standard is used to share data between meat producers and meat product manufacturers. Tracking the sourcing of meat along its entire chain is at the heart of this concept. The standard was established in 2001 to unite many partners and stakeholders to create a set of comprehensive and laudable standards for the sector. The group has issued several standards documents, and in 2014 it amalgamated with the GS1 US Meat and Poultry Workgroup~\citep{thesmar2019meat}.

In addition,~\citet{White2013} provided an overview of the International Consortium for Agricultural Systems Applications (ICAS) version 2.0 standards to simplify the exchange of information and software tools to record field studies on environmental conditions, such as weather and soil data and crop response measurements. Similarly,~\citet{Farrell2022AgriculturalV.1.2}, from the Agricultural Data Standards Guide Australia (ADSGA), which aims to make agricultural data findable, accessible, interoperable and reusable (FAIR), developed the Agricultural Research Federation (AgReFed) guidelines and steering policies for the management and handling of agricultural data to establish a common language to facilitate the creation, storage, ingestion, and use of agricultural data. On the other hand,~\citet{Bahlo2019TheReview} reviewed the literature on precision livestock farming (PLF) technologies in relation to the use of public data, open standards and interoperability. The review concluded that a new type of decision support tool is required, as well as agreement on data exchange schema, to demonstrate the value of shared data at the farm scale (commercial benefit) and regional scale (public good).

The International Committee for Animal Recording (ICAR) is an International Non-Governmental Organisation (INGO) that provides an open, secure network to share with, learn from and interact with fellow members and related stakeholders in global animal production~\citep{ICAR2013}. One of the ICAR members is the Integrity System Company (ISC), which is entirely owned and operated by Meat \& Livestock Australia. ISC set up an open source animal schema repository that contains the JSON standard for Animal Data Exchange (ADE)~\citep{Cooke2020}. The ISC schema objects were duplicated from the ICAR schema repository. ICAR aims to stimulate global animal production to be more sustainable and efficient than individual efforts in this area~\citep{ICAR2013}.

\begin{figure}[!ht]
\centering
\includegraphics[scale=0.6]{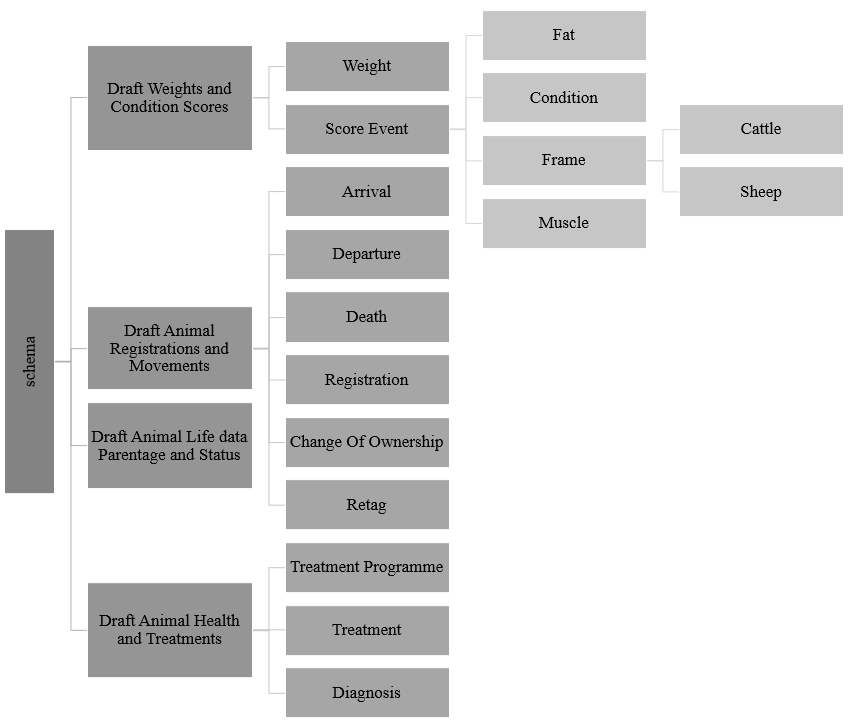}
\caption{Integrity system company (ISC) schema}
\label{fig:iscschema}
\end{figure}

 Furthermore, Rezare Systems, the National Livestock Language Committee and Meat Standards Australia, in collaboration with Meat and Livestock Australia (MLA) and Integrity Systems Company (ISC)~\citep{Patel2020} to develop strategies to assist the Australian beef meat industry in improving the efficient use of technology in terms of simplicity, traceability, increased production, and cost. The ISC scheme has four categories in a hierarchical structure, as shown in~\Cref{fig:iscschema}: 

\begin{enumerate}
\item \textbf{Animal identity, life data, and parentage.}
The system included details on the two-piece identity model created for the Farm Data Standards and approved by ICAR. This method offers an ID (a unique PIC code) and a scheme (for example, ``au.gov.ag.pic") for each identification. Many schemes can be used for places (PICs) and animals (e.g., NLIS tags). Alternative identifiers for an animal are possible (for example, a genetic identifier and an NLIS tag)~\citep{Patel2020}.

\item \textbf{Weights, condition scores, and related events.}
The system supplied information on the generalised model of individual animal events (observation or action) used by ICAR ADE. Based on an ``Event Core Resource" that records the animal, the event's date and time, location, and ID. The properties provided by the derived classes are specific to the event~\citep{Patel2020}.

\item \textbf{Health treatments.}
The system offered a structure to express a diagnosis, a single therapy, and a treatment plan that may contain a diagnosis and a series of therapies for animal health. Treatments had to be flexible enough to accommodate medical, non-medical, and combination therapies, and the batch of the product and the expiration dates had to be recorded~\citep{Patel2020}.

\item \textbf{Registration and movements.}
Also,~\citet{Patel2020} referred to the system's ability to give data on already-occurring registration and movement events that ICAR had identified, and these events were:
\begin{enumerate}
\item Recording livestock as found that were lost (implemented as a variation of registration).
\item Arrival (adding more livestock to the property).
\item Change of ownership (sale and purchase), separate from a movement.
\item Departure (the departure of livestock).
\item Registration (formal registration or induction of new livestock).
\item Death (such as those lost and those slaughtered for use on farms).
\item Retagging (replacing missing livestock tags).
\end{enumerate}
\end{enumerate}

\textbf{Limitations of existing schemas.}
Our study reveals that specific agricultural standards, such as AgroXML, have limited accessibility to the public. Some internal material needs to be updated, as with eDAPLOS. Even current standards, such as AgXML and mpXML, are primarily available in XML text format and are tailored towards supporting agricultural practices in Europe or the USA. These standards focus primarily on plants, crops, and machinery, with limited attention given to livestock events. Furthermore, even livestock standards prioritise dairy production over red meat production.

In contrast, the ICAR offers autonomous guidelines, standards and certification on animal identification, recording, and evaluation, which have been previously reported in the literature~\citep{Rosati2012,InternationalCommitteeforAnimalRecording-ICAR2014}. The ISC standard, designed to ensure data standardisation in the Australian livestock industry, follows the guidelines set forth by the ICAR. However, the information collected through the ICAR and ISC schemas must be improved as standalone information for consumers. It must provide a complete picture of the recorded livestock events, which is critical when sharing livestock event information with consumers. It is essential to acknowledge that consumers may need to analyse and process the received information rather than store it~\citep{Elly2013}.

Producers must share specific information about their livestock to enable informed consumer decision making. Schematic systems that incorporate this information facilitate transparent and informative data sharing between producers and consumers. The Line Information System Architecture (LISA) presented by~\citet{Theorin2017An4.0}, the message format in LISA is designed to be simple and flexible. It consists of a header and a body with an ordered key-value map. The header contains information related to the sending and routing of the message, while the body is used to build arbitrary hierarchical structures.~\citet{Subach2020StructuringSubach} generalised event structure involves a series of questions that describe a change in the system, including who performed the change, what object was changed, how it was affected, when it occurred, and where it happened. To represent contextual information,~\citet{Castelli2007} proposed the W4 Context Model, which accommodates incomplete information and adapts to context. The model uses a 4-field tuple to represent most world facts, Who, What, Where, and When. This representation can represent vast information about the world from various sources, such as sensors, tags, or web communities. As a result, this allows the data to be used flexibly and expressively.~\citet{Yuan2013,Yuan2015} proposed probabilistic models, W4 and Enhanced W4 (EW4), respectively, to discover the mobility behaviour of the individual user from spatial, temporal, and activity aspects. These models have potential applications in user profiles, location predictions, tweet recommendations, and activity prediction.

Interestingly, four critical elements of events, known as the four Ws, have been identified by~\citet{Allan1998} and other researchers~\citep{Mamo2021,Mohd2007,Zhou2017}. The four Ws are currently experiencing a resurgence in research on event modelling and mining~\citep{Rudnik2019}.~\citet{Mamo2021} propose a structured definition of events based on the four Ws to help automatically generate event knowledge and focus on Who and What, aligning them with other research areas to better track, model, and mine events.

Applying the W4 model to livestock event information would enable consumers to process, analyse, and make predictions based on the received information.~\Cref{tab:schemainformation} shows the application of the W4 model for a farmer to share with consumers to provide them with enough information. Furthermore,~\Cref{tab:schemainformation} indicates whether the ICAR/ISC schemas have applied the W4 model.

 \begin{table}[!ht]
 
\caption{Schema mandatory information}
\label{tab:schemainformation}
\centering
\resizebox{\textwidth}{!}{%
\begin{tabular}{llllp{7cm}}
 \hline
\multicolumn{2}{l}{Information} & ICAR & ISC & Note\\
\hline

``datetime": ``When?"& & \checkmark & \checkmark & Events capture date and time.\\

``source": ``Where?"& & x & x & In ICAR/ISC, there are not many details about the source of the event, for example, GPS location details for the device that triggered the event, device brand, or the company name.\\
``owner": ``Who?" & & x & x & The ICAR/ISC has no specific field for identifying the animal owner. Although it does have fields for organisations and people, these are not used to indicate the owner of an animal.\\ \multirow{2}{*}{``message": ``What?"}& ``item": ``Which?" & \checkmark & \checkmark & Animal details include identification number or tag number, date of birth, breed information, and gender.\\
&``event": ``Why?" & \checkmark & \checkmark & The details about the event include the animal's health and vaccination history, feeding regimen and nutrition information, weight or body condition score, information about any treatments (such as antibiotics or hormones) that the animal has received, and production data (such as milk yield or meat quality).\\
\hline
\end{tabular}
}
\end{table}

 On the other hand, LEI is a proposed schema that expands and supports different kinds of livestock events with different parameters by adopting and extending existing ICAR and ISC standards. In other words, LEI builds upon these standards. The LEI schema aims to provide a comprehensive and flexible schema to capture, share, consume, and organise data related to different types of livestock events. Furthermore, some important events are not captured, such as castration and weaning.~\Cref{tab:ISCICARevents} compares events recorded by the ISC with those represented in the ICAR schema, and the ICAR schema appears to have more events than the ISC schema.

\begin{table}[!ht]
 \caption{ISC vs ICAR events}
 \label{tab:ISCICARevents}
 \centering
  \begin{tabular}{llll}
 \hline
 No. & Event & ISC schema & ICAR schema\\
\hline
 
\rownumber & Weight & \checkmark & \checkmark \\
\rownumber & Fat & \checkmark & x \\
\rownumber & Condition & \checkmark & x \\
\rownumber & Frame & \checkmark & x \\
\rownumber & Muscle & \checkmark & x \\
\rownumber & Arrival & \checkmark & \checkmark \\
\rownumber & Departure & \checkmark & \checkmark \\
\rownumber & Registration & \checkmark & x \\
\rownumber & Change of ownership & \checkmark & x \\
\rownumber & Retag & \checkmark & x \\
\rownumber & Treatment program & \checkmark & \checkmark \\
\rownumber & Treatment & \checkmark & \checkmark \\
\rownumber & Diagnoses~ & \checkmark & \checkmark \\
\rownumber & Daily Milking Averages & x & \checkmark \\
\rownumber & Feed Intake & x & \checkmark \\
\rownumber & Milking Dry Off & x & \checkmark \\
\rownumber & Milking Visit & x & \checkmark \\
\rownumber & Abortion & x & \checkmark \\
\rownumber & Heat & x & \checkmark \\
\rownumber & Insemination & x & \checkmark \\
\rownumber & Parturition & x & \checkmark \\
\rownumber & Pregnancy Check & x & \checkmark. \\
\rownumber & Semen Straw & x & \checkmark \\
\rownumber & Status Observed~ & x & \checkmark \\
\rownumber & Lactation Status Observed & x & \checkmark \\
\rownumber & Birth & x & \checkmark\\
\hline
\end{tabular}
\end{table}

Therefore, the literature review highlights a gap in our understanding of the schema structure, the data standardisation for cattle events, and the parameters in single events. To address this gap, this article proposed a new schema called livestock event information (LEI) that standardises event modelling for cattle activities management. A single event must contain all data about livestock, producer, event details and the time when it happened, which are valuable to consumers.  The LEI schema was constructed in JSON format because it is an extension of ICAR and ISC, which had already been developed in JSON format.

\section{Livestock Event Information Schema}\label{sec:lei}
\subsection{Overview}

To elaborate more on the context, the proposed schema ``Livestock Event Information (LEI) Schema" is to provide a uniform and accurate way of recording events such as breeding, vaccinations, and the health status of the cattle. It is designed to make it easier for researchers, farmers, and industry professionals to access and use data, leading to more efficient and effective cattle management practices. The LEI schema will not only standardise the data, but will also provide a way to connect the data to other systems, such as those used for genetic evaluations and disease surveillance. Furthermore, the proposed schema is flexible and scalable to adapt to new and emerging technologies, such as IoT and RFID, which can be used to improve the accuracy and efficiency of data collection. Using the proposed schema and the data collected from it, researchers, farmers, and industry professionals will be better equipped to make data-driven decisions that ultimately will lead to a more productive and sustainable livestock industry.

As stated, we emphasise that the LEI schema is an extension of ICAR and ISC standards and builds upon them, but does not replace them. Therefore, the LEI schema cannot be used independently and requires that the ICAR and ISC schemas be in place. In other words, LEI proposes to enhance current standards rather than replace them. As a result, the LEI schema cannot function without ICAR or ISC schemas in place.~\Cref{tab:leievents} summarises 34 events captured by the LEI schema, along with their definitions.

\setcounter{rownumbers}{0}
\begin{longtable}[ht!]{l p{2.3cm} p{12cm}}
\caption{LEI events definition}
\label{tab:leievents}\\
 
 \hline
 No. & LEI Event & Definition\\
 \hline 
 \endfirsthead
 
\multicolumn{3}{c}%
{{ \tablename\ \thetable{} -- continued from previous page}} \\
\hline
No. & LEI Event & Definition\\
 \hline 
\endhead
\rownumber & Weight & The weight observation~\citep{Weber2020} in the units specified (usually kilograms). \\ 
\rownumber & Score & Body condition scoring is a management score designed to assess an animal's body reserves or fat accumulation~\citep{Qiao2021}. \\
\rownumber & Arrival & The arrival of an animal~\citep{Sellman2022} in a location (that is, farm, slaughterhouse, or any PIC different from the one in which the livestock was registered). \\ 
\rownumber & Departure & The departure of an animal~\citep{Sellman2022} from a location (that is, a farm, a slaughter yard, or any PIC different from the one where the livestock was registered). \\ 
\rownumber & Death & Death, kill or slaughter of an animal for a reason such as a disease, injury, or human consumption~\citep{MeatLivestockAustralia2021GlossaryAustralia}. \\ 
\rownumber & Registration & Registration of a new or found animal~\citep{nlis2016,Wismans1999}. \\ 
\rownumber & Retag & Replacement of an animal's tag due to loss, damage to the tag, or other circumstances that require retagging the animal~\citep{McGowan2014}. \\ 
\rownumber & Treatment program & Describes a course of treatments for one or more diagnoses~\citep{MeatLivestock2011}. \\ 
\rownumber & Treatment & Treatment of an animal with medicine and/or a procedure~\citep{MeatLivestock2011}. \\ 
\rownumber & Diagnosis & Diagnosis of an animal's health problem~\citep{Frost1997, Wang2023}. \\
\rownumber & Daily Milking Averages & Resource containing daily averages calculated from milking visits of a single animal~\citep{DepartmentofAgricultureFisheriesandForestry2020DairyAustralia}. \\ 
\rownumber & Feed Intake & Event for recording a feed intake~\citep{MeatLivestockAustralia2021GlossaryAustralia}. \\ 
\rownumber & Milking Dry Off & Records a description of each period when the livestock stopped producing milk~\citep{Hart1998} and the reason, also records a separate health treatment event. \\ 
\rownumber & Milking Visit & Event for recording milking visit~\citep{DepartmentofAgricultureFisheriesandForestry2020DairyAustralia}. \\
\rownumber & Abortion & The abortion event records an observation that abortion has occurred~\citep{Stegelmeier2020}. \\ 
\rownumber & Heat & Event for recording temperature~\citep{Murugeswari2022}. \\ 
\rownumber & Insemination & Event to record natural or artificial insemination~\citep{Benaissa2020}, including embryo transfer~\citep{Oliveira2020}. \\ 
\rownumber & Parturition & Event to record parturition (calving, lambing, kidding, fawning)~\citep{Smith2020}. \\ 
\rownumber & Pregnancy Check & Pregnancy diagnosis or check event~\citep{Smith2020}. \\ 
\rownumber & Semen Straw & Describes a semen straw~\citep{Benaissa2020}. \\ 
\rownumber & Status Observed & This event records an observed reproductive status without necessarily a pregnancy check, parturition, or other event. \\ 
\rownumber & Lactation Status Observed & This event records an observed lactation status without necessary parturition, drying off, or another event because calving and other factors can change the body condition during lactation and affect the health and fertility of dairy cows~\citep{Kuzuhara2015}. \\ 
\rownumber & Birth & Event for recording animal birth~\citep{Smith2020}. \\ 
\rownumber & Synchronisation & Insert a device into the female cattle's vagina for artificial insemination~\citep{Motavalli2017}. Synchronisation refers to the process of coordinating the reproductive cycles of a group of female animals, such as cattle, so that they are more likely to conceive at the same time. This can be done using hormones and can be used to improve the efficiency of breeding programs or to increase the number of calves born at a specific time of year. Synchronisation protocols can vary depending on the species and the specific goals of the breeding program.\\ 
\rownumber & Weaning & The calf is removed from the mother when it becomes old~\citep{MeatLivestockAustralia2021GlossaryAustralia} or for health problems. \\ 
\rownumber & Audit & Check the stock count. \\ 
\rownumber & Castrate & It is any action, surgical, chemical, or otherwise, by which a calf loses use of the testicles; in other words, desex~\citep{Ede2022}. \\ 
\rownumber & Pulse check & Assessed the heart rate by using a stethoscope and listening over the chest of the animal~\citep{Trevarthen2007}. \\ 
\rownumber & Respiration & Check the breaths per minute~\citep{Tuan2022}. \\ 
\rownumber & Find age by dentition & It is the characteristic arrangement, kind, and counting of the number of teeth in each species at a given age; thus, the eruption times and wear of the teeth are the main factors used to estimate the age of an animal~\citep{Whiting2013}. \\ 
\rownumber & Hoof trimming &It is trimming and balancing the hooves of animals and placing shoes on their hooves, if necessary. It is noticeable that after hoof trimming, cows walked better and started to keep more weight~\citep{Stoddard2017}.\\ 
\rownumber & Horn tipping & It is the removal of an adult animal's insensitive part of the horn resulting in a blunt horn end~\citep{Development2018}. \\ 
\rownumber & Dehorning & It is the removal of the horn of an animal by methods that destroy or remove keratin-producing cells and structures at the base of the horn~\citep{Reiche2020}. For example, genetically polled or dehorned cattle in groups showed less bruising at slaughter than horned cattle in groups.\\ 
\rownumber & Location & It is to locate the animal in the yard~\citep{Bishop-Hurley2007,IlyasAhmad2020} in the GeoLocation format~\citep{Eitzinger2019}. \\
\hline

\end{longtable}

~\Cref{sec:leiLayers} discusses the structure of the Livestock Event Information (LEI) schema. It explains the different layers of the schema.~\Cref{sec:leiImplementation} goes into more detail about how the LEI schema has been implemented to represent events. It explains how the properties and data elements within the schema are used to consistently and accurately describe the events.

 \subsection{Structure}\label{sec:leiLayers}

\begin{figure}[ht!]
\centering
\includegraphics[scale=0.32]{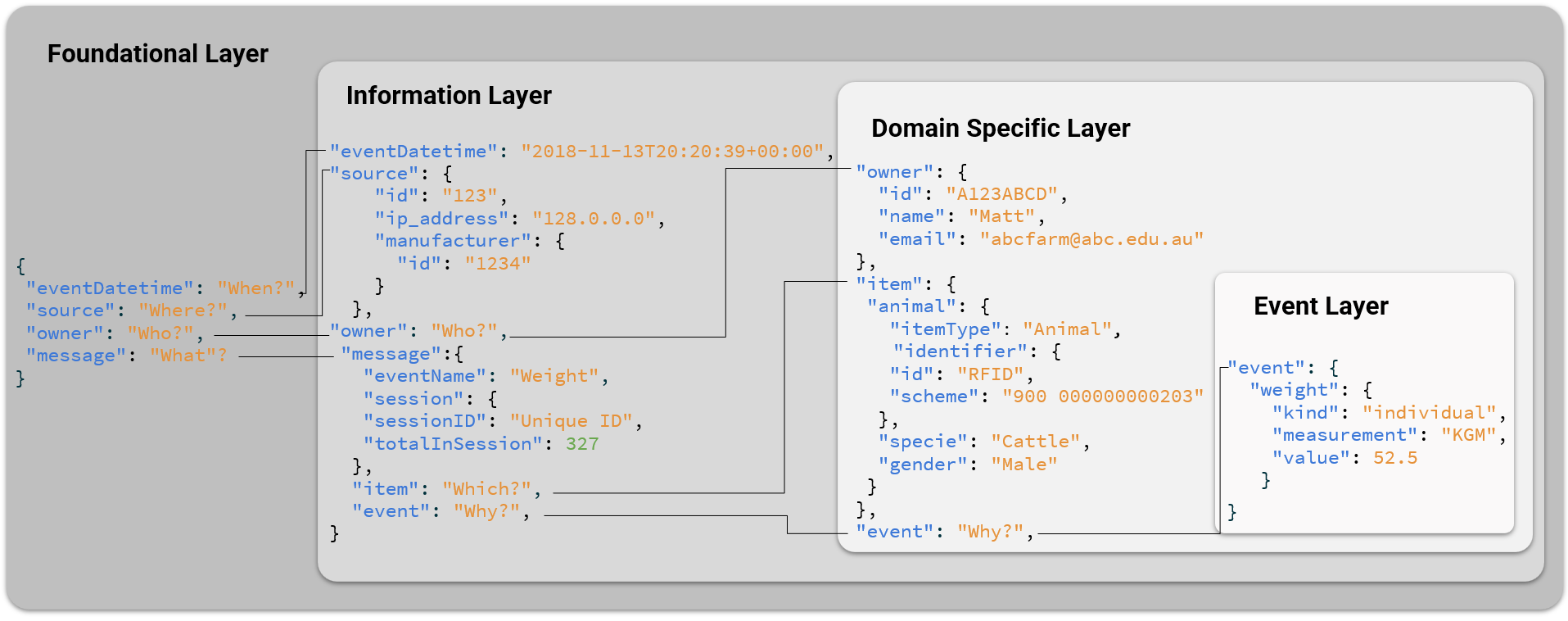}
\caption{LEI layers}
\label{fig:leilayers}
\end{figure}

The LEI schema is structured to address the questions of the W4 model: When, Where, Who and What? (that is, What=What?+Why?). Furthermore, the LEI is layered into four layers: foundational, information, domain-specific, and event layers. Each layer captures different types of information that are relevant to the event.~\Cref{fig:leilayers} shows the four layers of the schema.

As its name suggests, the foundational layer provides the receiver with basic contextual information about the event. This includes four essential properties: \textit{eventDatetime, source, owner,} and \textit{message}. The \textit{eventDatetime} property, which answers the question ``When?", contains the date and time stamp of when the event was recorded and is in ISO 8601 format~\citep{Wolf1997}. This information is crucial for tracking and documenting changes made to the data over time. The \textit{source} property, which answers the question ``Where?", identifies the cause of the event and includes information about the hardware or software that caused the event, such as an IP address. This information helped to understand the origin of the event and the device that recorded it. The \textit{owner's} property, which we questioned by asking ``Who?", includes information about the producer who raises the livestock. This information includes their property identification code (PIC), name, and address. This is important for tracking and identifying the owner of the animals. In addition, this information provides a way to understand the context of animal production. The \textit{message} property, which answers the question ``What?", indicates which events have been captured, such as breeding, vaccinations, or health status. Furthermore, it is divided into four sub-properties: \textit{eventName, session, item,} and \textit{event}. The \textit{eventName} property indicates the name of the captured event, and the data type for it is a string. The \textit{session} property is concerned with providing a unique number for the captured event and the number of animals in that session. The \textit{item's} property, which we questioned by asking ``Which?", contains information about live animals, specifically cattle, which are associated with the event. This information includes the animal's identification, such as RFID or NLISID, type, and description. Identification of the animal is essential for tracking the movements of the animal and accessing the animal's medical history or genetic evaluations. Additionally, the type and description of the animal can help to understand the characteristics of the animal.

The information layer provides high-level summary information about the event that has been captured, in other words, metadata about event information. It includes detailed information about the properties \textit{eventDatetime, source, eventName,} and \textit{session}. Both the \textit{owner} and \textit{item} properties are detailed in the domain-specific layer, which describes the animal and who owns it. Finally, the event layer of the LEI schema is designed to provide detailed information about the event that has been captured. The \textit{event's} property is designed to answer the question ``Why?" as it provides detailed information about the event that has been captured, including the reasons why the event took place, and what was captured. The \textit{event} property is an important component of the LEI schema. This information is specific to the event and can vary depending on the event being captured. For example, a weight event would include information about the weight of the animal, the method used to weigh the animal, and any relevant observations or notes. In contrast, a registration event would include information about the animal's registration number, the date of registration, and the reason for registration.

\subsection{Implementation} \label{sec:leiImplementation}

To facilitate event-based data sharing, the LEI schema extends the existing data standards ICAR and ISC by modifying or redefining properties within the schema. The modification process involves deleting some properties of the ICAR and ISC schemas that are not relevant to the specific use case of the LEI schema. This helps to streamline the data, making it more focused and relevant for the LEI schema. The redefinition process involves both deleting and adding properties to the ICAR and ISC schemas. This allows for a more detailed and specific way of capturing events, tailored to the specific use case of the LEI schema. It also ensures that the captured data is consistent and accurate.~\Cref{tab:leiicariscevents} summarises how the LEI events have been proposed, outlining the modifications and redefinitions made to the ICAR and ISC schemas. In addition to modifying and redefining events, the LEI schema also includes events built from scratch. These events are new and specific to the LEI schema and have not previously been included in the ICAR and ISC schemas.

\setcounter{rownumbers}{0}
\begin{ThreePartTable}
\begin{longtable}{l p{3.5cm}c p{10cm}}
\caption{Events in LEI that have been modified or redefined from ICAR or ISC or that have been proposed}
\label{tab:leiicariscevents}\\

\hline
No. &LEI Event & \begin{sideways}{Action}\; \end{sideways} & Note \\ \hline
\endfirsthead

\multicolumn{4}{c}%
{{ \tablename\ \thetable{} -- continued from previous page}} \\
\hline
No. &LEI Event & \begin{sideways}{Action}\; \end{sideways} & Note \\ \hline
\endhead
\rownumber&Weight & R & 2 properties have been deleted (traitLabel, timeOffFeed), added 2 properties (reason, date), and the ``weight" property redefined to have 3 properties instead of referencing to uncomplete ISC type. \\ 
\rownumber&Score & R & All scores (fat, frame, condition, muscles) have been merged into this event and added 1 property (date). \\
\rownumber&Arrival & R & Merge ``change of ownership" event on it, rename ``consignment" property to ``shipment" and added 1 property (date). \\
\rownumber&Departure & R & Merge ``change of ownership" event on it, rename ``consignment" property to ``shipment" and added 1 property (date). \\ 
\rownumber&Death & M & Deleted only the first part and added 1 property (date). \\
\rownumber&Registration & M & Deleted only the first part and added 1 property (date). \\
\rownumber&Retag & M & Deleted only the first part and added 1 property (date). \\ 
\rownumber&Treatment program & M & Deleted only the first part and added 1 property (date). \\ 
\rownumber&Treatment & M & Deleted only the first part and added 1 property (date). \\ 
\rownumber&Diagnosis & M & Deleted only the first part and added 1 property (date). \\ 
\rownumber&Daily Milking Averages & M & Deleted only the first part and added 1 property (date). \\ 
\rownumber&Feed Intake & M & Deleted only the first part and added 1 property (date). \\ 
\rownumber&Milking Dry Off & M & Deleted only the first part and added 1 property (date). \\ 
\rownumber&Milking Visit & M & Deleted only the first part and added 1 property (date). \\ 
\rownumber&Abortion & R & There were no properties, so 4 properties were added (reason, method, note, date). \\ 
\rownumber&Heat & M & Deleted only the first part and added 1 property (date). \\ 
\rownumber&Insemination & M & Deleted only the first part and added 1 property (date). \\ 
\rownumber&Parturition & M & Deleted only the first part and added 1 property (date). \\ 
\rownumber&Pregnancy Check & M & Deleted only the first part and added 1 property (date). \\ 
\rownumber&Semen Straw & M & Deleted only the first part and added 1 property (date). \\ 
\rownumber&Status Observed & M & Deleted only the first part and added 1 property (date). \\
\rownumber&Lactation Status Observed & M & Deleted only the first part and added 1 property (date). \\ 
\rownumber&Birth & R & Added 3 properties (EID, VID, date) and deleted ``animalDetail". \\
\rownumber&Synchronisation& P & Built from scratch. \\
\rownumber&Weaning & P & Built from scratch. \\
\rownumber&Audit & P & Built from scratch. \\
\rownumber&Castrate & P & Built from scratch. \\
\rownumber&Pulse check & P & Built from scratch. \\
\rownumber&Respiration & P & Built from scratch. \\
\rownumber&Find age by dentition & P & Built from scratch. \\
\rownumber&Hoof trimming & P & Built from scratch. \\
\rownumber&Horn tipping & P & Built from scratch. \\
\rownumber&Dehorning & P & Built from scratch. \\
\rownumber&Location & P & Built from scratch. \\
\hline
\end{longtable}
\begin{tablenotes}
\footnotesize
\item M (Modified), R (Redefined), P (Proposed)\\
\end{tablenotes}
\end{ThreePartTable}

The JSON (JavaScript Object Notation) format is a lightweight data interchange format that is easy for humans to read and write and for machines to parse and generate. It is based on a subset of the JavaScript Programming Language and is commonly used to transmit data between servers and web applications. JSON is typically smaller in file size than XML~\citep{Newman2021}, making it more efficient for data transmission and storage. Developers use JSON to request, access, and serialise data on the server side before transforming it into an object that can be used on the front end. Most of the time, small schemas can be made using primitive JSON data types such as strings~\citep{Michael2022Understanding2020-12}. However, as the system grows, developers may need to add more flexibility to these data types to create reusable object definitions or extend the current schema by adding or modifying properties~\citep{Marrs2017}. Developers can create a ``sub-schema" object definition in different documents but in the same directory containing the ``main schema" and then reference that object using the ``\$ref" keyword.

In the LEI schema, developers created ``sub-schemas" object definitions in different documents under the same directory of the ``main schema" and referenced them using the ``oneOf" and``\$ref" keywords. This allows the reuse of object definitions and reduces the number of duplicate JSON properties shared between diverse schemas. As mentioned, LEI is an extension of ICAR and ISC and references their objects using the ``\$ref" keywords. To implement the LEI schema, some fields need to be added, such as the sessionID and the number of animals in that session. Additionally, most of the existing ISC and ICAR events need a property to show the cattle owner for these events, so the predefined ``iscOrganisationType.json" type is used as the cattle producer.

~\Cref{tab:leidatatypes} shows the data types used in LEI from ICAR or ISC or defined in LEI and their definition. The LEI schema with examples can be accessed through the GitHub repository,\footnote{\url{https://github.com/mahirgamal/LEI-schema}} providing developers with a clear understanding of how the schema has been structured and implemented. 

\begin{figure}[!htb]
\centering
\includegraphics[width=\textwidth,height=\textheight,keepaspectratio]{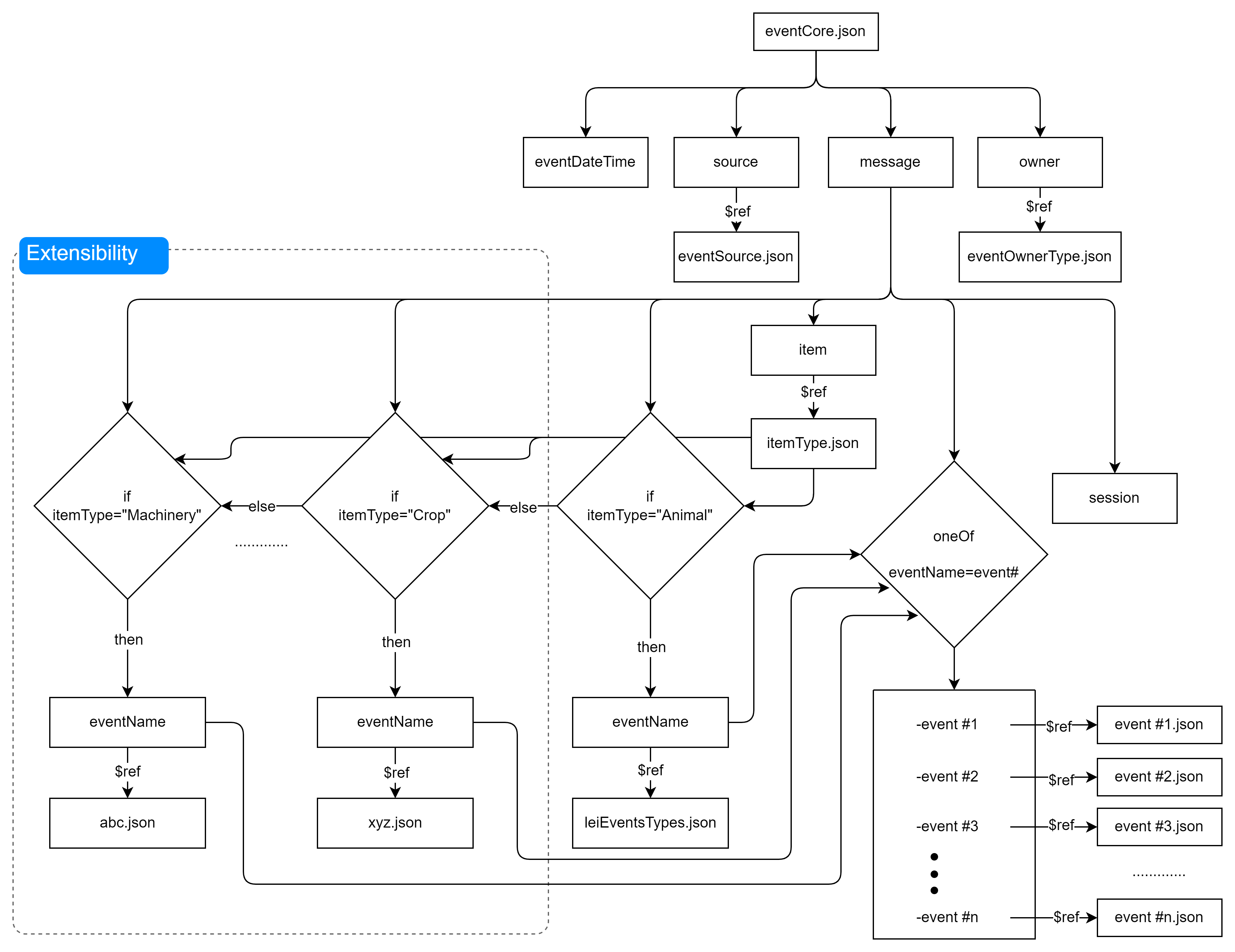}
\caption{LEI's visualisation and layout with embrace of extensibility}
\label{fig:leivisualise}
\end{figure}

~\Cref{fig:leivisualise} visualises the LEI schema with extensibility in future work. The ``eventCore.json" file is a crucial component in the organisation of data within the proposed data schema. It serves as the central hub that connects all properties and events sub-schemas together, allowing for a cohesive and consistent structure for data validation. This file outlines the various fields and their respective data types, as well as any constraints or validation rules that must be met. Using this single JSON schema eliminates the need to validate data against each individual event sub-schema. It also makes it easier to make changes and updates to the schema if needed, since all the information is in one place.

The schema starts by defining the ``\$schema" and ``\$id" fields, which specify the version of the JSON schema being used and the location of the schema file, respectively. The ``type" field is set to ``object", indicating that this schema defines an object with various properties. The ``description" field provides a brief explanation of what the schema represents.
The ``additionalProperties" field is set to ``false", which means that no additional properties that are not defined in the ``properties" field can be added to the object. The ``required" field lists the properties that must be present in the event object, in this case ``source", ``owner", ``eventDateTime", ``message", which are all found in the foundational layer.

In~\Cref{fig:leivisualise}, the JSON file, ``eventCore.json," includes several properties that are key to understanding the event data it contains. The ``source" property references the ``eventSource.json" file, which provides details about the hardware or software that triggered the event. Similarly, the ``owner" property references the ``eventOwnerType.json" file to indicate the producer or farmer responsible for the animal in question. 

The ``message" property is a complex object that must include specific properties, such as ``eventName," ``item," and ``event," as well as an optional ``session" property with ``sessionID" and ``totalInSession" sub-properties. The ``eventName" property is dependent on the ``itemType" property of the ``item" property and can be one of 34 different events. The ``item" property references the ``itemType.json" file, which defines the types of items (i.e., animals, crops, or machinery). The actual structure and properties of the object ``animal" are referenced from ``icarAnimalCoreResource.json" through ``\$ref".

Additionally, the ``message" property in ``eventCore.json" includes \emph{if-then-else} statements to check the value of the ``itemType" field, which is nested several levels deep within the object. If the ``itemType" is ``Animals," then the ``eventName" can only be selected from ``leiEventsTypes.json." This concept extends to other events related to farm fields such as crops and machinery, where ``eventName" can only be selected from specific JSON files such as ``xyz.json" or ``abc.json" respectively.

Finally, the ``message" property includes a ``oneOf" keyword that defines various options for the ``eventName" property. For example, if the ``eventName" is ``Weight," the ``event" must be of type object, and the required definition is found in the external JSON file ``leiWeightEvent.json." Similarly, if the ``eventName" is ``Score," the ``event" must be of type object and is defined in the external JSON file ``leiScoreEvent.json.", and so on.

Furthermore, to accomplish the task of validating the JSON event message in accordance with the JSON event schema, we developed a web application. This web application consists of one web page with two text areas. JSON event schema on the left-hand side and JSON event message on the right-hand side.~\Cref{fig:validotionerror} shows if the validation is correct and~\Cref{fig:validotionnoerror} shows if there is an error in the event message.~\Cref{lst:schemavalidationcode} shows the validation function code that receives two parameters: data (i.e. event message) and schema. It returns a message indicating whether the validation was correct or incorrect. We use Eclipse IDE\footnote{\url{https://www.eclipse.org/ide/}} and Maven\footnote{\url{https://maven.apache.org/}} technology to extend Java libraries such as the JSON schema validator 1.0.63,\footnote{A JSON schema validator that supports draft v4, v6, v7, v2019-09 and v2020-12 \url{https://mvnrepository.com/artifact/com.networknt/json-schema-validator/1.0.63}} jackson.databind 2.12.1,\footnote{General data-binding functionality for Jackson: works on core streaming API \url{https://mvnrepository.com/artifact/com.fasterxml.jackson.core/jackson-databind/2.12.1}} and jackson.core 2.12.1.\footnote{Core Jackson processing abstractions, implementation for JSON \url{https://mvnrepository.com/artifact/com.fasterxml.jackson.core/jackson-core/2.12.1}}

\section{Evaluation}\label{sec:evaluation}
\subsection{Evaluation metrics}

~\Citet{Gomez2021} proposed a set of structural metrics that allowed for a comparison of alternatives for the abstraction of JSON-compatible schemas. These metrics reflect the complexity of the structure and are intended to be used as decision criteria for schema analysis and the design process. These metrics are grouped into five categories: \textit{existence} of types and collections, nesting \textit{depth}, the \textit{width} of properties, \textit{referencing} rate, and \textit{redundancy}.~\Cref{tab:structuralmetrics} summarises those categories and the structural metrics under each category. These metrics are evaluated in the context of collections\footnote{A JSON collection is a group of JSON documents that are stored together in a database or other data storage system. Each document in the collection is a separate JSON object. The collection as a whole is typically used to represent a specific set of data, such as customer records.} (donated as $\varphi$), types\footnote{In JSON, a ``type" refers to the kind of data that a particular attribute or element in a JSON document can contain.} (donated as $t$), or the entire schema (donated as $x$).

\setcounter{rownumbers}{0}
\begin{table}[!ht]
\caption{Structural metrics~\citep{Gomez2021}}
\label{tab:structuralmetrics}
\centering
\resizebox{\textwidth}{!}{%
\begin{tabular}{lllllll}

\hline
 No.&Category&Metric&Description&Schema $x$ &Collection $\varphi$&Type $t$\\
\hline

\multirow{3}{*}{\rownumber} & \multirow{3}{*}{Existence} & \textit{colExistence}&Existence of a collection&&\checkmark& \\
& & \textit{docExistence}& Existence of a document type in a collection&&\checkmark&\checkmark \\
&&\textit{nbrCol}&Number of collections&\checkmark&&\\ 

\hline

\multirow{5}{*}{\rownumber}&\multirow{5}{*}{Depth}&\textit{colDepth}&Maximal depth of a collection& &\checkmark&\\
& &\textit{globalDepth}&Maximal depth of a schema&\checkmark& &\\
& &\textit{docDepthInCol}&Level where a document type is in a collection &&\checkmark&\checkmark \\ 
& &\textit{maxDocDepth}&The deepest level where a document type appears&\checkmark& &\\ 
& &\textit{minDocDepth}&The shallowest level where a document type appears&\checkmark& &\\

 \hline
 
\multirow{5}{*}{\rownumber}&\multirow{5}{*}{Width}& \textit{nbrAtomicAttributes}&Number of attributes of type atomic& & &\checkmark \\ 
& &\textit{nbrDocAttributes}&Number of attributes of type document& & &\checkmark \\
& &\textit{nbrArrayAtomicAttributes}&Number of attributes of type array of atomic values& & &\checkmark\\ 
& &\textit{nbrArrayDocAttributes}&Number of attributes of type array of documents& & &\checkmark\\ 
& &\textit{docWidth}&``Width" of a document type& &\checkmark&\checkmark\\ 

 \hline
 
\rownumber&Referencing&\textit{refLoad} & Number of times that a collection is referenced&& \checkmark&\\

 \hline

\multirow{2}{*}{\rownumber}&\multirow{2}{*}{Redundancy}& \textit{docCopiesInCol}&Estimated nbr of copies of a document type t in a collection& &\checkmark&\checkmark\\ 
& &\textit{docTypeCopies}& Number of times a type is present in the scheme&\checkmark& &\\

 \hline
\end{tabular}
}
\end{table} 

The given set of~\Cref{eq:colExistence,eq:docExistence,eq:nbrCol} describes the existence of a document or a type $t$ in a schema. The choice to create a collection for a type of document is primarily driven by the need for quick or frequent access to the documents of that type or to documents of nested types. On the other hand, nesting one type of document into another may be driven by the fact that the information in these documents is frequently accessed together. To reduce the complexity of collections, it may also be desirable to limit the number of places where a particular type of document is nested.~\citet{Gomez2021} proposed using metrics to determine the existence of a document or a type $t$ in a schema. Two cases are considered: (1) the existence of a collection of type $t$ and (2) the presence of documents of type $t$ nested within other documents. These cases are quantified using the metrics $colExistence(t)$ and $docExistence(\varphi, t)$, respectively.

$colExistence(t)$ measures the presence of a node of type $t$ at level 0 in the schema. It is defined as:

\begin{equation}\label{eq:colExistence}
colExistence (t)=\left\{\begin{array}{cc}
1 & \text { node exists in schema } \mathrm{x} \text { at level } 0 \\
0
\end{array}\right.
\end{equation}

$docExistence(\varphi, t)$ quantifies the nesting of documents of type $t$ and is represented in the graph by a node $*EMBt$. It is defined as:

\begin{equation} \label{eq:docExistence}
docExistence(\varphi, t)=\left\{\begin{array}{lll}
1 & t \in \varphi & \quad \text { node } * E M B t \text { exists in the paths child of node } \varphi \text { in } x\\
0 & t \notin \varphi
\end{array} \right.
\end{equation}

Finally, the metric $nbrCol()$ indicates the number of collections present in the schema, which is equal to the number of child nodes of the root node in the graph and is defined as:

\begin{equation} \label{eq:nbrCol}
nbrCol()=n \quad \text {number of child nodes of root}
\end{equation}

The given set of~\Cref{eq:depth,eq:colDepth,eq:globalDepth,eq:docDepthInCol,eq:docDepth,eq:maxdocDepth,eq:mindocDepth} describes the metrics to measure the level of nesting in a document collection. These metrics are used to assess the costs associated with accessing information that is deeply nested, as well as the costs of restructuring the data to make it more accessible. The deeper the documents are embedded, the higher the cost to access them, unless intermediary information is also required. The complexity induced by the embedded data can be evaluated using a set of metrics.

The metric $colDepth$ measures the level of the most deeply embedded document in a collection. This is calculated by finding the maximum number of $EMB$ nodes in the child paths of a node. The formula is as follows:

\begin{equation} \label{eq:depth}
depth(p)=n \quad \text { number of nodes } E M B \text { in path } p
\end{equation}

\begin{equation} \label{eq:colDepth}
colDepth(\varphi)=\max (depth(p_i)) \quad: p_i \text { is a valid child path of node } \varphi
\end{equation}

The $globalDepth$ metric indicates the deepest nesting level of all collections in a schema. This metric is calculated by taking the maximum value of $colDepth$ among all collections in the schema.

\begin{equation} \label{eq:globalDepth}
globalDepth (x)=\max (col Depth(\varphi_i)) \quad: \forall \text { collection } \varphi_i \in x
\end{equation}

The $docDepthInCol$ metric indicates the level at which a certain type of document appears in a collection. This metric is calculated by finding the deepest level where a document of this type is nested and counting the number of $EMB$ nodes between the root and the $*EMBt$.

\begin{equation}\label{eq:docDepthInCol}
docDepthInCol(\varphi, t)=\left\{\begin{array}{cc}
0 & : \text { the node child of } \varphi \text { has not children } \\
\max \left(docDepth(p_i, t)\right) & : p_i \text { is a valid root-leaf child path of node } \varphi
\end{array}\right.
\end{equation}

\begin{equation}\label{eq:docDepth}
docDepth(p, t)=n \text { number of nodes } E M B \text { between root and } * E M B t
\end{equation}

The $maxDocDepth$ and $minDocDepth$ metrics indicate the deepest and shallowest levels, respectively, where a document type appears in the schema. These metrics are calculated by taking the maximum and minimum values, respectively, of the $docDepthInCol$ metric for each collection in the schema where the type appears.

\begin{equation}\label{eq:maxdocDepth}
\max DocDepth(t)=\max \left(DocDepthInCol(\varphi_i, t\right)) \quad: \varphi_i \in x \wedge t \in \varphi_i
\end{equation}

\begin{equation}\label{eq:mindocDepth}
\min DocDepth(t)=\min \left(Doc DepthInCol(\varphi_i), t\right) \quad: \varphi_i \in x \wedge t \in \varphi_i
\end{equation}

~\Cref{eq:nbrAtomicAttributes,eq:nbrDocAttributes,eq:nbrArrayAtomicAttributes,eq:nbrArrayDocAttributes,eq:docWidth} are used to identify the width of the schema, which is the number of attributes of different types present at a specific level in the schema.~\Cref{eq:nbrAtomicAttributes,eq:nbrDocAttributes,eq:nbrArrayAtomicAttributes,eq:nbrArrayDocAttributes} measure the number of attributes of different types (atomic, document, array of atomic values, and array of documents, respectively) present in a given node.~\Cref{eq:docWidth} computes the width of a node taking into account the number of attributes of different types and their weight \textit{cfAtom} =1, \textit{cfDoc} =2, \textit{cfblAtom} =1 et \textit{cfTblDoc} =3).

\begin{equation} \label{eq:nbrAtomicAttributes}
 nbrAtomicAttributes (t, \varphi)=n \quad \mathrm{n}: \text {number of attributes of type atomic present in type } \mathrm{t}
\end{equation}

\begin{equation}\label{eq:nbrDocAttributes}
nbrDocAttributes (t, \varphi)=n \quad \mathrm{n}: \text {number of attributes of type document}
\end{equation}

\begin{equation}\label{eq:nbrArrayAtomicAttributes}
nbrArrayAtomicAttributes (t, \varphi)=n \quad \mathrm{n}: \text {number of attributes of type array of atomic values}
\end{equation}

\begin{equation}\label{eq:nbrArrayDocAttributes}
nbrArrayDocAttributes (t, \varphi)=n \quad \mathrm{n}: \text {number of attributes of type array of documents}
\end{equation}

\begin{equation} \label{eq:docWidth}
\begin{aligned}
docWidth (t, \varphi)= & c f \text { Atom } * \text { nbrAtomic Attributes }(t, \varphi)+ \\
& c f \text { Doc } * \text { nbrDocAttributes }(t, \varphi)+ \\
& c f \text { Tbl Atom } * \text { nbrArrayAtomicAttributes }(t, \varphi)+ \\
& c f \text { TblDoc } * \text { nbrArrayDocAttributes }(t, \varphi)
\end{aligned}
\end{equation}

~\Cref{eq:refLoad} measures the reference of a given node by counting the number of nodes that reference other nodes in the schema.

\begin{equation}\label{eq:refLoad}
refLoad(\varphi)=n \quad n: \text { number of nodes } * R E F t \text { where } t \text { is child of } \varphi
\end{equation}

The set of~\Cref{eq:docCopiesInCol,eq:card,eq:docTypeCopies} is used to estimate the redundancy in the schema. The goal is to estimate the potential for data redundancy during schema design to inform developers. Although data redundancy can provide faster access and avoid certain operations, it has negative impacts, such as an increased memory footprint, more difficulty in enforcing coherency, and reduced maintainability of the application.

~\Cref{eq:docCopiesInCol,eq:card} are used to identify redundancy by counting the number of copies of a document in the schema.

\begin{equation}\label{eq:docCopiesInCol}
docCopiesInCol(t, \varphi)=\left\{\begin{array}{cl}
0 & : t \notin \varphi \text { docExistence }(\varphi, t)=0 \\
1 & : \text { the child node of } \varphi \text { has not children } \\
\prod \operatorname{card}\left(r_{r o l}, t\right) & : \begin{array}{c}
r_{r o l} \text { father of a node } E M B \\
\text { of the path } \varphi \text { and } * E M B t
\end{array}
\end{array}\right.
\end{equation}

\begin{equation}\label{eq:card}
card(r, \varepsilon)=n \quad n: \text { cardinality of } r \text { on the } \varepsilon
\end{equation}

~\Cref{eq:docTypeCopies} indicates the number of times a document type is used in the schema and reflects the potential number of structures that can store documents of type $t$. This metric is calculated using the metric of $docExistence$ already introduced in~\Cref{eq:docExistence}.

\begin{equation}\label{eq:docTypeCopies}
\text {docTypeCopies}(t)=\sum_{i=1}^{|x|} docExistence \left(\varphi_i, t\right)
\end{equation}

\subsection{Evaluation scenario}

\begin{figure}[!ht]
\centering
\begin{subfigure}{\textwidth}
\includegraphics[scale=0.48, angle=0]{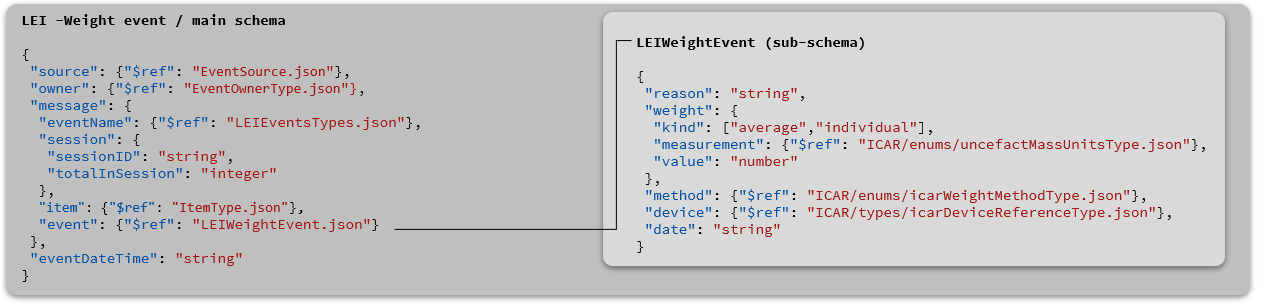}
\caption{LEI Weight event's JSON schema}
\label{fig:leiicariscweightevent:lei}
\end{subfigure}
\begin{subfigure}{1\textwidth}
\includegraphics[scale=0.48, angle=0]{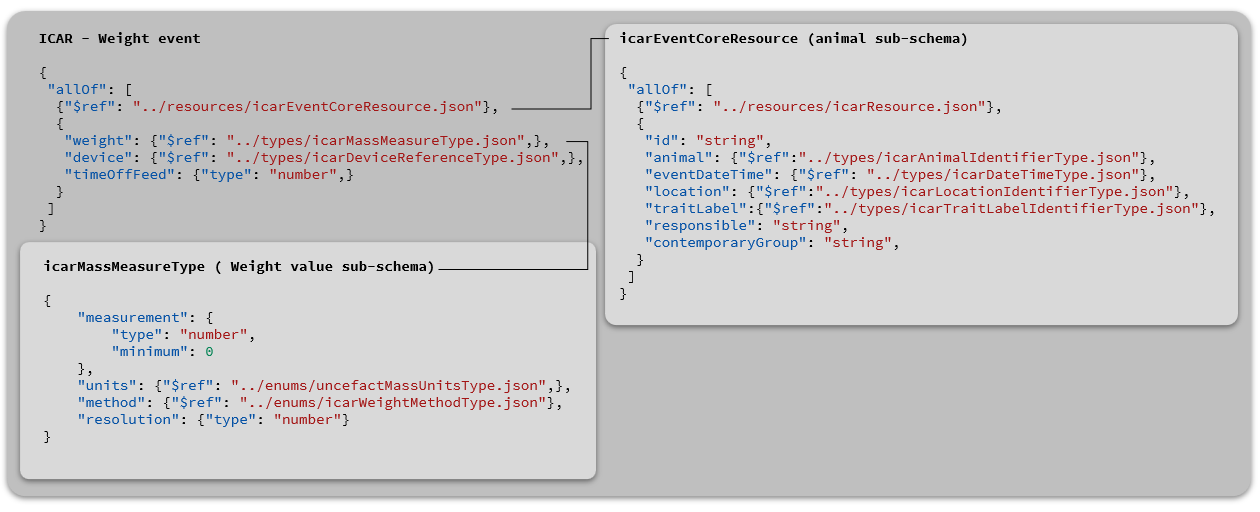}
\caption{ICAR Weight event's JSON schema}
\label{fig:leiicariscweightevent:icar}
\end{subfigure}
\begin{subfigure}{1\textwidth}
\includegraphics[scale=0.48, angle=0]{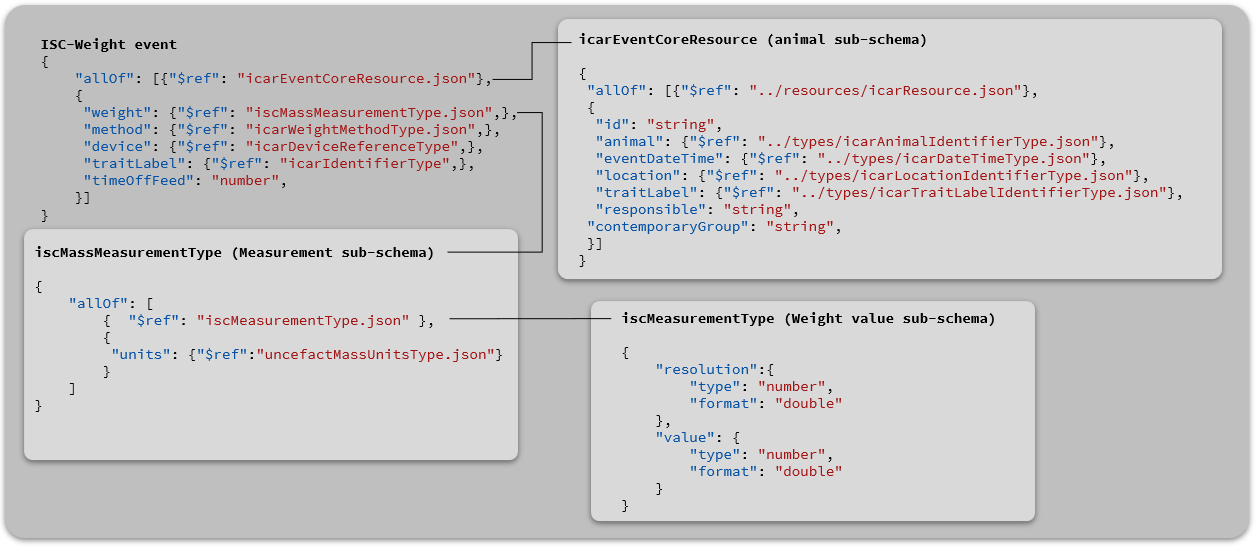}
\caption{ISC Weight event's JSON schema}
\label{fig:leiicariscweightevent:isc}
\end{subfigure}
\caption{Weight Event data for LEI, ICAR, and ISC schemas}
\label{fig:leiicariscweightevent}
\end{figure}

According to the findings of~\citep{Gomez2021}, an event schema must be selected to compare LEI, ICAR, and ISC to ensure integrity. This chosen event must be consistent across all three schemas. The selection process excluded events that were not present in either ICAR or ISC or events that were not fully mature, thereby ensuring that the comparison is based on well-established and mature events. Ultimately, the \textit{weight} event was selected as the evaluation factor due to its presence in the three schemas, providing a fair evaluation through a common basis for comparison.~\Cref{fig:leiicariscweightevent:lei,fig:leiicariscweightevent:icar,fig:leiicariscweightevent:isc} show the \textit{weight} event schema of LEI, ICAR, and ISC, respectively. 

To facilitate the sharing of livestock data and ensure that all necessary information is available to consumers, an evaluation scenario has been established using a set of criteria and the corresponding metrics. Of the 16 available metrics, 10 were selected for evaluation.~\Cref{tab:criteriametrics} illustrates the criteria that have been chosen, which encompasses efficient access to the \textit{weight} event and its properties, consistency of data, and ease of managing instances of the \textit{weight} event's properties, such as \textit{source}, \textit{session}, \textit{owner}, \textit{uncefactMassUnits}, \textit{eventDateTime}, and \textit{eventName}. 

\setcounter{rownumbers}{0}
\begin{table}[htb]
\caption{Criteria and metrics used in the evaluation scenario}
\label{tab:criteriametrics}
\centering
\begin{tabular}{lll}
\hline
 Criterion & Description & Metric\\
\hline
Criterion \rownumber & Favour the existence of the collection \textbf{\textit{weight}} event & \textit{colExistence} \\ 
Criterion \rownumber & Favour copies of documents \textbf{t\textit{source}} & \textit{docCopies} \\ 
Criterion \rownumber & Favour copies of documents \textbf{t\textit{session}} & \textit{docCopies} \\ 
Criterion \rownumber & Favour copies of documents \textbf{t\textit{owner}} & \textit{docCopies} \\ 
Criterion \rownumber & Favour references to the collection \textbf{\textit{uncefactMassUnitsType}} & \textit{refLoad} \\ 
Criterion \rownumber & Reduce the complexity of the collection \textbf{t\textit{weight}} event & \textit{docWidth} \\ 
Criterion \rownumber & Favour the nesting of \textbf{t\textit{eventDateTime}} in \textbf{\textit{weight}} event & \textit{docDepthInCol} \\
Criterion \rownumber & Favour the existence of the collection \textbf{t\textit{eventName}} in \textit{weight} & \textit{docExistence} \\ \hline
\end{tabular}
\end{table}

To determine the priority criteria for the structuring of the data in the evaluation scenario, an analysis of access priorities and additional information was conducted. These criteria include the significance of the \textit{weight} event as a data source (Criterion 1) in the three schemas, the ease of managing instances of \textit{weight} data (Criterion 6), and the accessibility of the \textit{eventDateTime} property through the \textit{weight} event (Criterion 7). Additionally, the consistency of \textit{weight's} \textit{source}, \textit{session}, and \textit{owner} were deemed to be of the utmost importance (Criteria 2, 3, and 4), therefore the objective was to minimise the number of copies in the \textit{weight} event. On the other hand, accessibility of \textit{eventName} was deemed a priority (Criterion 8). Lastly, the reference to \textit{uncefactMassUnitsType} was deemed a priority (Criterion 5).

Once the relevant criteria have been established, the associated metrics were identified on the basis of the structural aspects involved. Criteria 1 and 8, which depend on the presence of \textit{weight} as a collection and \textit{eventName} as type, are associated with the metrics \textit{colExistence} and \textit{docExistance}, respectively. To ensure consistency in the information of \textit{weight} across possible copies of \textit{source}, \textit{session}, and \textit{owner}, the metric \textit{docCopies} is used (criterion 2, 3, and 4). To determine the presence of \textit{eventDateTime} type within the \textit{weight} collection, the metric \textit{docDepthInCol} was proposed, which allows the assessment of nesting of the type of document within a collection (criterion 7). The metric \textit{docWidth} was used to determine the number of attributes in the \textit{weight} collection (criterion 6), while the metric \textit{refLoad} was used to count the number of references to \textit{uncefactMassUnitsType} (criterion 5).

\subsection{Results}\label{sec:results}
To assess the metric\textit{colExistence(tweight)}, the value of 1 was assigned to the LEI, ICAR, and ISC schemas as they all contain the weight event. For the metrics \textit{docCopiest(tsource)}, \textit{docCopiest(tsession)}, and \textit{docCopiest(towner)}, a value of 1 was assigned to the LEI schema, while the fields for ICAR and ISC were left empty since they do not have \textit{source}, \textit{session}, and \textit{owner} types. For \textit{refLoad (uncefactMassUnitsType)}, the value of 1 was assigned to all schemas, as each schema has only one reference to the \textit{uncefactMassUnitsType} type.

To evaluate the metric \textit{docWidth (weight, l1)}, which indicates the number of attributes for the collection of weight at the first level,~\Cref{eq:docWidth} was applied to the LEI schema. The calculation results showed that LEI has \textit{source}, \textit{owner}, and \textit{message} as document types and \textit{eventDateTime} as an atomic type. The calculation can be represented as:
\[nbrAtomicAttributes (t, \varphi)=1\]
\[nbrDocAttributes(t, \varphi)=3\]
\[nbrArrayAtomicAttributes(t, \varphi)=0\]
\[nbrArrayDocAttributes(t, \varphi)=0\]
\[cfAtom =1, cfDoc =2, cfblAtom =1 et cfTblDoc =3\]
\[docWidth(t, \varphi)=1*1+2*3+1*0+3*0=6\]
It should also be noted that ``allOf" is a type of inheritance of all properties in the JSON schema. The width calculation for ICAR can be represented as:
\[nbrAtomicAttributes (t, \varphi)=4\]
\[nbrDocAttributes(t, \varphi)=6\]
\[nbrArrayAtomicAttributes(t, \varphi)=0\]
\[nbrArrayDocAttributes(t, \varphi)=0\]
\[cfAtom =1, cfDoc =2, cfblAtom =1 et cfTblDoc =3\]
\[docWidth(t, \varphi)=1*4+2*6+1*0+3*0=16\]
And for ISC:
\[nbrAtomicAttributes (t, \varphi)=4\]
\[nbrDocAttributes(t, \varphi)=8\]
\[nbrArrayAtomicAttributes(t, \varphi)=0\]
\[nbrArrayDocAttributes(t, \varphi)=0\]
\[cfAtom =1, cfDoc =2, cfblAtom =1 et cfTblDoc =3\]
\[docWidth(t, \varphi)=1*4+2*8+1*0+3*0=20\]

For the metric \textit{docDepthInCol(teventDateTime,weight)}, a value of 1 was assigned to all schemas since \textit{eventDateTime} is located at a depth of 1. For the metric \textit{docExistence(teventName,weight)}, a value of 1 was assigned to the LEI schema as it has the \textit{eventName} present. The absence of \textit{eventName} in the ICAR and ISC schemas was indicated by empty fields. These results are summarised in~\Cref{tab:subsetmetricsLEIICARISC}.

\setcounter{rownumbers}{0}
\begin{table}[!ht]
\caption{Subset of metrics of schema LEI, ICAR, and ISC}
\label{tab:subsetmetricsLEIICARISC}
\centering
\begin{tabular}{lllll}
\hline 
Criterion & Metric & LEI & ICAR & ISC \\ 

\hline
Criterion \rownumber & \textit{colExistence}(t\textit{weight}) & 1 & 1 & 1\\ 
Criterion \rownumber & \textit{docCopiest}(t\textit{source}) & 1 & - & -\\
Criterion \rownumber & \textit{docCopiest}(t\textit{session}) & 1 & - & -\\ 
Criterion \rownumber & \textit{docCopiest}(t\textit{owner}) & 1 & - & -\\ 
Criterion \rownumber & \textit{refLoad}(\textit{uncefactMassUnitsType}) & 1 & 1 & 1\\ 
Criterion \rownumber & \textit{docWidth}(\textit{weight,l1}) & 6 & 12 & 20\\ 
Criterion \rownumber & \textit{docDepthInCol}(t\textit{eventDateTime,wight}) & 1 & 1 & 1\\ 
Criterion \rownumber & \textit{docExistence}(\textit{teventName},weight) & 1 & - & -\\
\hline 
\end{tabular}
\end{table}

Each criterion is formalised as a function whose value is maximised or minimised. The values of the criteria were normalised between 0 and 1 except for criterion 6, which evaluates the ``width" of a document by combining a factor with the property types that it may contain. So, in the sixth criterion, LEI possesses 6 attributes, ICAR possesses 12 attributes, and ISC possesses 20 attributes at level 1. Furthermore, the empty box translated to 0. These numbers are translated as potential value (1/ number of collections) and can be found in~\Cref{tab:criteriaevaluationLEIICARISC}. 

\setcounter{rownumbers}{0}
\begin{table}[!ht]
\caption{Criteria evaluation on LEI, ICAR and ISC}
\label{tab:criteriaevaluationLEIICARISC}
\centering
\begin{tabular}{lllll}
\hline 

Criterion& & LEI & ICAR & ISC \\ 
\hline

Criterion \rownumber & {\(f_{c1}(s) = colExistenceWeight \; (s)\)} & 1 & 1 & 1 \\ 
Criterion \rownumber & {\(f_{c2}(s) = docCopiestSource \; ^{max} \; (s)\)} & 1 & 0 & 0 \\  
Criterion \rownumber & {\(f_{c3}(s) = docCopiestSession \; ^{max} \; (s)\)} & 1 & 0 & 0 \\ 
Criterion \rownumber & {\(f_{c4}(s) = docCopiestOwner \; ^{max} \; (s)\)} & 1 & 0 & 0 \\ 
Criterion \rownumber & {\(f_{c5}(s) = refLoaduncefactMassUnitsType \; s^{max} \; (s)\)} & 1 & 1 & 1 \\ 
Criterion \rownumber & {\(f_{c6}(s) = levelWidthWeightL1 \;s^{min} \; (s)\)} & 0.17 & 0.08 & 0.05 \\ 
Criterion \rownumber & {\(f_{c7}(s) = docEventDateTimeInWeight \;s ^{min} \; (s)\)} & 1 & 1 & 1 \\ 
Criterion \rownumber & {\(f_{c8}(s) = docExistenceEventName \; (s)\)} & 1 & 0 & 0 \\
\hline 
\end{tabular}
\end{table}

To maintain consistency with the approach presented by~\citet{Gomez2021}, the following five cases are proposed for comparison:
\begin{itemize}
\item In case 1, we assigned the same weight score to each of the eight criteria, which came out to \(\frac{100}{8} = 12.5\) total points. 
\item In case 2, the criteria pertaining to the simplicity of utilising the \textit{weight} extension were given priority.
\item In case 3, we gave referring to the \textit{uncefactMassUnits} property additional priority.
\item In case 4, we assigned additional priority to the existence of the document of type \textit{eventName} property.
\item In case 5, we gave the \textit{source}, the \textit{session}, and the \textit{owner} equal weight in terms of importance.
\end{itemize}

~\Cref{tab:setofweight} summarises the weight scores assigned to each of the eight criteria in five different cases. In case 1, each criterion is given an equal weight score of 12.5 points. In case 2, the criteria pertaining to the simplicity of using the \textit{weight} extension are given priority, with criteria 1 given 50 points. In case 3, referring to the \textit{uncefactMassUnits} property is given additional priority, with criterion 5 given 20 points. In case 4, the existence of a document of type \textit{eventName} property is given additional priority, with criterion 8 given 20 points. In case 5, the \textit{source}, the \textit{session}, and the \textit{owner} are given equal weight in terms of importance, with criteria 2, 3, and 4 each given 6.67 points. The total weight score for each case is 100 points.

\begin{table}[!ht]
\caption{Weight scores assigned to criteria in five different cases}
\label{tab:setofweight}
\centering
\begin{tabular}{llllll}
\hline

Criterion & Case 1 & Case 2 & Case 3 & Case 4 & Case 5\\
\hline
Criterion 1 & 12.50 & 50.00 & 30.00 & 30.00 & 30.00 \\ 
Criterion 2 & 12.50 & 0 & 0 & 0 & 6.667 \\ 
Criterion 3 & 12.50 & 0 & 0 & 0 & 6.667 \\ 
Criterion 4 & 12.50 & 0 & 0 & 0 & 6.667 \\ 
Criterion 5 & 12.50 & 0 & 20.00 & 0 & 0 \\
Criterion 6 & 12.50 & 15.00 & 15.00 & 15.00 & 15.00 \\ 
Criterion 7 & 12.50 & 35.00 & 35.00 & 35.00 & 35.00 \\ 
Criterion 8 & 12.50 & 0 & 0 & 20.00 & 0 \\ 
\hline
\end{tabular}
\end{table}

The evaluation of a \textit{weight} event schema, represented by the function $schemaEvaluation$, is calculated as the weighted sum of the criteria, where the weight of each criterion reflects its importance. A high evaluation score indicates that the \textit{weight} event schema is well suited to the requirements under consideration. The formula for evaluating the schemas is as follows:
\begin{equation} \label{eq:schemaEvaluation}
schemaEvaluation \; (s)=\sum_{i=1}^{carteria} weight \; _{criterion \; i}*f_{ci} \; (s)
\end{equation}

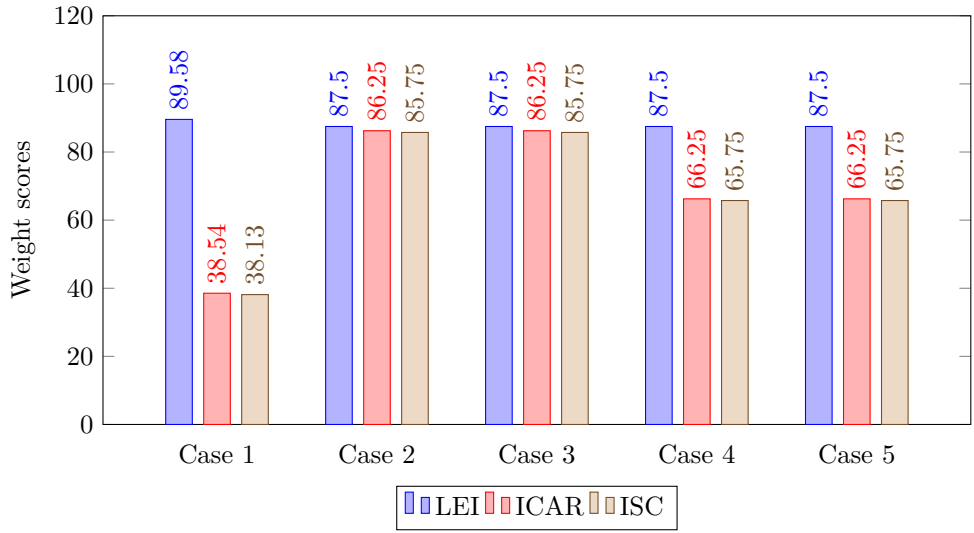
\begin{figure}[!ht]
\centering
\begin{tikzpicture}
\begin{axis}[
ybar,
bar width=0.35cm, 
symbolic x coords={Case 1, Case 2, Case 3, Case 4, Case 5},
xtick=data,
ymin=0,
ymax=120,
ylabel={Weight scores},
xlabel={},
legend style={at={(0.5,-0.15)},anchor=north,legend columns=-1},
nodes near coords,
every node near coord/.append style={rotate=90,anchor=west},
xtick style={draw=none},
ytick distance=20,
enlarge x limits={abs=1.5cm},
width=13cm,
height=7 cm]
\addplot+[bar shift=-0.5cm] coordinates {
    (Case 1, 89.58)
    (Case 2,87.50)
    (Case 3,87.50)
    (Case 4,87.50)
    (Case 5,87.50)
};

\addplot+[bar shift=0cm] coordinates {
    (Case 1,38.54)
    (Case 2,86.25)
    (Case 3,86.25)
    (Case 4,66.25)
    (Case 5,66.25)
};
\addplot+[bar shift=0.5cm] coordinates {
    (Case 1,38.13)
    (Case 2,85.75)
    (Case 3,85.75)
    (Case 4,65.75)
    (Case 5,65.75)

};
\legend{LEI, ICAR, ISC}
\end{axis}
\end{tikzpicture}
\caption{Comparison of schema evaluation scores for three schemas across five cases}
\label{fig:schemasevaluationbycases}
\end{figure}

After applying the formula \ref{eq:schemaEvaluation},~\Cref{fig:schemasevaluationbycases} displays the overall total for the three schemas and the five cases linked to each schema. The findings obtained by LEI are superior to those obtained by ICAR and ISC in terms of the schema structure for existence, referencing, depth, and width.

The figure compares the three schemas broken down into five individual cases. Since each of the eight criteria is given the same weight in case 1, it is abundantly evident that LEI achieves superior results. In addition, the weights of the three schemas are extremely close to one another in case 2 (the situation in which the \textit{weight} property of the schemas was given priority). Similarly, the weight scores of the three schemas are extremely close to one another in case 3, which assesses the reference to the \textit{uncefactMassUni} property. With significant weight scores, LEI prevailed over ICAR and ISC in case 4 by giving \textit{eventName} assistance more priority. Similarly, case 5 equated the relevance of the \textit{source}, \textit{session}, and \textit{owner} properties to have equal weight.

In short, it is unequivocally demonstrated that LEI achieves better results in cases 1, 2, 3, 4, and 5. To put it in perspective, ICAR came in at a respectable second place. At the same time, ISC finished last and had the lowest weight score.

\section{Case Study}\label{sec:casestudy}
We conducted a case study to assess the applicability and usefulness of the LEI schema compared to the ICAR and ISC. The case study is based on a hypothetical experience in livestock management on the CSU farm. Our case study comprises 14 scenarios, each of which contains one or more events. These scenarios were chosen to cover the different types of livestock events.

There are four producers and one abattoir in the case study. Producer A’s PIC is A123ABCD. He is a breeder with 105 breeding cows with 100 calves at foot and 100 yearling stock (50 heifers and 50 steers). Producer A also has three bulls for the next breeding season in 2021. Producer B is a backgrounder with a PIC of B123ABCD. He buys weaned cattle and raises them until they are big enough to go to a feedlot. He has 200 steers on the property right now. Producer C has a PIC of C123ABCD and runs a small feedlot selling steers to a local domestic processor. He has 100 steers in the feedlot right now. Producer D’s PIC is D123ABCD, and he runs a small cattle stud as a seed stock producer. He keeps 100 cows, 50 heifers, 50 bull calves, and 30 bulls on his land. The following scenarios can be formed from the above use case description:

\textbf{Scenario \#1 (departure for sale, arrival for purchase, and change of ownership):} Producer A sold 50 steers to producer B on January 17, 2021, and the animals arrived at producer B's property on January 19, 2021. In this scenario, three events are involved in the sale of 50 steers from producer A to producer B. The first event is the \textit{departure} event for sale, which documents that the cattle have left producer A's farm for the purpose of being sold. The second event is the arrival event for \textit{purchase}, which records that the cattle have arrived at producer B's farm as they have been purchased. The third event is the \textit{change of ownership} event, which documents the transfer of ownership from producer A to producer B. These three events are crucial to tracking the movement and ownership of cattle and maintaining proper record keeping.~\Cref{departure1,arrival1} show the JSON code for the departure and arrival events, respectively, which also include the change of ownership event.

\nolinenumbers
\begin{tcolorbox}[
breakable,
  arc=3mm,
  outer arc=3mm,
  boxrule=0.0mm,
  boxsep=0mm,
  left=3mm,
  right=3mm,
  top=3mm,
  bottom=3mm,
]
\begin{lstlisting}[language=json,caption= {Movement event - Departure for sale},label={departure1}]
{
  "eventDateTime": "2021-01-17T14:13:23Z",
  "source": {
    "id": "CSUFarm-Computer",
    "ip_address": "128.0.0.0",
    "manufacturer": {
      "id": "DELL"
    }
  },
  "owner": {
    "id": "A123ABCD",
    "name": "Producer A Beef",
    "email": "prodA@beef.com",
    "givenName": "Aaron",
    "familyName": "Farmer"
  },
  "message": {
    "eventName": "Departure",
    "item": {
      "itemType": "Animal",
      "animal": {
        "identifier": {
          "id": "982 123456789101",
          "scheme": "rfid"
        },
        "specie": "Cattle",
        "gender": "Male"
      }
    },
    "session": {
      "sessionID": "un1Que1d1",
      "totalInSession": 50
    },
    "event": {
      "departureKind": "Sale",
      "departureReason": "Sale",
      "shipment": {
        "id": {
          "scheme": "sale invoice",
          "id": "123456789"
        },
        "origin": {
          "id": "A123ABCD",
          "name": "Producer A Beef",
          "email": "prodA@beef.com"
        },
        "destination": {
          "id": "B123ABCD",
          "name": "Producer B 4 Beef",
          "email": "b4beef@beef.com"
        },
        "vehicles": [
          "ABC123",
          "CBA321"
        ]
      },
      "changeOfOwnership": {
        "lot": {
          "lotId": "1256"
        },
        "purchaser": {
          "id": "B123ABCD",
          "name": "Producer B 4 Beef",
          "email": "b4beef@beef.com"
        },
        "transactionCost": {
          "currency": "AUD",
          "totalCost": 42250
        }
      }
    }
  }
}
\end{lstlisting}
\end{tcolorbox}

\textbf{Scenario \#2 (Movement for agistment):} Producer A received 50 weaned heifers from producer D on January 30, 2021, which had been transferred from producer D's property on January 27. In this scenario, the purpose of this movement is agistment, which refers to removing animals from a property for various reasons, such as fire restoration, lack of feed, drought, or finishing animals for sale on better quality feed. The scenario involves two movement events: the \textit{departure} event for agistment, which shows that the cattle left producer D's farm, and the \textit{arrival} event for agistment, which shows that the cattle arrived at producer A's farm.~\Cref{departure2,arrival2} provide the JSON code for these events. This scenario is similar to scenario \#1 but the purpose of the movement is different. In this case, the cattle are moved to another farm for agriculture, but the ownership is not transferred, while in scenario \#1, the cattle are moved for sale and change of ownership.

\textbf{Scenario \#3 (Death for injury, status observed, and weight):}Producer C purchased 100 steers from producer B on February 2, 2021. On arrival in the paddock of producer C on 5 February 2021, one of the steers had been injured during transport and had to be euthanised. Producer C then provides producer B with a report containing the induction weights of each animal that serves as the basis for its evaluation. There are three events in this scenario. First, the \textit{death} of an animal, which occurred due to an injury while transported from property B to property C, and the animal was euthanised by producer C using a humane method. Second, the \textit{observed status} event occurred, so producer C observed that one of the incoming steers was injured Third, the \textit{weight} event using scales located in cattle yards in property C.~\Cref{death1,statusobserved,weight} show the JSON code for these three events, respectively.

\textbf{Scenario \#4 (Audit):} Producer D discovered three steers on his farm that were not registered with his PIC and did not belong to him during a stock audit on February 14, 2021. An \textit{audit} event counts the animals on the farm to find which animals belong to the farm and which do not. The audit was carried out at the cattle yards by scanning each animal's individual tag in the race. Thus, all animals were recorded and tags of animals that did not belong to that property were identified. The JSON code for the audit event is shown in~\Cref{audit}.

\textbf{Scenario \#5 (Death for slaughter):} Twenty-five steers were slaughtered in the abattoir (PIC code E123ABCD) on March 1, 2021. This scenario is similar to the first event of scenario \# 3, death for injury, but the purpose of death is different. In scenario \#3, the death was due to the animal being euthanised after an injury that occurred during transport, while in this scenario, the \textit{death} is for human consumption.~\Cref{death2} provides the JSON code for this event. 

\textbf{Scenario \#6 (Synchronisation):} As the time to breed has arrived, producer A planned to perform artificial insemination on 50 heifers in his herd. On 1 October 2021, the cue mate (that is, the brand of the device) was implanted in the heifers after consulting with a representative of the company that specialises in artificial breeding, Artificial Breeding Services. The \emph{Synchronisation} event occurs before the artificial insemination event (see scenario \#7), and farmers frequently use artificial insemination when they have many livestock (that is, 500 head). The JSON code for this event is in~\Cref{synchronisation}.

\textbf{Scenario \#7 (Artificial insemination):} Producer A completed the AI for the 50 heifers on October 11, 2021. In this scenario, the artificial \textit{insemination} event records the process of inserting sperm directly into the uterus of the heifers. This information is captured using a JSON code, which is listed as~\Cref{insemination1}. Artificial insemination is a common method used in the livestock industry to increase breeding efficiency and improve herd genetics. 

\textbf{Scenario \#8 (Natural insemination):} At the same time, on October 1, 2021, producer A arranged the natural mating of 1 bull and 20 cows. This scenario is similar to scenario \#7, where artificial insemination is used, but in this case, the \textit{insemination} is done through the bull. The JSON code for this event can be found in~\Cref{insemination2}.

\textbf{Scenario \#9 (Pregnancy check):} On December 20, 2021, almost ten weeks later, producer A performed pregnancy checks. Producer A discovered 10 pregnant heifers and 10 pregnant cows. The \emph{pregnancy check} event always follows the insemination event, which indicates the number of days of pregnancy and the sex of the embryo. This information is recorded using a JSON code, which can be found in~\Cref{pergnancycheck}.

\textbf{Scenario \#10 (Parturition and birth):} On 29 September 2022, the 10 heifers belonging to Producer A gave birth to five calves and five heifers. In addition, the 10 cows gave birth to eight calves and two heifers.
Two events were captured in this scenario: the first is \emph{parturition} which is when a cow gives birth, and this event is used to determine the pedigree of a born calf. The second is \emph{birth}, which is about the newborn calf, and often the producer at the beginning uses a local identification code or tag to mark the new calf. We refer to it in the schema as VID, which stands for visual identification code. The relevant JSON code for both events can be found in~\Cref{birth,parturition}, respectively.

\textbf{Scenario \#11 (Registration):} Producer A earmarked 20 young animals with RFID tags in the range 900 000000000203-900 000000000222 on September 30, 2022.
In this scenario, a \emph{registration} event is captured. The farmer complies with the NLIS regulation by ordering new RFIDs and linking them to the VIDs of newly born calves or older RFIDs, if any have been lost. The JSON code for this event can be found in~\Cref{registration}.

\textbf{Scenario \#12 (Treatment):} On 29 December 2022, producer A injected Dectomax (batch number 1122346T, expiration date April 4, 2023) into each of the calves on his farm. As a result, all animals are subject to a 42-day withholding period. In this scenario, and because it includes information about the animals, their vaccines, the doses they received, the date of expiration, and the person who administered the injections, the \emph{treatment} event is regarded as one of the essential events. The farmer's priority while purchasing animals from another farmer is to consider this kind of event since it impacts the animal's health. The JSON code for this event can be found in~\Cref{treatment}.

\textbf{Scenario \#13 (Castration):} Producer A surgically \emph{castrated} 5 calves with a scalpel on 29 January 2023, while the remaining eight calves were castrated with rubber rings. Producer A performed the castration.
In practice, \textit{castration} involves removing the testicles of male calves. Castration is one of the husbandry procedures performed during calf marking. This scenario contains a wealth of information, including the type of castration, the methods, and the performer. Accordingly, this type of event provides health-related information that affects the animal's health status. The JSON code for this event can be found in~\Cref{castrated}.

\textbf{Scenario \#14 (Weaning):} Producer A weaned five calves on 29 April 2023. The weaning date of a calf depends on various factors, such as the breed of the calf, the producer's management practices, and the operation's production goals. The \emph{weaning} event in this scenario occurs due to age. The JSON code for this event can be found in~\Cref{weaning}.

\setcounter{rownumbers}{0}
\begin{table}[!ht]
\begin{ThreePartTable}
\caption{Comparison of livestock event information captured by LEI, ICAR, and ISC schemas for 14 case study scenarios}
\label{tab:casestudyresults}
\centering
\begin{tabular}{clccc}

\hline
\multicolumn{1}{l}{No.} & \multicolumn{1}{l}{Case study event} & \multicolumn{1}{c}{ICAR} & \multicolumn{1}{c}{ISC} & \multicolumn{1}{c}{LEI}\\ 
\hline

\rownumber & Departure & $\sim$ & $\sim$ & \checkmark\\
\rownumber & Arrival & $\sim$ & $\sim$ & \checkmark\\
\rownumber & Death & $\sim$ & $\sim$ & \checkmark \\
\rownumber & Status observed & $\sim$ & x & \checkmark\\
\rownumber & Weight & $\sim$ & $\sim$ & \checkmark\\
\rownumber & Audit & x & x & \checkmark\\
\rownumber & Synchronisation & x & x & \checkmark\\
\rownumber & Insemination & $\sim$ & x & \checkmark\\
\rownumber & Pregnancy check & $\sim$ & x & \checkmark\\
\rownumber & Birth & $\sim$ & x & \checkmark\\
\rownumber & Parturition & $\sim$ & x & \checkmark\\
\rownumber & Registration & $\sim$ & $\sim$ & \checkmark\\
\rownumber & Weaning & x & x & \checkmark\\
\rownumber & Treatment & $\sim$ & $\sim$ & \checkmark\\
\rownumber & Castration & x & x & \checkmark\\
\hline

\end{tabular}
\begin{tablenotes}
\footnotesize
\item [$\sim$] The event is captured in the schema but is missing information such as owner and source
\item [X] The event not supported by the schema
\item [\checkmark] The event is fully captured in the schema with all the required information
\end{tablenotes}
\end{ThreePartTable}
\end{table}
The findings of the case study, presented in~\Cref{tab:casestudyresults}, present a comparative evaluation of the capacity of the LEI, ICAR, and ISC schemas to capture livestock event information. It is noted that LEI is superior to ICAR and ISC in capturing the necessary information for record keeping and tracking in all highlighted scenarios. LEI can fully capture various events, including movements (arrival and departure), deaths, status observed, weights, audits, synchronisation, insemination (AI and natural insemination), pregnancy checks, births, parturition, registration, weaning, treatments, and castration. In contrast, ICAR and ISC can only partially capture these events as they lack information about the event's source and the cattle owner's properties or do not capture them at all. For example, in the case of death and weight events, LEI can fully capture them by including the source of the event, the date and time of the event, the owner of the livestock, the livestock themselves, and the other necessary details of the event. On the other hand, ICAR and ISC partially support these events because the event's source and the cattle owner's properties are missing. The same is true in ICAR for other events such as insemination, pregnancy check, birth, parturition, and registration. In terms of audit, synchronisation, weaning, and castration events, LEI fully captures these events, while ICAR and ISC do not, as these events do not exist in their schema. The same is true in ISC for the status observed, insemination, pregnancy check, birth, and parturition events do not exist in the schema. Additionally, LEI fully captures the treatment event, while ICAR and ISC only partially do, as they do not have information about the source of the event and the cattle owner's properties. Overall, the results of the case study show the advantages of LEI over ICAR and ISC in terms of its ability to fully capture various events with the necessary information.

\section{Conclusion}\label{sec:conclusion}

In this paper, we performed an in-depth analysis of the ICAR and ISC livestock schemas. We expanded these two schemas and proposed a new schema (LEI) that recorded 34 events related to cattle, from calving to death. We standardised the information regarding these events to facilitate data sharing and communication among the various parties (e.g., producers and consumers).

We evaluated the LEI schema against the ICAR and ISC standards using a case study that compared the ability of the schemas to capture different events, such as movements, deaths, status observed, weights, audits, insemination, pregnancy checks, etc. We also used structural metrics to measure the properties of the existence, depth, width, reference, and redundancy of the schemas using the schema of the ``weight" event as a common denominator. The results showed that LEI captured more information, had a more balanced structure, and improved data quality and usability than ICAR and ISC. LEI enhances the current standards rather than replacing them, as it builds upon ICAR and ISC standards.

The results of this study will open up new opportunities for future research in the field of data standardisation and event modelling in agricultural data sharing. Furthermore, we plan to extend our schema to include more agricultural fields in the future.

\section*{Funding Sources}
This project was supported by funding from Food Agility CRC Ltd, funded under the Commonwealth Government CRC Program. The CRC Program supports industry-led collaborations between industry, researchers, and the community, and this manuscript was funded by the Gulbali Institute Accelerated Publication Scheme (GAPS).

\section*{Data availability}
LEI schema is available in \url{https://github.com/mahirgamal/LEI-schema}.

\section*{Acknowledgment}
We thank Dave Swain, Will Swain (TerraCipher) and Emmaline Lear (Office of Research Services and Graduate Studies, CSU) for their valuable comments and reviews.

\bibliographystyle{elsarticle-harv}
\bibliography{library1}

\begin{thebibliography}{85}
\expandafter\ifx\csname natexlab\endcsname\relax\def\natexlab#1{#1}\fi
\providecommand{\url}[1]{\texttt{#1}}
\providecommand{\href}[2]{#2}
\providecommand{\path}[1]{#1}
\providecommand{\DOIprefix}{doi:}
\providecommand{\ArXivprefix}{arXiv:}
\providecommand{\URLprefix}{URL: }
\providecommand{\Pubmedprefix}{pmid:}
\providecommand{\doi}[1]{\href{http://dx.doi.org/#1}{\path{#1}}}
\providecommand{\Pubmed}[1]{\href{pmid:#1}{\path{#1}}}
\providecommand{\bibinfo}[2]{#2}
\ifx\xfnm\relax \def\xfnm[#1]{\unskip,\space#1}\fi
\bibitem[{Allan et~al.(1998)Allan, Papka and Lavrenko}]{Allan1998}
\bibinfo{author}{Allan, J.}, \bibinfo{author}{Papka, R.}, \bibinfo{author}{Lavrenko, V.}, \bibinfo{year}{1998}.
\newblock \bibinfo{title}{On-line new event detection and tracking}, in: \bibinfo{booktitle}{Proceedings of the 21st annual international ACM SIGIR conference on Research and development in information retrieval}, pp. \bibinfo{pages}{37--45}.
\newblock \DOIprefix\doi{10.1145/290941.290954}.
\bibitem[{Alshuqayran et~al.(2016)Alshuqayran, Ali and Evans}]{Alshuqayran2016}
\bibinfo{author}{Alshuqayran, N.}, \bibinfo{author}{Ali, N.}, \bibinfo{author}{Evans, R.}, \bibinfo{year}{2016}.
\newblock \bibinfo{title}{A systematic mapping study in microservice architecture}, in: \bibinfo{booktitle}{2016 IEEE 9th International Conference on Service-Oriented Computing and Applications (SOCA)}, \bibinfo{organization}{IEEE}. pp. \bibinfo{pages}{44--51}.
\newblock \DOIprefix\doi{10.1109/SOCA.2016.15}.
\bibitem[{Antonino et~al.(2022)Antonino, Capilla, Pelliccione, Schnicke, Espen, Kuhn and Schmid}]{Antonino2022}
\bibinfo{author}{Antonino, P.O.}, \bibinfo{author}{Capilla, R.}, \bibinfo{author}{Pelliccione, P.}, \bibinfo{author}{Schnicke, F.}, \bibinfo{author}{Espen, D.}, \bibinfo{author}{Kuhn, T.}, \bibinfo{author}{Schmid, K.}, \bibinfo{year}{2022}.
\newblock \bibinfo{title}{A quality 4.0 model for architecting industry 4.0 systems}.
\newblock \bibinfo{journal}{Advanced Engineering Informatics} \bibinfo{volume}{54}, \bibinfo{pages}{101801}.
\newblock \DOIprefix\doi{10.1016/j.aei.2022.101801}.
\bibitem[{{Australian Beef Sustainability Framework}(2022)}]{Anon2022}
\bibinfo{author}{{Australian Beef Sustainability Framework}}, \bibinfo{year}{2022}.
\newblock \bibinfo{title}{{Australian Beef Sustainability 2022 Annual Update}}.
\newblock \bibinfo{type}{Technical Report}. Australian Beef Sustainability Framework.
\newblock \URLprefix \url{https://www.sustainableaustralianbeef.com.au/globalassets/beef-sustainability/documents/absf_update_2022_web.pdf}.
\bibitem[{{Australian Government}(2021)}]{AustralianGovernment2021RegulatoryDepartment}
\bibinfo{author}{{Australian Government}}, \bibinfo{year}{2021}.
\newblock \bibinfo{title}{{Attorney-General's Department--Regulatory powers}}.
\newblock \bibinfo{howpublished}{\url{https://www.ag.gov.au/legal-system/administrative-law/regulatory-powers}}.
\newblock \bibinfo{note}{Accessed: 2022-05-01}.
\bibitem[{Aydin and {Nafiz Aydin}(2022)}]{Aydin2022DesignDatab}
\bibinfo{author}{Aydin, S.}, \bibinfo{author}{{Nafiz Aydin}, M.}, \bibinfo{year}{2022}.
\newblock \bibinfo{title}{Design and implementation of a smart beehive and its monitoring system using microservices in the context of iot and open data}.
\newblock \bibinfo{journal}{Computers and Electronics in Agriculture} \bibinfo{volume}{196}, \bibinfo{pages}{106897}.
\newblock \DOIprefix\doi{10.1016/J.COMPAG.2022.106897}.
\bibitem[{Bahlo and Dahlhaus(2021)}]{Bahlo2021LivestockAustralia}
\bibinfo{author}{Bahlo, C.}, \bibinfo{author}{Dahlhaus, P.}, \bibinfo{year}{2021}.
\newblock \bibinfo{title}{Livestock data--is it there and is it fair? a systematic review of livestock farming datasets in australia}.
\newblock \bibinfo{journal}{Computers and Electronics in Agriculture} \bibinfo{volume}{188}, \bibinfo{pages}{106365}.
\newblock \DOIprefix\doi{10.1016/j.compag.2021.106365}.
\bibitem[{Bahlo et~al.(2019)Bahlo, Dahlhaus, Thompson and Trotter}]{Bahlo2019TheReview}
\bibinfo{author}{Bahlo, C.}, \bibinfo{author}{Dahlhaus, P.}, \bibinfo{author}{Thompson, H.}, \bibinfo{author}{Trotter, M.}, \bibinfo{year}{2019}.
\newblock \bibinfo{title}{The role of interoperable data standards in precision livestock farming in extensive livestock systems: A review}.
\newblock \bibinfo{journal}{Computers and electronics in agriculture} \bibinfo{volume}{156}, \bibinfo{pages}{459--466}.
\newblock \DOIprefix\doi{10.1016/j.compag.2018.12.007}.
\bibitem[{Benaissa et~al.(2020)Benaissa, Tuyttens, Plets, Trogh, Martens, Vandaele, Joseph and Sonck}]{Benaissa2020}
\bibinfo{author}{Benaissa, S.}, \bibinfo{author}{Tuyttens, F.A.M.}, \bibinfo{author}{Plets, D.}, \bibinfo{author}{Trogh, J.}, \bibinfo{author}{Martens, L.}, \bibinfo{author}{Vandaele, L.}, \bibinfo{author}{Joseph, W.}, \bibinfo{author}{Sonck, B.}, \bibinfo{year}{2020}.
\newblock \bibinfo{title}{Calving and estrus detection in dairy cattle using a combination of indoor localization and accelerometer sensors}.
\newblock \bibinfo{journal}{Computers and electronics in agriculture} \bibinfo{volume}{168}, \bibinfo{pages}{105153}.
\newblock \DOIprefix\doi{10.1016/j.compag.2019.105153}.
\bibitem[{Bishop-Hurley et~al.(2007)Bishop-Hurley, Swain, Anderson, Sikka, Crossman and Corke}]{Bishop-Hurley2007}
\bibinfo{author}{Bishop-Hurley, G.}, \bibinfo{author}{Swain, D.}, \bibinfo{author}{Anderson, D.}, \bibinfo{author}{Sikka, P.}, \bibinfo{author}{Crossman, C.}, \bibinfo{author}{Corke, P.}, \bibinfo{year}{2007}.
\newblock \bibinfo{title}{Virtual fencing applications: Implementing and testing an automated cattle control system}.
\newblock \bibinfo{journal}{Computers and Electronics in Agriculture} \bibinfo{volume}{56}, \bibinfo{pages}{14--22}.
\newblock \DOIprefix\doi{10.1016/j.compag.2006.12.003}.
\bibitem[{Car et~al.(2012)Car, Christen, Hornbuckle and Moore}]{Car2012}
\bibinfo{author}{Car, N.J.}, \bibinfo{author}{Christen, E.W.}, \bibinfo{author}{Hornbuckle, J.W.}, \bibinfo{author}{Moore, G.A.}, \bibinfo{year}{2012}.
\newblock \bibinfo{title}{Using a mobile phone short messaging service (sms) for irrigation scheduling in australia--farmers’ participation and utility evaluation}.
\newblock \bibinfo{journal}{Computers and electronics in agriculture} \bibinfo{volume}{84}, \bibinfo{pages}{132--143}.
\newblock \DOIprefix\doi{10.1016/j.compag.2012.03.003}.
\bibitem[{Castelli et~al.(2007)Castelli, Rosi, Mamei and Zambonelli}]{Castelli2007}
\bibinfo{author}{Castelli, G.}, \bibinfo{author}{Rosi, A.}, \bibinfo{author}{Mamei, M.}, \bibinfo{author}{Zambonelli, F.}, \bibinfo{year}{2007}.
\newblock \bibinfo{title}{A simple model and infrastructure for context-aware browsing of the world}, in: \bibinfo{booktitle}{Fifth Annual IEEE International Conference on Pervasive Computing and Communications (PerCom'07)}, \bibinfo{organization}{IEEE}. pp. \bibinfo{pages}{229--238}.
\newblock \DOIprefix\doi{10.1109/PERCOM.2007.4}.
\bibitem[{Chang et~al.(2022)Chang, Fogarty, Moraes, Garc{\'\i}a-Guerra, Swain and Trotter}]{Chang2022}
\bibinfo{author}{Chang, A.Z.}, \bibinfo{author}{Fogarty, E.S.}, \bibinfo{author}{Moraes, L.E.}, \bibinfo{author}{Garc{\'\i}a-Guerra, A.}, \bibinfo{author}{Swain, D.L.}, \bibinfo{author}{Trotter, M.G.}, \bibinfo{year}{2022}.
\newblock \bibinfo{title}{Detection of rumination in cattle using an accelerometer ear-tag: A comparison of analytical methods and individual animal and generic models}.
\newblock \bibinfo{journal}{Computers and Electronics in Agriculture} \bibinfo{volume}{192}, \bibinfo{pages}{106595}.
\newblock \DOIprefix\doi{10.1016/j.compag.2021.106595}.
\bibitem[{Cooke(2020)}]{Cooke2020}
\bibinfo{author}{Cooke, A.}, \bibinfo{year}{2020}.
\newblock \bibinfo{title}{{About ICAR and ADE}}.
\newblock \bibinfo{howpublished}{\url{https://github.com/adewg/ICAR/wiki/About-ICAR-and-ADE}}.
\newblock \bibinfo{note}{Accessed: 2022-01-03}.
\bibitem[{{Department of Agriculture Fisheries and Forestry}(2020)}]{DepartmentofAgricultureFisheriesandForestry2020DairyAustralia}
\bibinfo{author}{{Department of Agriculture Fisheries and Forestry}}, \bibinfo{year}{2020}.
\newblock \bibinfo{title}{{Dairy in Australia}}.
\newblock \bibinfo{howpublished}{\url{https://www.agriculture.gov.au/agriculture-land/farm-food-drought/meat-wool-dairy/dairy}}.
\newblock \bibinfo{note}{Accessed: 2022-01-20}.
\bibitem[{{Department of Primary Industries and Regional Development}(2018)}]{Development2018}
\bibinfo{author}{{Department of Primary Industries and Regional Development}}, \bibinfo{year}{2018}.
\newblock \bibinfo{title}{{Horn tipping of livestock}}.
\newblock \bibinfo{type}{Technical Report} \bibinfo{number}{July}. Department of Primary Industries and Regional Development - Government of Western Australia.
\newblock \URLprefix \url{https://www.agric.wa.gov.au/sites/gateway/files/Horn tipping of livestock_1.pdf}.
\bibitem[{Durrant et~al.(2021)Durrant, Markovic, Matthews, May, Leontidis and Enright}]{Durrant2021}
\bibinfo{author}{Durrant, A.}, \bibinfo{author}{Markovic, M.}, \bibinfo{author}{Matthews, D.}, \bibinfo{author}{May, D.}, \bibinfo{author}{Leontidis, G.}, \bibinfo{author}{Enright, J.}, \bibinfo{year}{2021}.
\newblock \bibinfo{title}{How might technology rise to the challenge of data sharing in agri-food?}
\newblock \bibinfo{journal}{Global Food Security} \bibinfo{volume}{28}, \bibinfo{pages}{100493}.
\newblock \DOIprefix\doi{10.1016/j.gfs.2021.100493}.
\bibitem[{Ede et~al.(2022)Ede, Nogues, von Keyserlingk and Weary}]{Ede2022}
\bibinfo{author}{Ede, T.}, \bibinfo{author}{Nogues, E.}, \bibinfo{author}{von Keyserlingk, M.A.}, \bibinfo{author}{Weary, D.M.}, \bibinfo{year}{2022}.
\newblock \bibinfo{title}{Pain in the hours following surgical and rubber ring castration in dairy calves: Evidence from conditioned place avoidance}.
\newblock \bibinfo{journal}{JDS communications} \bibinfo{volume}{3}, \bibinfo{pages}{421--425}.
\newblock \DOIprefix\doi{10.3168/jdsc.2022-0241}.
\bibitem[{Eitzinger et~al.(2019)Eitzinger, Cock, Atzmanstorfer, Binder, L{\"a}derach, Bonilla-Findji, Bartling, Mwongera, Zurita and Jarvis}]{Eitzinger2019}
\bibinfo{author}{Eitzinger, A.}, \bibinfo{author}{Cock, J.}, \bibinfo{author}{Atzmanstorfer, K.}, \bibinfo{author}{Binder, C.R.}, \bibinfo{author}{L{\"a}derach, P.}, \bibinfo{author}{Bonilla-Findji, O.}, \bibinfo{author}{Bartling, M.}, \bibinfo{author}{Mwongera, C.}, \bibinfo{author}{Zurita, L.}, \bibinfo{author}{Jarvis, A.}, \bibinfo{year}{2019}.
\newblock \bibinfo{title}{Geofarmer: A monitoring and feedback system for agricultural development projects}.
\newblock \bibinfo{journal}{Computers and electronics in agriculture} \bibinfo{volume}{158}, \bibinfo{pages}{109--121}.
\newblock \DOIprefix\doi{10.1016/j.compag.2019.01.049}.
\bibitem[{Elly and Epafra~Silayo(2013)}]{Elly2013}
\bibinfo{author}{Elly, T.}, \bibinfo{author}{Epafra~Silayo, E.}, \bibinfo{year}{2013}.
\newblock \bibinfo{title}{Agricultural information needs and sources of the rural farmers in tanzania: A case of iringa rural district}.
\newblock \bibinfo{journal}{Library review} \bibinfo{volume}{62}, \bibinfo{pages}{547--566}.
\newblock \DOIprefix\doi{10.1108/LR-01-2013-0009}.
\bibitem[{Farooq et~al.(2019)Farooq, Riaz, Abid, Abid and Naeem}]{Farooq2019}
\bibinfo{author}{Farooq, M.S.}, \bibinfo{author}{Riaz, S.}, \bibinfo{author}{Abid, A.}, \bibinfo{author}{Abid, K.}, \bibinfo{author}{Naeem, M.A.}, \bibinfo{year}{2019}.
\newblock \bibinfo{title}{A survey on the role of iot in agriculture for the implementation of smart farming}.
\newblock \bibinfo{journal}{Ieee Access} \bibinfo{volume}{7}, \bibinfo{pages}{156237--156271}.
\newblock \DOIprefix\doi{10.1109/ACCESS.2019.2949703}.
\bibitem[{Farrell et~al.(2022)Farrell, Lawrence, Schilling, David and Amorim}]{Farrell2022AgriculturalV.1.2}
\bibinfo{author}{Farrell, R.}, \bibinfo{author}{Lawrence, W.}, \bibinfo{author}{Schilling, R.K.}, \bibinfo{author}{David, R.}, \bibinfo{author}{Amorim, A.}, \bibinfo{year}{2022}.
\newblock \bibinfo{title}{{Agricultural Data Standards Guide - Australia (ADSGA): Promoting FAIR data for the agricultural industry, v.1.2}}.
\newblock \bibinfo{type}{Technical Report}. Agricultural Data Standards Guide - Australia (ADSGA).
\newblock \URLprefix \url{https://axistech.co/wp-content/uploads/2022/04/Agricultural-Data-Standards-Guide-Australia-ADSGA-promoting-FAIR-data-for-the-agricultural-industry-v1.2-2.pdf}.
\bibitem[{F{\`e}vre et~al.(2006)F{\`e}vre, Bronsvoort, Hamilton and Cleaveland}]{Fevre2006AnimalDiseases}
\bibinfo{author}{F{\`e}vre, E.M.}, \bibinfo{author}{Bronsvoort, B.M.d.C.}, \bibinfo{author}{Hamilton, K.A.}, \bibinfo{author}{Cleaveland, S.}, \bibinfo{year}{2006}.
\newblock \bibinfo{title}{Animal movements and the spread of infectious diseases}.
\newblock \bibinfo{journal}{Trends in microbiology} \bibinfo{volume}{14}, \bibinfo{pages}{125--131}.
\newblock \DOIprefix\doi{10.1016/J.TIM.2006.01.004}.
\bibitem[{Frost et~al.(1997)Frost, Schofield, Beaulah, Mottram, Lines and Wathes}]{Frost1997}
\bibinfo{author}{Frost, A.}, \bibinfo{author}{Schofield, C.}, \bibinfo{author}{Beaulah, S.}, \bibinfo{author}{Mottram, T.}, \bibinfo{author}{Lines, J.}, \bibinfo{author}{Wathes, C.}, \bibinfo{year}{1997}.
\newblock \bibinfo{title}{A review of livestock monitoring and the need for integrated systems}.
\newblock \bibinfo{journal}{Computers and electronics in agriculture} \bibinfo{volume}{17}, \bibinfo{pages}{139--159}.
\newblock \DOIprefix\doi{10.1016/s0168-1699(96)01301-4}.
\bibitem[{G{\'o}mez et~al.(2021)G{\'o}mez, Roncancio and Casallas}]{Gomez2021}
\bibinfo{author}{G{\'o}mez, P.}, \bibinfo{author}{Roncancio, C.}, \bibinfo{author}{Casallas, R.}, \bibinfo{year}{2021}.
\newblock \bibinfo{title}{Analysis and evaluation of document-oriented structures}.
\newblock \bibinfo{journal}{Data \& Knowledge Engineering} \bibinfo{volume}{134}, \bibinfo{pages}{101893}.
\newblock \DOIprefix\doi{10.1016/j.datak.2021.101893}.
\bibitem[{Hardin~IV et~al.(2022)Hardin~IV, Barnes, Delhom, Wanjura and Ward}]{Hardin2022InternetProcessing}
\bibinfo{author}{Hardin~IV, R.G.}, \bibinfo{author}{Barnes, E.M.}, \bibinfo{author}{Delhom, C.D.}, \bibinfo{author}{Wanjura, J.D.}, \bibinfo{author}{Ward, J.K.}, \bibinfo{year}{2022}.
\newblock \bibinfo{title}{Internet of things: Cotton harvesting and processing}.
\newblock \bibinfo{journal}{Computers and Electronics in Agriculture} \bibinfo{volume}{202}, \bibinfo{pages}{107294}.
\newblock \DOIprefix\doi{10.1016/j.compag.2022.107294}.
\bibitem[{Hart et~al.(1998)Hart, Larcombe, Sherlock and Smith}]{Hart1998}
\bibinfo{author}{Hart, R.P.}, \bibinfo{author}{Larcombe, M.T.}, \bibinfo{author}{Sherlock, R.A.}, \bibinfo{author}{Smith, L.A.}, \bibinfo{year}{1998}.
\newblock \bibinfo{title}{Optimisation techniques for a computer simulation of a pastoral dairy farm}.
\newblock \bibinfo{journal}{Computers and electronics in agriculture} \bibinfo{volume}{19}, \bibinfo{pages}{129--153}.
\newblock \DOIprefix\doi{10.1016/S0168-1699(97)00039-2}.
\bibitem[{{ICAR}(2013)}]{ICAR2013}
\bibinfo{author}{{ICAR}}, \bibinfo{year}{2013}.
\newblock \bibinfo{title}{{International Committee for Animal Recording}}.
\newblock \bibinfo{howpublished}{\url{http://www.icar.org/ http://icar.org/}}.
\newblock \bibinfo{note}{Accessed: 2022-08-03}.
\bibitem[{Iftikhar and Pedersen(2011)}]{Iftikhar2011}
\bibinfo{author}{Iftikhar, N.}, \bibinfo{author}{Pedersen, T.B.}, \bibinfo{year}{2011}.
\newblock \bibinfo{title}{Flexible exchange of farming device data}.
\newblock \bibinfo{journal}{Computers and Electronics in Agriculture} \bibinfo{volume}{75}, \bibinfo{pages}{52--63}.
\newblock \DOIprefix\doi{10.1016/j.compag.2010.09.010}.
\bibitem[{Iglesias and East(2015)}]{Iglesias2015}
\bibinfo{author}{Iglesias, R.}, \bibinfo{author}{East, I.}, \bibinfo{year}{2015}.
\newblock \bibinfo{title}{Cattle movement patterns in australia: An analysis of the nlis database 2008--2012}.
\newblock \bibinfo{journal}{Australian veterinary journal} \bibinfo{volume}{93}, \bibinfo{pages}{394--403}.
\newblock \DOIprefix\doi{10.1111/avj.12377}.
\bibitem[{Ilyas and Ahmad(2020)}]{IlyasAhmad2020}
\bibinfo{author}{Ilyas, Q.M.}, \bibinfo{author}{Ahmad, M.}, \bibinfo{year}{2020}.
\newblock \bibinfo{title}{Smart farming: An enhanced pursuit of sustainable remote livestock tracking and geofencing using iot and gprs}.
\newblock \bibinfo{journal}{Wireless communications and mobile computing} \bibinfo{volume}{2020}, \bibinfo{pages}{1--12}.
\newblock \DOIprefix\doi{10.1155/2020/6660733}.
\bibitem[{{Integrity Systems Company}(2021)}]{IntegritySystemsCompany2021a}
\bibinfo{author}{{Integrity Systems Company}}, \bibinfo{year}{2021}.
\newblock \bibinfo{title}{{Livestock Production Assurance (LPA)}}.
\newblock \bibinfo{howpublished}{\url{https://www.integritysystems.com.au/on-farm-assurance/livestock-product-assurance/}}.
\newblock \bibinfo{note}{Accessed: 2022-07-10}.
\bibitem[{{International Committee for Animal Recording - ICAR}(2014)}]{InternationalCommitteeforAnimalRecording-ICAR2014}
\bibinfo{author}{{International Committee for Animal Recording - ICAR}}, \bibinfo{year}{2014}.
\newblock \bibinfo{title}{{International Agreement of Recording Practices ICAR Recording Guidelines}}.
\newblock \bibinfo{type}{Technical Report}. International Committee for Animal Recording - ICAR.
\newblock \URLprefix \url{https://pecuaria.pt/docs/Guidelines_2014.pdf}.
\bibitem[{{International Organization for Standardization}(2016)}]{ISO2016}
\bibinfo{author}{{International Organization for Standardization}}, \bibinfo{year}{2016}.
\newblock \bibinfo{title}{Animal welfare management—general requirements and guidance for organizations in the food supply chain}.
\newblock \bibinfo{howpublished}{\url{https://www.iso.org/standard/64749.html}}.
\newblock \bibinfo{note}{Accessed: 2022-08-03}.
\bibitem[{{International Organization for Standardization}(2017)}]{ios2017}
\bibinfo{author}{{International Organization for Standardization}}, \bibinfo{year}{2017}.
\newblock \bibinfo{title}{{ISO and agriculture}}.
\newblock \bibinfo{type}{Technical Report}. International Organization for Standardization.
\newblock \URLprefix \url{https://www.iso.org/files/live/sites/isoorg/files/store/en/PUB100317.pdf}.
\bibitem[{Karwowski(2010)}]{karwowski2010ontologies}
\bibinfo{author}{Karwowski, W.}, \bibinfo{year}{2010}.
\newblock \bibinfo{title}{Ontologies and agricultural information management standards}.
\newblock \bibinfo{journal}{Information systems in managment VI/sci. ed. Piotr Ja{\l}owiecki, Arkadiusz Or{\l}owski, Warsaw, WULS Press} , \bibinfo{pages}{47--56}\URLprefix \url{http://isim.wzim.sggw.pl/resources/ISIM_VI_2010.pdf}.
\bibitem[{Kuzuhara et~al.(2015)Kuzuhara, Kawamura, Yoshitoshi, Tamaki, Sugai, Ikegami, Kurokawa, Obitsu, Okita, Sugino et~al.}]{Kuzuhara2015}
\bibinfo{author}{Kuzuhara, Y.}, \bibinfo{author}{Kawamura, K.}, \bibinfo{author}{Yoshitoshi, R.}, \bibinfo{author}{Tamaki, T.}, \bibinfo{author}{Sugai, S.}, \bibinfo{author}{Ikegami, M.}, \bibinfo{author}{Kurokawa, Y.}, \bibinfo{author}{Obitsu, T.}, \bibinfo{author}{Okita, M.}, \bibinfo{author}{Sugino, T.}, et~al., \bibinfo{year}{2015}.
\newblock \bibinfo{title}{A preliminarily study for predicting body weight and milk properties in lactating holstein cows using a three-dimensional camera system}.
\newblock \bibinfo{journal}{Computers and Electronics in Agriculture} \bibinfo{volume}{111}, \bibinfo{pages}{186--193}.
\newblock \DOIprefix\doi{10.1016/j.compag.2014.12.020}.
\bibitem[{Mamo et~al.(2021)Mamo, Azzopardi and Layfield}]{Mamo2021}
\bibinfo{author}{Mamo, N.}, \bibinfo{author}{Azzopardi, J.}, \bibinfo{author}{Layfield, C.}, \bibinfo{year}{2021}.
\newblock \bibinfo{title}{Who? what? event tracking needs event understanding.}, in: \bibinfo{booktitle}{KDIR}, pp. \bibinfo{pages}{139--146}.
\newblock \DOIprefix\doi{10.5220/0010650500003064}.
\bibitem[{Manoj et~al.(2023)Manoj, Makkithaya and Narendra}]{T.2022}
\bibinfo{author}{Manoj, T.}, \bibinfo{author}{Makkithaya, K.}, \bibinfo{author}{Narendra, V.}, \bibinfo{year}{2023}.
\newblock \bibinfo{title}{A trusted iot data sharing and secure oracle based access for agricultural production risk management}.
\newblock \bibinfo{journal}{Computers and Electronics in Agriculture} \bibinfo{volume}{204}, \bibinfo{pages}{107544}.
\newblock \DOIprefix\doi{10.1016/j.compag.2022.107544}.
\bibitem[{Marrs(2017)}]{Marrs2017}
\bibinfo{author}{Marrs, T.}, \bibinfo{year}{2017}.
\newblock \bibinfo{title}{JSON at work: practical data integration for the web}.
\newblock \bibinfo{publisher}{O'Reilly Media, Inc.}
\newblock \URLprefix \url{https://www.oreilly.com/library/view/json-at-work/9781491982389/}.
\bibitem[{Marshall et~al.(2020)Marshall, Dezuanni, Burgess, Thomas and Wilson}]{Marshall2020}
\bibinfo{author}{Marshall, A.}, \bibinfo{author}{Dezuanni, M.}, \bibinfo{author}{Burgess, J.}, \bibinfo{author}{Thomas, J.}, \bibinfo{author}{Wilson, C.K.}, \bibinfo{year}{2020}.
\newblock \bibinfo{title}{Australian farmers left behind in the digital economy--insights from the australian digital inclusion index}.
\newblock \bibinfo{journal}{Journal of Rural Studies} \bibinfo{volume}{80}, \bibinfo{pages}{195--210}.
\newblock \DOIprefix\doi{10.1016/j.jrurstud.2020.09.001}.
\bibitem[{McGowan et~al.(2014)McGowan, Fordyce, O'Rourke, Barnes, Morton, Menzies, Jephcott, McCosker, Smith, Perkins et~al.}]{McGowan2014}
\bibinfo{author}{McGowan, M.R.}, \bibinfo{author}{Fordyce, G.}, \bibinfo{author}{O'Rourke, P.K.}, \bibinfo{author}{Barnes, T.S.}, \bibinfo{author}{Morton, J.}, \bibinfo{author}{Menzies, D.}, \bibinfo{author}{Jephcott, S.}, \bibinfo{author}{McCosker, K.D.}, \bibinfo{author}{Smith, D.}, \bibinfo{author}{Perkins, N.R.}, et~al., \bibinfo{year}{2014}.
\newblock \bibinfo{title}{Northern australia beef fertility project: Cash cow}.
\newblock \bibinfo{journal}{Meat \& Livestock Australia} \bibinfo{volume}{364}, \bibinfo{pages}{67--68}.
\newblock \URLprefix \url{https://www.mla.com.au/contentassets/6428c467b1904744b287e539b50f17e7/b.nbp.0382_final_report.pdf}.
\bibitem[{{Meat {\&} Livestock Australia}(2021)}]{MeatLivestockAustralia2021GlossaryAustralia}
\bibinfo{author}{{Meat {\&} Livestock Australia}}, \bibinfo{year}{2021}.
\newblock \bibinfo{title}{{Meat {\&} Livestock Australia--Glossary}}.
\newblock \bibinfo{howpublished}{\url{https://www.mla.com.au/general/glossary/}}.
\newblock \bibinfo{note}{Accessed: 2022-06-01}.
\bibitem[{{Meat and Livestock Australia}(2011)}]{MeatLivestock2011}
\bibinfo{author}{{Meat and Livestock Australia}}, \bibinfo{year}{2011}.
\newblock \bibinfo{title}{{Herd health and welfare Introduction}}.
\newblock \bibinfo{type}{Technical Report}. Meat and Livestock Australia.
\newblock \URLprefix \url{https://mbfp.mla.com.au/siteassets/herd-health-welfare/herd-health-and-welfare-1.pdf}.
\bibitem[{Michael et~al.(2022)Michael, Kjeldaas, Lang, Johnson, Abric, Hutton, Wright, Tubbs, Carpenter, Gore, Branner, Karr, Worth, Bray, {Fenhl}, {Forevermatt}, {Goldaxe}, Andrews, {Herv{\'{e}}}, {Hongwei}, Claven, Rouwhorst, Kobit, Kurmis, Blackman and Jacques}]{Michael2022Understanding2020-12}
\bibinfo{author}{Michael, D.}, \bibinfo{author}{Kjeldaas, A.}, \bibinfo{author}{Lang, A.}, \bibinfo{author}{Johnson, A.D.}, \bibinfo{author}{Abric, A.}, \bibinfo{author}{Hutton, B.}, \bibinfo{author}{Wright, B.}, \bibinfo{author}{Tubbs, B.}, \bibinfo{author}{Carpenter, C.}, \bibinfo{author}{Gore, C.M.}, \bibinfo{author}{Branner, D.}, \bibinfo{author}{Karr, D.M.}, \bibinfo{author}{Worth, D.}, \bibinfo{author}{Bray, E.M.}, \bibinfo{author}{{Fenhl}}, \bibinfo{author}{{Forevermatt}}, \bibinfo{author}{{Goldaxe}}, \bibinfo{author}{Andrews, H.}, \bibinfo{author}{{Herv{\'{e}}}}, \bibinfo{author}{{Hongwei}}, \bibinfo{author}{Claven, J.}, \bibinfo{author}{Rouwhorst, K.}, \bibinfo{author}{Kobit, M.}, \bibinfo{author}{Kurmis, O.}, \bibinfo{author}{Blackman, S.}, \bibinfo{author}{Jacques, V.}, \bibinfo{year}{2022}.
\newblock \bibinfo{title}{{Understanding JSON Schema Release 2020-12}}.
\newblock \bibinfo{publisher}{Space Telescope Science Institute}.
\bibitem[{Mohd(2007)}]{Mohd2007}
\bibinfo{author}{Mohd, M.}, \bibinfo{year}{2007}.
\newblock \bibinfo{title}{Named entity patterns across news domains}, in: \bibinfo{booktitle}{BCS IRSG Symposium: Future Directions in Information Access}, pp. \bibinfo{pages}{30--36}.
\newblock \URLprefix \url{https://www.scienceopen.com/document_file/71209c27-4790-47ca-943c-29d463d560fe/ScienceOpen/001_Mohd.pdf}.
\bibitem[{Motavalli et~al.(2017)Motavalli, Dirandeh, Deldar and Colazo}]{Motavalli2017}
\bibinfo{author}{Motavalli, T.}, \bibinfo{author}{Dirandeh, E.}, \bibinfo{author}{Deldar, H.}, \bibinfo{author}{Colazo, M.}, \bibinfo{year}{2017}.
\newblock \bibinfo{title}{Evaluation of shortened timed-ai protocols for resynchronization of ovulation in multiparous holstein dairy cows}.
\newblock \bibinfo{journal}{Theriogenology} \bibinfo{volume}{95}, \bibinfo{pages}{187--192}.
\newblock \DOIprefix\doi{10.1016/j.theriogenology.2017.03.003}.
\bibitem[{Murugeswari et~al.(2022)Murugeswari, Murugan, Rajathi and Kumar}]{Murugeswari2022}
\bibinfo{author}{Murugeswari, S.}, \bibinfo{author}{Murugan, K.}, \bibinfo{author}{Rajathi, S.}, \bibinfo{author}{Kumar, M.S.}, \bibinfo{year}{2022}.
\newblock \bibinfo{title}{Monitoring body temperature of cattle using an innovative infrared photodiode thermometer}.
\newblock \bibinfo{journal}{Computers and Electronics in Agriculture} \bibinfo{volume}{198}, \bibinfo{pages}{107120}.
\newblock \DOIprefix\doi{10.1016/j.compag.2022.107120}.
\bibitem[{Nash et~al.(2009a)Nash, Dreger, Schwarz, Bill and Werner}]{Nash2009a}
\bibinfo{author}{Nash, E.}, \bibinfo{author}{Dreger, F.}, \bibinfo{author}{Schwarz, J.}, \bibinfo{author}{Bill, R.}, \bibinfo{author}{Werner, A.}, \bibinfo{year}{2009}a.
\newblock \bibinfo{title}{Development of a model of data-flows for precision agriculture based on a collaborative research project}.
\newblock \bibinfo{journal}{Computers and electronics in Agriculture} \bibinfo{volume}{66}, \bibinfo{pages}{25--37}.
\newblock \DOIprefix\doi{10.1016/j.compag.2008.11.005}.
\bibitem[{Nash et~al.(2009b)Nash, Korduan and Bill}]{Nash2009}
\bibinfo{author}{Nash, E.}, \bibinfo{author}{Korduan, P.}, \bibinfo{author}{Bill, R.}, \bibinfo{year}{2009}b.
\newblock \bibinfo{title}{Applications of open geospatial web services in precision agriculture: a review}.
\newblock \bibinfo{journal}{Precision agriculture} \bibinfo{volume}{10}, \bibinfo{pages}{546--560}.
\newblock \DOIprefix\doi{10.1007/s11119-009-9134-0}.
\bibitem[{{National Farmers Federation}(2019)}]{nff}
\bibinfo{author}{{National Farmers Federation}}, \bibinfo{year}{2019}.
\newblock \bibinfo{title}{{2030 Roadmap Australian Agriculture’s Plan for a {\$}100 Billion Industry}}.
\newblock \bibinfo{type}{Technical Report}. National Farmers Federation.
\newblock \URLprefix \url{https://nff.org.au/wp-content/uploads/2020/02/NFF_Roadmap_2030_FINAL.pdf}.
\bibitem[{{National Livestock Identification System}(2009)}]{Victoria2009}
\bibinfo{author}{{National Livestock Identification System}}, \bibinfo{year}{2009}.
\newblock \bibinfo{title}{{Australia's system for livestock identification and traceability, NLIS Database}}.
\newblock \bibinfo{type}{Technical Report}. National Livestock Identification System Ltd.
\newblock \URLprefix \url{https://www.nlis.mla.com.au/NLISDocuments/NLIS Cattle brochure (Mar 09).pdf}.
\bibitem[{Newman(2021)}]{Newman2021}
\bibinfo{author}{Newman, S.}, \bibinfo{year}{2021}.
\newblock \bibinfo{title}{{Building Microservices, 2nd Edition}}.
\bibitem[{{NLIS Cattle Advisory Committee}(2016)}]{nlis2016}
\bibinfo{author}{{NLIS Cattle Advisory Committee}}, \bibinfo{year}{2016}.
\newblock \bibinfo{title}{{NLIS (Cattle) Traceability Standards}}.
\newblock \bibinfo{type}{Technical Report}. National Livestock Identification System Ltd.
\newblock \URLprefix \url{https://www.nlis.com.au/Files/1/PDF/NLIS Cattl Traceability Standards watermark.pdf}.
\bibitem[{{NSW Government}(2022)}]{NSWGovernment2022NationalServices}
\bibinfo{author}{{NSW Government}}, \bibinfo{year}{2022}.
\newblock \bibinfo{title}{{National Livestock Identification Scheme - Local Land Services}}.
\newblock \bibinfo{howpublished}{\url{https://www.lls.nsw.gov.au/help-and-advice/livestock-health-and-production/the-national-livestock-identification-scheme}}.
\newblock \bibinfo{note}{Accessed: 2022-12-20}.
\bibitem[{{NSW Government Department of Industry}(2008)}]{nlis2008}
\bibinfo{author}{{NSW Government Department of Industry}}, \bibinfo{year}{2008}.
\newblock \bibinfo{title}{{National Livestock Identification System}}.
\newblock \bibinfo{howpublished}{\url{https://www.dpi.nsw.gov.au/animals-and-livestock/nlis}}.
\newblock \bibinfo{note}{Accessed: 2022-06-05}.
\bibitem[{Oliveira et~al.(2020)Oliveira, da~Silva~Feuchard, de~Freitas, da~Silva~Rosa, dos Reis~Camargo and Saraiva}]{Oliveira2020}
\bibinfo{author}{Oliveira, C.S.}, \bibinfo{author}{da~Silva~Feuchard, V.L.}, \bibinfo{author}{de~Freitas, C.}, \bibinfo{author}{da~Silva~Rosa, P.M.}, \bibinfo{author}{dos Reis~Camargo, A.J.}, \bibinfo{author}{Saraiva, N.Z.}, \bibinfo{year}{2020}.
\newblock \bibinfo{title}{In-straw warming protocol improves survival of vitrified embryos and allows direct transfer in cattle}.
\newblock \bibinfo{journal}{Cryobiology} \bibinfo{volume}{97}, \bibinfo{pages}{222--225}.
\newblock \DOIprefix\doi{10.1016/j.cryobiol.2020.02.007}.
\bibitem[{Patel et~al.(2020)Patel, Nagaraj and Cooke}]{Patel2020}
\bibinfo{author}{Patel, S.}, \bibinfo{author}{Nagaraj, A.}, \bibinfo{author}{Cooke, A.}, \bibinfo{year}{2020}.
\newblock \bibinfo{title}{{Common Animal Data Framework}}.
\newblock \bibinfo{type}{Technical Report}. Meat and Livestock Australia Limited.
\newblock \URLprefix \url{https://www.mla.com.au/contentassets/4ffcd2f6805441cea005799a3f842806/common-animal-data-framework.pdf}.
\bibitem[{Qiao et~al.(2021)Qiao, Kong, Clark, Lomax, Su, Eiffert and Sukkarieh}]{Qiao2021}
\bibinfo{author}{Qiao, Y.}, \bibinfo{author}{Kong, H.}, \bibinfo{author}{Clark, C.}, \bibinfo{author}{Lomax, S.}, \bibinfo{author}{Su, D.}, \bibinfo{author}{Eiffert, S.}, \bibinfo{author}{Sukkarieh, S.}, \bibinfo{year}{2021}.
\newblock \bibinfo{title}{Intelligent perception for cattle monitoring: A review for cattle identification, body condition score evaluation, and weight estimation}.
\newblock \bibinfo{journal}{Computers and Electronics in Agriculture} \bibinfo{volume}{185}, \bibinfo{pages}{106143}.
\newblock \DOIprefix\doi{10.1016/j.compag.2021.106143}.
\bibitem[{Reiche et~al.(2020)Reiche, Dohme-Meier and Terlouw}]{Reiche2020}
\bibinfo{author}{Reiche, A.M.}, \bibinfo{author}{Dohme-Meier, F.}, \bibinfo{author}{Terlouw, E.C.}, \bibinfo{year}{2020}.
\newblock \bibinfo{title}{Effects of horn status on behaviour in fattening cattle in the field and during reactivity tests}.
\newblock \bibinfo{journal}{Applied Animal Behaviour Science} \bibinfo{volume}{231}, \bibinfo{pages}{105081}.
\newblock \DOIprefix\doi{10.1016/j.applanim.2020.105081}.
\bibitem[{Rosati(2012)}]{Rosati2012}
\bibinfo{author}{Rosati, A.}, \bibinfo{year}{2012}.
\newblock \bibinfo{title}{{ICAR standards and guidelines on animal identification and performance recording and its role as ISO Registration Authority for RFID}}.
\newblock \bibinfo{type}{Technical Report} \bibinfo{number}{4}. ICAR.
\newblock \URLprefix \url{https://www.icar.org/wp-content/uploads/2015/09/Rosati-.pdf}.
\bibitem[{Rudnik et~al.(2019)Rudnik, Ehrhart, Ferret, Teyssou, Troncy and Tannier}]{Rudnik2019}
\bibinfo{author}{Rudnik, C.}, \bibinfo{author}{Ehrhart, T.}, \bibinfo{author}{Ferret, O.}, \bibinfo{author}{Teyssou, D.}, \bibinfo{author}{Troncy, R.}, \bibinfo{author}{Tannier, X.}, \bibinfo{year}{2019}.
\newblock \bibinfo{title}{Searching news articles using an event knowledge graph leveraged by wikidata}, in: \bibinfo{booktitle}{Companion proceedings of the 2019 world wide web conference}, pp. \bibinfo{pages}{1232--1239}.
\newblock \DOIprefix\doi{10.1145/3308560.3316761}.
\bibitem[{Santos and Riyuiti(2012)}]{Santos2012}
\bibinfo{author}{Santos, C.}, \bibinfo{author}{Riyuiti, A.}, \bibinfo{year}{2012}.
\newblock \bibinfo{title}{An overview of the use of metadata in agriculture}.
\newblock \bibinfo{journal}{IEEE Latin America Transactions} \bibinfo{volume}{10}, \bibinfo{pages}{1265--1267}.
\newblock \DOIprefix\doi{10.1109/TLA.2012.6142471}.
\bibitem[{Schmitz et~al.(2009)Schmitz, Martini, Kunisch and M{\"o}singer}]{Schmitz2009}
\bibinfo{author}{Schmitz, M.}, \bibinfo{author}{Martini, D.}, \bibinfo{author}{Kunisch, M.}, \bibinfo{author}{M{\"o}singer, H.J.}, \bibinfo{year}{2009}.
\newblock \bibinfo{title}{agroxml enabling standardized, platform-independent internet data exchange in farm management information systems}, in: \bibinfo{booktitle}{Metadata and semantics}. \bibinfo{publisher}{Springer}, pp. \bibinfo{pages}{463--468}.
\newblock \DOIprefix\doi{10.1007/978-0-387-77745-0}.
\bibitem[{Sellman et~al.(2022)Sellman, Beck-Johnson, Hallman, Miller, Bonner, Portacci, Webb and Lindstr{\"o}m}]{Sellman2022}
\bibinfo{author}{Sellman, S.}, \bibinfo{author}{Beck-Johnson, L.M.}, \bibinfo{author}{Hallman, C.}, \bibinfo{author}{Miller, R.S.}, \bibinfo{author}{Bonner, K.A.O.}, \bibinfo{author}{Portacci, K.}, \bibinfo{author}{Webb, C.T.}, \bibinfo{author}{Lindstr{\"o}m, T.}, \bibinfo{year}{2022}.
\newblock \bibinfo{title}{Modeling us cattle movements until the cows come home: Who ships to whom and how many?}
\newblock \bibinfo{journal}{Computers and Electronics in Agriculture} \bibinfo{volume}{203}, \bibinfo{pages}{107483}.
\newblock \DOIprefix\doi{10.1016/j.compag.2022.107483}.
\bibitem[{Shahinfar and Kahn(2018)}]{Shahinfar2018}
\bibinfo{author}{Shahinfar, S.}, \bibinfo{author}{Kahn, L.}, \bibinfo{year}{2018}.
\newblock \bibinfo{title}{Machine learning approaches for early prediction of adult wool growth and quality in australian merino sheep}.
\newblock \bibinfo{journal}{Computers and electronics in agriculture} \bibinfo{volume}{148}, \bibinfo{pages}{72--81}.
\newblock \DOIprefix\doi{10.1016/j.compag.2018.03.001}.
\bibitem[{Smith et~al.(2020)Smith, McNally, Little, Ingham and Schmoelzl}]{Smith2020}
\bibinfo{author}{Smith, D.}, \bibinfo{author}{McNally, J.}, \bibinfo{author}{Little, B.}, \bibinfo{author}{Ingham, A.}, \bibinfo{author}{Schmoelzl, S.}, \bibinfo{year}{2020}.
\newblock \bibinfo{title}{Automatic detection of parturition in pregnant ewes using a three-axis accelerometer}.
\newblock \bibinfo{journal}{Computers and Electronics in Agriculture} \bibinfo{volume}{173}, \bibinfo{pages}{105392}.
\newblock \DOIprefix\doi{10.1016/j.compag.2020.105392}.
\bibitem[{Stegelmeier et~al.(2020)Stegelmeier, Davis and Clayton}]{Stegelmeier2020}
\bibinfo{author}{Stegelmeier, B.L.}, \bibinfo{author}{Davis, T.Z.}, \bibinfo{author}{Clayton, M.J.}, \bibinfo{year}{2020}.
\newblock \bibinfo{title}{Plant-induced reproductive disease, abortion, and teratology in livestock}.
\newblock \bibinfo{journal}{Veterinary Clinics of North America: Food Animal Practice} \bibinfo{volume}{36}, \bibinfo{pages}{735--743}.
\newblock \DOIprefix\doi{10.1016/j.cvfa.2020.08.004}.
\bibitem[{Stoddard and Cramer(2017)}]{Stoddard2017}
\bibinfo{author}{Stoddard, G.C.}, \bibinfo{author}{Cramer, G.}, \bibinfo{year}{2017}.
\newblock \bibinfo{title}{A review of the relationship between hoof trimming and dairy cattle welfare}.
\newblock \bibinfo{journal}{Veterinary Clinics: Food Animal Practice} \bibinfo{volume}{33}, \bibinfo{pages}{365--375}.
\newblock \DOIprefix\doi{10.1016/j.cvfa.2017.02.012}.
\bibitem[{Subach(2020)}]{Subach2020StructuringSubach}
\bibinfo{author}{Subach, Y.}, \bibinfo{year}{2020}.
\newblock \bibinfo{title}{{Structuring events in a messaging system}}.
\newblock \bibinfo{howpublished}{\url{https://yurisubach.com/2020/04/10/structuring-events/}}.
\newblock \bibinfo{note}{Accessed: 2022-02-03}.
\bibitem[{Subirats-Coll et~al.(2022)Subirats-Coll, Kolshus, Turbati, Stellato, Mietzsch, Martini and Zeng}]{Subirats-Coll2022}
\bibinfo{author}{Subirats-Coll, I.}, \bibinfo{author}{Kolshus, K.}, \bibinfo{author}{Turbati, A.}, \bibinfo{author}{Stellato, A.}, \bibinfo{author}{Mietzsch, E.}, \bibinfo{author}{Martini, D.}, \bibinfo{author}{Zeng, M.}, \bibinfo{year}{2022}.
\newblock \bibinfo{title}{Agrovoc: The linked data concept hub for food and agriculture}.
\newblock \bibinfo{journal}{Computers and Electronics in Agriculture} \bibinfo{volume}{196}, \bibinfo{pages}{105965}.
\newblock \DOIprefix\doi{10.1016/j.compag.2020.105965}.
\bibitem[{Theorin et~al.(2017)Theorin, Bengtsson, Provost, Lieder, Johnsson, Lundholm and Lennartson}]{Theorin2017An4.0}
\bibinfo{author}{Theorin, A.}, \bibinfo{author}{Bengtsson, K.}, \bibinfo{author}{Provost, J.}, \bibinfo{author}{Lieder, M.}, \bibinfo{author}{Johnsson, C.}, \bibinfo{author}{Lundholm, T.}, \bibinfo{author}{Lennartson, B.}, \bibinfo{year}{2017}.
\newblock \bibinfo{title}{An event-driven manufacturing information system architecture for industry 4.0}.
\newblock \bibinfo{journal}{International journal of production research} \bibinfo{volume}{55}, \bibinfo{pages}{1297--1311}.
\newblock \DOIprefix\doi{10.1080/00207543.2016.1201604}.
\bibitem[{Thesmar and Stevens(2019)}]{thesmar2019meat}
\bibinfo{author}{Thesmar, H.S.}, \bibinfo{author}{Stevens, S.K.}, \bibinfo{year}{2019}.
\newblock \bibinfo{title}{Meat and poultry traceability--its history and continuing challenges}.
\newblock \bibinfo{journal}{Food Traceability: From Binders to Blockchain} , \bibinfo{pages}{71--80}\DOIprefix\doi{10.1007/978-3-030-10902-8_6}.
\bibitem[{Trevarthen(2007)}]{Trevarthen2007}
\bibinfo{author}{Trevarthen, A.}, \bibinfo{year}{2007}.
\newblock \bibinfo{title}{The national livestock identification system: the importance of traceability in e-business}.
\newblock \bibinfo{journal}{Journal of Theoretical and Applied Electronic Commerce Research} \bibinfo{volume}{2}, \bibinfo{pages}{49--62}.
\newblock \DOIprefix\doi{10.3390/jtaer2010005}.
\bibitem[{Tuan et~al.(2022)Tuan, Rustia, Hsu and Lin}]{Tuan2022}
\bibinfo{author}{Tuan, S.A.}, \bibinfo{author}{Rustia, D.J.A.}, \bibinfo{author}{Hsu, J.T.}, \bibinfo{author}{Lin, T.T.}, \bibinfo{year}{2022}.
\newblock \bibinfo{title}{Frequency modulated continuous wave radar-based system for monitoring dairy cow respiration rate}.
\newblock \bibinfo{journal}{Computers and Electronics in Agriculture} \bibinfo{volume}{196}, \bibinfo{pages}{106913}.
\newblock \DOIprefix\doi{10.1016/j.compag.2022.106913}.
\bibitem[{Wang et~al.(2023)Wang, Shen, Zhang, Gao, Zhang, Xiaohui, Du and Qiu}]{Wang2023}
\bibinfo{author}{Wang, H.}, \bibinfo{author}{Shen, W.}, \bibinfo{author}{Zhang, Y.}, \bibinfo{author}{Gao, M.}, \bibinfo{author}{Zhang, Q.}, \bibinfo{author}{Xiaohui, A.}, \bibinfo{author}{Du, H.}, \bibinfo{author}{Qiu, B.}, \bibinfo{year}{2023}.
\newblock \bibinfo{title}{Diagnosis of dairy cow diseases by knowledge-driven deep learning based on the text reports of illness state}.
\newblock \bibinfo{journal}{Computers and Electronics in Agriculture} \bibinfo{volume}{205}, \bibinfo{pages}{107564}.
\newblock \DOIprefix\doi{10.1016/j.compag.2022.107564}.
\bibitem[{Weber et~al.(2020)Weber, de~Lima~Weber, da~Silva~Oliveira, Astolfi, Menezes, de~Andrade~Porto, Rezende, de~Moraes, Matsubara, Mateus et~al.}]{Weber2020}
\bibinfo{author}{Weber, V.A.M.}, \bibinfo{author}{de~Lima~Weber, F.}, \bibinfo{author}{da~Silva~Oliveira, A.}, \bibinfo{author}{Astolfi, G.}, \bibinfo{author}{Menezes, G.V.}, \bibinfo{author}{de~Andrade~Porto, J.V.}, \bibinfo{author}{Rezende, F.P.C.}, \bibinfo{author}{de~Moraes, P.H.}, \bibinfo{author}{Matsubara, E.T.}, \bibinfo{author}{Mateus, R.G.}, et~al., \bibinfo{year}{2020}.
\newblock \bibinfo{title}{Cattle weight estimation using active contour models and regression trees bagging}.
\newblock \bibinfo{journal}{Computers and electronics in agriculture} \bibinfo{volume}{179}, \bibinfo{pages}{105804}.
\newblock \DOIprefix\doi{10.1016/j.compag.2020.105804}.
\bibitem[{Whitacre et~al.(2014)Whitacre, Mark and Griffin}]{Whitacre2014}
\bibinfo{author}{Whitacre, B.E.}, \bibinfo{author}{Mark, T.B.}, \bibinfo{author}{Griffin, T.W.}, \bibinfo{year}{2014}.
\newblock \bibinfo{title}{How connected are our farms?}
\newblock \bibinfo{journal}{Choices} \bibinfo{volume}{29}, \bibinfo{pages}{1--9}.
\newblock \URLprefix \url{https://ageconsearch.umn.edu/record/188271/files/cmsarticle_392.pdf}.
\bibitem[{White et~al.(2013)White, Hunt, Boote, Jones, Koo, Kim, Porter, Wilkens and Hoogenboom}]{White2013}
\bibinfo{author}{White, J.W.}, \bibinfo{author}{Hunt, L.}, \bibinfo{author}{Boote, K.J.}, \bibinfo{author}{Jones, J.W.}, \bibinfo{author}{Koo, J.}, \bibinfo{author}{Kim, S.}, \bibinfo{author}{Porter, C.H.}, \bibinfo{author}{Wilkens, P.W.}, \bibinfo{author}{Hoogenboom, G.}, \bibinfo{year}{2013}.
\newblock \bibinfo{title}{Integrated description of agricultural field experiments and production: The icasa version 2.0 data standards}.
\newblock \bibinfo{journal}{Computers and electronics in agriculture} \bibinfo{volume}{96}, \bibinfo{pages}{1--12}.
\newblock \DOIprefix\doi{10.1016/j.compag.2013.04.003}.
\bibitem[{Whiting et~al.(2013)Whiting, Brown, Browne, Hadley and Knowles}]{Whiting2013}
\bibinfo{author}{Whiting, K.J.}, \bibinfo{author}{Brown, S.N.}, \bibinfo{author}{Browne, W.J.}, \bibinfo{author}{Hadley, P.J.}, \bibinfo{author}{Knowles, T.G.}, \bibinfo{year}{2013}.
\newblock \bibinfo{title}{The anterior tooth development of cattle presented for slaughter: an analysis of age, sex and breed}.
\newblock \bibinfo{journal}{animal} \bibinfo{volume}{7}, \bibinfo{pages}{1323--1331}.
\newblock \DOIprefix\doi{10.1017/S1751731113000499}.
\bibitem[{Wismans(1999)}]{Wismans1999}
\bibinfo{author}{Wismans, W.}, \bibinfo{year}{1999}.
\newblock \bibinfo{title}{Identification and registration of animals in the european union}.
\newblock \bibinfo{journal}{Computers and electronics in agriculture} \bibinfo{volume}{24}, \bibinfo{pages}{99--108}.
\newblock \DOIprefix\doi{10.1016/S0168-1699(99)00040-X}.
\bibitem[{Wolf and Wicksteed(1998)}]{Wolf1997}
\bibinfo{author}{Wolf, M.}, \bibinfo{author}{Wicksteed, C.}, \bibinfo{year}{1998}.
\newblock \bibinfo{title}{Date and time formats}.
\newblock \bibinfo{journal}{W3C} \URLprefix \url{http://www.w3.org/TR/NOTE-datetime}.
\bibitem[{Yuan et~al.(2013)Yuan, Cong, Ma, Sun and Thalmann}]{Yuan2013}
\bibinfo{author}{Yuan, Q.}, \bibinfo{author}{Cong, G.}, \bibinfo{author}{Ma, Z.}, \bibinfo{author}{Sun, A.}, \bibinfo{author}{Thalmann, N.M.}, \bibinfo{year}{2013}.
\newblock \bibinfo{title}{Who, where, when and what: discover spatio-temporal topics for twitter users}, in: \bibinfo{booktitle}{Proceedings of the 19th ACM SIGKDD international conference on Knowledge discovery and data mining}, pp. \bibinfo{pages}{605--613}.
\newblock \DOIprefix\doi{10.1145/2487575.2487576}.
\bibitem[{Yuan et~al.(2015)Yuan, Cong, Zhao, Ma and Sun}]{Yuan2015}
\bibinfo{author}{Yuan, Q.}, \bibinfo{author}{Cong, G.}, \bibinfo{author}{Zhao, K.}, \bibinfo{author}{Ma, Z.}, \bibinfo{author}{Sun, A.}, \bibinfo{year}{2015}.
\newblock \bibinfo{title}{Who, where, when, and what: A nonparametric bayesian approach to context-aware recommendation and search for twitter users}.
\newblock \bibinfo{journal}{ACM Transactions on Information Systems (TOIS)} \bibinfo{volume}{33}, \bibinfo{pages}{1--33}.
\newblock \DOIprefix\doi{10.1145/2699667}.
\bibitem[{Zhou et~al.(2017)Zhou, Chen, Zhang and He}]{Zhou2017}
\bibinfo{author}{Zhou, D.}, \bibinfo{author}{Chen, L.}, \bibinfo{author}{Zhang, X.}, \bibinfo{author}{He, Y.}, \bibinfo{year}{2017}.
\newblock \bibinfo{title}{Unsupervised event exploration from social text streams}.
\newblock \bibinfo{journal}{Intelligent Data Analysis} \bibinfo{volume}{21}, \bibinfo{pages}{849--866}.
\newblock \DOIprefix\doi{10.3233/IDA-160048}.

\end{thebibliography}

\appendix

\section{Table}

\begin{longtable}{lc p{7.9cm}}
\caption{Datatype used in LEI events}
\label{tab:leidatatypes}\\
\hline
Datatype & Source & Note \\ \hline
\endfirsthead

\multicolumn{3}{c}%
{{ \tablename\ \thetable{} -- continued from previous page}} \\
\hline
Datatypes & Source & Note \\ \hline
\endhead
iscTransactionCostType & ISC & Describes the animal transaction cost. \\
icarAnimalCoreResource & ICAR & Schema to represent animal details. \\
icarAnimalIdentifierType & ICAR & Unique animal scheme and identifier combination. \\
iscDiagnosisType & ISC & Diagnosis description for an animal. \\
iscOrganisationType & ISC & Organisation or farm owner details. \\
icarFeedDurationType & ICAR & The length of feeding time. \\
icarAnimalSpecieType & ICAR & Enumeration for species of animal, we added 2 additional items (``Camel", ``Kangaroo"). \\
icarDateTimeType & ICAR & ISO8601 date and time. \\
icarBreedIdentifierType & ICAR & Identifies a breed using a scheme and ID. \\
icarBreedFractionsType & ICAR & Type of the proportion of the denominator that this breed comprises. \\
icarMilkCharacteristicsType & ICAR & Milk characteristics of the quarter. \\
icarMilkingRemarksType & ICAR & Enumeration for different possible milking remarks. \\
icarReproCalvingEaseType & ICAR & Enumeration for calving ease. \\
uncefactMassUnitsType & ICAR & Enumeration for mass units for the weight from UN/CEFACT trade facilitation recommendation (Kilogram, Gram, Pound, Metric Ton, Microgram, Milligram, Ounce, Pound net.) \\ 
icarArrivalReasonType & ICAR & Enumeration for animal arrival reason to a specific property (farm, slaughter yard). \\ 
iscConsignmentType & ISC & Details of stock movement shipment should include origin, destination, loading and unloading times, as well as driver information. \\
icarProductionPurposeType & ICAR & An enumeration defines the primary product for which this animal is bred or kept. If animals are kept for breeding or live trade (sale), specify the end purpose of that breeding/trade (meat, milk, wool). 1 extra value (Pet) had been added. \\
icarAnimalStatusType & ICAR & An enumeration defines the status of the animal either absolutely and/or with respect to the location on which it is recorded. Off-farm means that the animal is no longer recorded at the location. \\
icarAnimalReproductionStatusType & ICAR & Enumeration for different possible reproduction statuses of an animal. \\
icarAnimalLactationStatusType & ICAR & Enumeration for different possible lactation statuses of an animal. \\
icarParentageType & ICAR & Use this type to define a parent of an animal. \\
icarRegistrationReasonType & ICAR & Enumeration for registration reason: born or registered (induct existing animal). \\
iscWithdrawalType & ISC & Withholding period with an end date that applies to the specific food chain due to a task that occurred to the animal, such as treatment administered. \\
icarTraitAmountType & ICAR & Type for measuring by kilogram or pound. \\
icarDeathReasonType & ICAR & Coded reasons for death, including disease, complications of gestation, and consumption by humans or animals. \\
icarDepartureKindType & ICAR & Enumeration for the kind of departure. Type of destination or transfer, including the agistment, which refers to the case where cattle are taken in to feed on pastureland upon contract. \\
icarDepartureReasonType & ICAR & Enumeration for the cause of departure. \\
icarConsumedFeedType & ICAR & Provide the consumed feed and the amount the animal was entitled to. \\
icarConsumedRationType & ICAR & Indicate both the amount of food consumed and available to the animal. \\
icarReproHeatDetectionMethodType & ICAR & Enumeration for the method of detecting the heat of an animal. \\
icarReproHeatCertaintyType & ICAR & Enumeration for the certainty of specific heat. \\
icarDeviceReferenceType & ICAR & The details of a device (model, which represents manufacturer, model, hardware, software version and serial number) that performs a specific task, for example, weight. \\
icarReproHeatSignType & ICAR & Enumeration for the signs of heat (Slime, Clear slime, Interested in other animals, Standing under, Bawling, Blood). \\
icarReproHeatIntensityType & ICAR & Enumeration for the method of insemination (Very weak, Weak, Normal, Strong, Very strong). \\
icarReproInseminationType & ICAR & Enumeration for the insemination method (natural service, run with bull, insemination, implantation). \\
icarReproSemenStrawResource & ICAR & Describes a semen straw. \\
icarReproEmbryoResource & ICAR & Describes an implanted embryo. \\
icarLocationIdentifierType & ICAR & Location identifier. \\
iscLotAssessmentType & ISC & Assessment for the lot to change ownership in order of sale or purchase. \\
icarMilkDurationType & ICAR & The time spent milking the milk. \\
icarMilkingTypeCode & ICAR & The type of milking (manual or automated). \\
icarMilkingMilkWeightType & ICAR & The amount of milked milk. \\
icarQuarterMilkingType & ICAR & Provides the milking result and optionally sampling and characteristic details. \\
icarAnimalMilkingSampleType & ICAR & Describes the details of a milk sample taken from the quarter for laboratory analyses. \\
icarReproPregnancyResultType & ICAR & Enumeration for the result of pregnancy diagnosis (empty/pregnant). \\
icarReproSemenPreservationType & ICAR & Enumeration for the method of semen preservation (liquid, usually with extender, frozen). \\
icarAnimalGenderType & ICAR & Enumeration for the sex of the animal. \\
iscMedicineReferenceType & ISC & Basic information about medicine. \\
iscDoseType & ISC & Quantity of dose administered. \\
iscCourseSummaryType & ISC & Medicine course summary. \\
icarWeightMethodType & ICAR & The method by which the weight is observed. Includes loadcell (loadbars), girth (tape), assessed (visually), walk-over weighing, prediction, imaging (camera / IR), front-end weight correlated with the whole body, and group average (pen/sample weigh). \\
icarDeathDisposalMethodType & ICAR & Coded disposal methods include approved service, consumption by humans or animals, etc. \\ 
leiAbortionMethodTypes & LEI & Enumeration for the abortion methods. \\
leiLotDetailType & LEI & The lot details to change ownership in order of sale or purchase. \\
leiScoresTypes & LEI & Enumeration for the score's name. \\
leiEventsTypes & LEI & Enumeration for all event names. \\
leiAbortionReasonTypes & LEI & Enumeration for abortion reason. \\
leiCastrateMethod & LEI & Enumeration for the castration methods. \\
leiDehorningMethod & LEI & Enumeration for the dehorning tools. \\
leiWeaningMethod & LEI & Enumeration for the weaning method. \\
leiWeaningReason & LEI & Enumeration for the wean reason. \\
eventOwnerType & LEI & It defines that the object should have a property called ``id" which is required.
It also defines the object as a combination of two types, represented by the ``allOf" keyword. The object must have all the properties and constraints defined in the schema referenced by ``\$ref": ``ISC/types/iscOrganisationType.json" and ``\$ref": ``ISC/types/iscPersonType.json"
This means that the object should have all the properties defined in the files ``iscOrganisationType.json" and ``iscPersonType.json".
The schema describes a data owner, which can be a cattle producer, a crop farmer, or any other type of owner and should have a unique identifier. \\
eventSource & LEI & It defines that the object should be a combination of two types, represented by the ``allOf" keyword. This means that the object should have all the properties defined in the ``icarDeviceResource.json" file and the properties defined in the second element of the allOf array. The second element of the allOf array has a property called ``ip\_address". The ``ip\_address" property is a string type. It should be in the format ``ipv4".
This schema describes the device or software from which the event originates. \\

itemType & LEI & It defines that the object should have a property called ``itemType" which is a string type, and it is referred to another JSON file, ``itemsTypes.json".
It also defines the object as one of three types, represented by the ``oneOf" keyword. Each type has its own set of required properties and constraints on the values of those properties. For example, if the ``itemType" is ``Animals", the object must have a property called ``animal" which is an object type, and it is referred to another JSON file ``ICAR/resources/icarAnimalsCoreResource.json".\\

\hline
\end{longtable}

\section{Code}

\nolinenumbers

\begin{tcolorbox}[
breakable,
  arc=3mm,
  outer arc=3mm,
  boxrule=0.0mm,
  boxsep=0mm,
  left=3mm,
  right=3mm,
  top=3mm,
  bottom=3mm,
]
\begin{lstlisting}[language=Java, caption={Java function for LEI schema validation},label={lst:schemavalidationcode}]
public void validation (String data, String schema) {  
    // create instance of the ObjectMapper class  
    ObjectMapper objectMapper = new ObjectMapper();  
    // create an instance of the JsonSchemaFactory using version flag  
    JsonSchemaFactory schemaFactory = JsonSchemaFactory.getInstance( SpecVersion.VersionFlag.V201909 );  
    
    // store the JSON data in InputStream 
    try {  
    // read data from the stream and store it into JsonNode  
    JsonNode json = objectMapper.readTree(data);  
    // get schema from the schemaStream and store it into JsonSchema  
    JsonSchema jschema = schemaFactory.getSchema(schema) ; 
    // create set of validation message and store result in it  
    Set<ValidationMessage> validationResult = jschema.validate( json );  
        // show the validation errors   
        if (validationResult.isEmpty()) {
            // show custom message if there is no validation error   
        	System.out.println("There is no validation errors");  
        } else {  
        	System.out.println("There are errors");  
            // show all the validation error  
            validationResult.forEach(vm -> System.out.println(vm.getMessage()));
        }  
    } catch(Exception e) {e.printStackTrace();}
}
\end{lstlisting}
\end{tcolorbox}

\begin{tcolorbox}[
breakable,
  arc=3mm,
  outer arc=3mm,
  boxrule=0.0mm,
  boxsep=0mm,
  left=3mm,
  right=3mm,
  top=3mm,
  bottom=3mm,
]
\begin{lstlisting}[language=json,caption= {Movement event - Arrival for agistment},label={arrival2}]
{
 "eventDateTime": "2021-01-30T14:13:23Z",
 "source": {
   "id": "123",
   "ip_address": "1.1.1.1",
   "manufacturer": {
     "id": "Apple"
   }
 },
 "owner": {
   "id": "A123ABCD",
   "name": "Producer A Beef",
   "email": "prodA@beef.com",
   "givenName": "Aaron",
   "familyName": "Farmer"
 },
 "message": {
   "eventName": "Arrival",
   "item": {
     "itemType": "Animal",
     "animal": {
       "identifier": {
         "id": "982 123456789101",
         "scheme": "rfid"
       },
       "specie": "Cattle",
       "gender": "Male"
     }
   },
   "session": {
     "sessionID": "un1Que1d2",
     "totalInSession": 50
   },
   "event": {
     "arrivalReason": "Agistment",
     "shipment": {
       "id": {
         "scheme": "agreement",
         "id": "123456789"
       },
       "origin": {
         "id": "D123ABCD",
         "name": "Pro Ducer",
         "email": "proDucer@beef.com"         
       },
       "destination": {
         "id": "A123ABCD",
         "name": "Producer A Beef",
         "email": "prodA@beef.com"
       },
       "vehicles": [
         "ABC123"
       ]
     },
     "changeOfOwnership": null
   }
  }
}
\end{lstlisting}
\end{tcolorbox}

\begin{tcolorbox}[
breakable,
  arc=3mm,
  outer arc=3mm,
  boxrule=0.0mm,
  boxsep=0mm,
  left=3mm,
  right=3mm,
  top=3mm,
  bottom=3mm,
]
\begin{lstlisting}[language=json,caption= {Movement event - Arrival for purchase},label={arrival1}]
{
 "eventDateTime": "2021-01-19T14:13:23Z",
 "source": {
   "id": "123",
   "ip_address": "128.0.0.0",
   "manufacturer": {
     "id": "Apple"
   }
 },
 "owner": {
   "id": "B123ABCD",
   "name": "Producer B 4 Beef",
   "email": "b4beef@beef.com",
   "givenName": "Bob",
   "familyName": "Farmer"
 },
 "message": {
   "eventName": "Arrival",
   "item": {
     "itemType": "Animal",
     "animal": {
       "identifier": {
         "id": "982 123456789101",
         "scheme": "rfid"
       },
       "specie": "Cattle",
       "gender": "Male"
     }
   },
   "session": {
     "sessionID": "un1Que1d1",
     "totalInSession": 50
   },
   "event": {
     "arrivalReason": "Purchase",
     "shipment": {
       "id": {
         "scheme": "sale invoice",
         "id": "123456789"
       },
       "origin": {
         "id": "A123ABCD",
         "name": "Producer A Beef",
         "email": "prodA@beef.com"
       },
       "destination": {
         "id": "B123ABCD",
         "name": "Producer B 4 Beef",
         "email": "b4beef@beef.com"
       },
       "vehicles": [
         "ABC123",
         "CBA321"
       ]
     },
     "changeOfOwnership": {
       "lot": {
         "lotId": "1256"
       },
       "purchaser": {
         "id": "B123ABCD",
         "name": "Producer B 4 Beef",
         "email": "b4beef@beef.com"
       },
       "transactionCost": {
         "currency": "AUD",
         "totalCost": 42250
       }
     }
   }
  }
}
\end{lstlisting}
\end{tcolorbox}

\begin{tcolorbox}[
breakable,
  arc=3mm,
  outer arc=3mm,
  boxrule=0.0mm,
  boxsep=0mm,
  left=3mm,
  right=3mm,
  top=3mm,
  bottom=3mm,
]
\begin{lstlisting}[language=json,caption= {Movement event - Departure for agistment},label={departure2}]
{
  "eventDateTime": "2021-01-27T14:13:23Z",
  "source": {
    "id": "Prod D Computer",
    "ip_address": "128.0.0.0",
    "manufacturer": {
      "id": "HP"
    }
  },
  "owner": {
    "id": "D123ABCD",
    "name": "Pro Ducer",
    "email": "proDucer@beef.com",
    "givenName": "Dorothy",
    "familyName": "Farmer"
  },
  "message": {
    "eventName": "Departure",
    "item": {
      "itemType": "Animal",
      "animal": {
        "identifier": {
          "id": "982 123456789101",
          "scheme": "rfid"
        },
        "specie": "Cattle",
        "gender": "Male"
      }
    },
    "session": {
      "sessionID": "un1Que1d2",
      "totalInSession": 50
    },
    "event": {
      "departureKind": "Agistment",
      "departureReason": "Production",
      "shipment": {
        "id": {
          "scheme": "agreement",
          "id": "123456789"
        },
        "origin": {
          "id": "D123ABCD",
          "name": "Pro Ducer",
          "email": "proDucer@beef.com"
        },
        "destination": {
          "id": "A123ABCD",
          "name": "Producer A Beef",
          "email": "prodA@beef.com"
        },
        "vehicles": [
          "ABC123"
        ]
      },
      "changeOfOwnership": null
    }
  }
}
\end{lstlisting}
\end{tcolorbox}

\begin{tcolorbox}[
breakable,
  arc=3mm,
  outer arc=3mm,
  boxrule=0.0mm,
  boxsep=0mm,
  left=3mm,
  right=3mm,
  top=3mm,
  bottom=3mm,
]
\begin{lstlisting}[language=json,caption= {Audit event},label={audit}]
{
  "eventDateTime": "2021-02-14T16:13:23Z",
  "source": {
    "id": "123",
    "ip_address": "128.0.0.0",
    "manufacturer": {
      "id": "Apple"
    }
  },
  "owner": {
    "id": "C123ABCD",
    "name": "Pro Cattle",
    "email": "proCattle@beef.com",
    "givenName": "Carol",
    "familyName": "Cattle"
  },
  "message": {
    "eventName": "Audit",
    "item": {
      "itemType": "Animal",
      "animal": {
        "identifier": {
          "id": "",
          "scheme": ""
        },
        "specie": "Cattle",
        "gender": "Unknown"
      }
    },
    "event": [
      {
        "specie": "Cattle",
        "breed": "Cows",
        "stock": 100,
        "note": ""
      },
      {
        "specie": "Cattle",
        "breed": "Heifers",
        "stock": 0,
        "note": ""
      },
      {
        "specie": "Cattle",
        "breed": "Steers",
        "stock": 3,
        "note": "found 3"
      },
      {
        "specie": "Cattle",
        "breed": "Bulls",
        "stock": 28,
        "note": ""
      },
      {
        "specie": "Cattle",
        "breed": "Bulls calves",
        "stock": 50,
        "note": ""
      }
    ]
  }
}
\end{lstlisting}
\end{tcolorbox}

\begin{tcolorbox}[
breakable,
  arc=3mm,
  outer arc=3mm,
  boxrule=0.0mm,
  boxsep=0mm,
  left=3mm,
  right=3mm,
  top=3mm,
  bottom=3mm,
]
\begin{lstlisting}[language=json,caption= {Birth event},label={birth}]
{
 "eventDateTime": "2022-09-29T13:35:25Z",
 "source": {
   "id": "Aarons Mobile",
   "ip_address": "1.1.1.1",
   "manufacturer": {
     "id": "Apple"
   }
 },
 "owner": {
   "id": "A123ABCD",
   "name": "Producer A Beef",
   "email": "prodA@beef.com",
   "givenName": "Aaron",
   "familyName": "Farmer"
 },
 "message": {
   "eventName": "Birth",
   "item": {
     "itemType": "Animal",
     "animal": {
       "identifier": {
         "id": "1-29sep2022",
         "scheme": "VID"
       },
       "specie": "Cattle",
       "gender": "Male"
     }
   },
   "session": {
     "sessionID": "78975",
     "totalInSession": 20
   },
   "event": {
     "registrationReason": "Born",
     "EID": "",
     "VID": "1-29sep2022"
   }
 }
}
\end{lstlisting}
\end{tcolorbox}

\begin{tcolorbox}[
breakable,
  arc=3mm,
  outer arc=3mm,
  boxrule=0.0mm,
  boxsep=0mm,
  left=3mm,
  right=3mm,
  top=3mm,
  bottom=3mm,
]
\begin{lstlisting}[language=json,caption= {Castrated event},label={castrated}]
{
 "eventDateTime": "2023-01-29T13:35:25Z",
 "source": {
   "id": "Aarons Computer",
   "ip_address": "1.1.1.1",
   "manufacturer": {
     "id": "Apple"
   }
 },
 "owner": {
   "id": "A123ABCD",
   "name": "Producer A Beef",
   "email": "prodA@beef.com",
   "givenName": "Aaron",
   "familyName": "Farmer"
 },
 "message": {
   "eventName": "Castrate",
   "item": {
     "itemType": "Animal",
     "animal": {
       "identifier": {
         "id": "900 000000000203",
         "scheme": "rfid"
       },
       "specie": "Cattle",
       "gender": "Male"
     }
   },
   "session": {
     "sessionID": "Dec-2022-Calves-Castration",
     "totalInSession": 13
   },
   "event": {
     "castrateMethod": "Surgical",
     "note": "Removed using a scalpel",
     "responsible": "Producer A 0400 111 111"
   }
 }
}
\end{lstlisting}
\end{tcolorbox}

\begin{tcolorbox}[
breakable,
  arc=3mm,
  outer arc=3mm,
  boxrule=0.0mm,
  boxsep=0mm,
  left=3mm,
  right=3mm,
  top=3mm,
  bottom=3mm,
]
\begin{lstlisting}[language=json,caption= {Death event - Injured},label={death1}]
{
  "eventDateTime": "2021-02-05T16:13:23Z",
  "source": {
    "id": "123",
    "ip_address": "128.0.0.0",
    "manufacturer": {
      "id": "Apple"
    }
  },
  "owner": {
    "id": "C123ABCD",
    "name": "Pro Cattle",
    "email": "proCattle@beef.com",
    "givenName": "Carol",
    "familyName": "Cattle"
  },
  "message": {
    "eventName": "Death",
    "item": {
      "itemType": "Animal",
      "animal": {
        "identifier": {
          "id": "600 000000000199",
          "scheme": "rfid"
        },
        "specie": "Cattle",
        "gender": "Female"
      }
    },
    "session": {
      "sessionID": "5",
      "totalInSession": 1
    },
    "event": {
      "deathReason": "Injured",
      "explanation": "injured during transportation",
      "disposalMethod": "OnPremise",
      "disposalOperator": "Vessna Vet",
      "disposalReference": "receipt123"
    }
  }
}
\end{lstlisting}
\end{tcolorbox}

\begin{tcolorbox}[
breakable,
  arc=3mm,
  outer arc=3mm,
  boxrule=0.0mm,
  boxsep=0mm,
  left=3mm,
  right=3mm,
  top=3mm,
  bottom=3mm,
]
\begin{lstlisting}[language=json,caption= {Death event - Consumption},label={death2}]
{
  "eventDateTime": "2021-03-01T09:53:23Z",
  "source": {
    "id": "TEYS Online System",
    "ip_address": "128.0.0.0"
  },
  "owner": {
    "id": "E123ABCD",
    "name": "Eden Aba",
    "email": "ed@beef.com",
    "givenName": "Eden",
    "familyName": "Aba"
  },
  "message": {
    "eventName": "Death",
    "item": {
      "itemType": "Animal",
      "animal": {
        "identifier": {
          "id": "982 123456789101",
          "scheme": "rfid"
        },
        "specie": "Cattle",
        "gender": "Male"
      }
    },
    "session": {
      "sessionID": "un1q31d",
      "totalInSession": 25
    },
    "event": {
      "deathReason": "Consumption",
      "explanation": "Slaughter",
      "disposalMethod": "ApprovedService",
      "disposalOperator": "TEYS",
      "disposalReference": "s147987"
    }
  }
}
\end{lstlisting}
\end{tcolorbox}

\begin{tcolorbox}[
breakable,
  arc=3mm,
  outer arc=3mm,
  boxrule=0.0mm,
  boxsep=0mm,
  left=3mm,
  right=3mm,
  top=3mm,
  bottom=3mm,
]
\begin{lstlisting}[language=json,caption= {Insemination event - Artificial insemination},label={insemination1}]
{
  "eventDateTime": "2021-10-11T07:30:23Z",
  "source": {
    "id": "Aarons Computer",
    "ip_address": "1.1.1.1",
    "manufacturer": {
      "id": "Apple"
    }
  },
  "owner": {
    "id": "A123ABCD",
    "name": "Producer A Beef",
    "email": "prodA@beef.com",
    "givenName": "Aaron",
    "familyName": "Farmer"
  },
  "message": {
    "eventName": "Insemination",
    "item": {
      "itemType": "Animal",
      "animal": {
        "identifier": {
          "id": "982 123456789101",
          "scheme": "rfid"
        },
        "specie": "Cattle",
        "gender": "Female"
      }
    },
    "session": {
      "sessionID": "11871198",
      "totalInSession": 55
    },
    "event": {
      "rank": 3,
      "inseminationType": "Insemination",
      "sireIdentifiers": [
        {
          "id": "982 111111111111",
          "scheme": "rfid"
        }
      ],
      "sireOfficialName": "Big Bull",
      "sireURI": "02347234-UKE98234982734987",
      "straw": {
        "id": {
          "id": "CHA8SDCKJ",
          "scheme": "IMV Identification"
        },
        "isSexedSemen": true,
        "sexedGender": "Female",
        "preservationType": "Frozen"
      },
      "eventEndDateTime": "2021-10-11T07:30:23Z",
      "semenFromFarmStocks": true,
      "farmContainer": "125a 2021-10-03"
    }
  }
}
\end{lstlisting}
\end{tcolorbox}

\begin{tcolorbox}[
breakable,
  arc=3mm,
  outer arc=3mm,
  boxrule=0.0mm,
  boxsep=0mm,
  left=3mm,
  right=3mm,
  top=3mm,
  bottom=3mm,
]
\begin{lstlisting}[language=json,caption= {Insemination event - Run with bull},label={insemination2}]
{
  "eventDateTime": "2021-10-01T07:30:23Z",
  "source": {
    "id": "Prod A Computer",
    "ip_address": "128.0.0.0",
    "manufacturer": {
      "id": "HP"
    }
  },
  "owner": {
    "id": "A123ABCD",
    "name": "Producer A Beef",
    "email": "prodA@beef.com",
    "givenName": "Aaron",
    "familyName": "Farmer"
  },
  "message": {
    "eventName": "Insemination",
    "item": {
      "itemType": "Animal",
      "animal": {
        "identifier": {
          "id": "200 000000000005",
          "scheme": "rfid"
        },
        "specie": "Cattle",
        "gender": "Female"
      }
    },
    "session": {
      "sessionID": "178917",
      "totalInSession": 20
    },
    "event": {
      "rank": 3,
      "inseminationType": "RunWithBull",
      "sireIdentifiers": [
        {
          "id": "200 000000000200",
          "scheme": "rfid"
        }
      ],
      "sireOfficialName": "",
      "sireURI": "",
      "straw": {
        
      },
      "eventEndDateTime": "2021-10-1T07:30:23Z",
      "semenFromFarmStocks": true,
      "farmContainer": "",
      "embryo": {
        
      }
    }
  }
}
\end{lstlisting}
\end{tcolorbox}

\begin{tcolorbox}[
breakable,
  arc=3mm,
  outer arc=3mm,
  boxrule=0.0mm,
  boxsep=0mm,
  left=3mm,
  right=3mm,
  top=3mm,
  bottom=3mm,
]
\begin{lstlisting}[language=json,caption= {Parturition event},label={parturition}]
{
  "eventDateTime": "2022-09-29T13:35:25Z",
  "source": {
    "id": "Aarons Mobile",
    "ip_address": "1.1.1.1",
    "manufacturer": {
      "id": "Apple"
    }
  },
  "owner": {
    "id": "A123ABCD",
    "name": "Producer A Beef",
    "email": "prodA@beef.com",
    "givenName": "Aaron",
    "familyName": "Farmer"
  },
  "message": {
    "eventName": "Parturition",
    "item": {
      "itemType": "Animal",
      "animal": {
        "identifier": {
          "id": "982 123456789101",
          "scheme": "rfid"
        },
        "specie": "Cattle",
        "gender": "Female"
      }
    },
    "session": {
      "sessionID": "6454456",
      "totalInSession": 20
    },
    "event": {
      "isEmbryoImplant": false,
      "damParity": 1,
      "liveProgeny": 1,
      "totalProgeny": 1,
      "calvingEase": "DifficultVeterinaryCare",
      "progeny": [
        {
          "identifier": {
            "id": "1-29sep2022",
            "scheme": "VID"
          },
          "specie": "Cattle",
          "gender": "Male"
        }
      ]
    }
  }
}
\end{lstlisting}
\end{tcolorbox}

\begin{tcolorbox}[
breakable,
  arc=3mm,
  outer arc=3mm,
  boxrule=0.0mm,
  boxsep=0mm,
  left=3mm,
  right=3mm,
  top=3mm,
  bottom=3mm,
]
\begin{lstlisting}[language=json,caption= {Pregnancy check event},label={pergnancycheck}]
{
  "eventDateTime": "2021-12-20T12:35:25Z",
  "source": {
    "id": "Aarons Computer",
    "ip_address": "1.1.1.1",
    "manufacturer": {
      "id": "Apple"
    }
  },
  "owner": {
    "id": "A123ABCD",
    "name": "Producer A Beef",
    "email": "prodA@beef.com",
    "givenName": "Aaron",
    "familyName": "Farmer"
  },
  "message": {
    "eventName": "PregnancyCheck",
    "item": {
      "itemType": "Animal",
      "animal": {
        "identifier": {
          "id": "200 000000000005",
          "scheme": "rfid"
        },
        "specie": "Cattle",
        "gender": "Female"
      }
    },
    "session": {
      "sessionID": "6454456",
      "totalInSession": 20
    },
    "event": {
      "checkMethod": "Palpation",
      "result": "Pregnant",
      "foetalAge": 21,
      "foetusCount": 1,
      "foetusCountMale": 1,
      "foetusCountFemale": 0,
      "exceptions": [
        "normal",
        "due around September to October of 2022"
      ]
    }
  }
}
\end{lstlisting}
\end{tcolorbox}

\begin{tcolorbox}[
breakable,
  arc=3mm,
  outer arc=3mm,
  boxrule=0.0mm,
  boxsep=0mm,
  left=3mm,
  right=3mm,
  top=3mm,
  bottom=3mm,
]
\begin{lstlisting}[language=json,caption= {Registration event},label={registration}]
{
  "eventDateTime": "2022-09-30T13:35:25Z",
  "source": {
    "id": "Aarons Mobile",
    "ip_address": "1.1.1.1",
    "manufacturer": {
      "id": "Apple"
    }
  },
  "owner": {
    "id": "A123ABCD",
    "name": "Producer A Beef",
    "email": "prodA@beef.com",
    "givenName": "Aaron",
    "familyName": "Farmer"
  },
  "message": {
    "eventName": "Registration",
    "item": {
      "itemType": "Animal",
      "animal": {
        "identifier": {
          "id": "900 000000000203",
          "scheme": "rfid"
        },
        "specie": "Cattle",
        "gender": "Male"
      }
    },
    "session": {
      "sessionID": "14785",
      "totalInSession": 20
    },
    "event": {
      "registrationReason": "Born"
    }
  }
}
\end{lstlisting}
\end{tcolorbox}

\begin{tcolorbox}[
breakable,
  arc=3mm,
  outer arc=3mm,
  boxrule=0.0mm,
  boxsep=0mm,
  left=3mm,
  right=3mm,
  top=3mm,
  bottom=3mm,
]
\begin{lstlisting}[language=json,caption= {Status observed event},label={statusobserved}]
{
  "eventDateTime": "2021-02-05T14:13:23Z",
  "source": {
    "id": "123",
    "ip_address": "128.0.0.0",
    "manufacturer": {
      "id": "Apple"
    }
  },
  "owner": {
    "id": "C123ABCD",
    "name": "Pro Cattle",
    "email": "proCattle@beef.com",
    "givenName": "Carol",
    "familyName": "Cattle"
  },
  "message": {
    "eventName": "StatusObserved",
    "item": {
      "itemType": "Animal",
      "animal": {
        "identifier": {
          "id": "600 000000000199",
          "scheme": "rfid"
        },
        "specie": "Cattle",
        "gender": "Male"
      }
    },
    "session": {
      "sessionID": "123",
      "totalInSession": 100
    },
    "event": {
      "observedStatus": "Injured",
      "note": "found 1 of the steers was injured on the truck and needs to be euthanized RFID: 600 000000000199",
      "responsible": "Luke"
    }
  }
}
\end{lstlisting}
\end{tcolorbox}

\begin{tcolorbox}[
breakable,
  arc=3mm,
  outer arc=3mm,
  boxrule=0.0mm,
  boxsep=0mm,
  left=3mm,
  right=3mm,
  top=3mm,
  bottom=3mm,
]
\begin{lstlisting}[language=json,caption= {Synchronisation event},label={synchronisation}]
{
  "eventDateTime": "2021-10-01T07:30:23Z",
  "source": {
    "id": "123",
    "ip_address": "1.1.1.1",
    "manufacturer": {
      "id": "Apple"
    }
  },
  "owner": {
    "id": "A123ABCD",
    "name": "Producer A Beef",
    "email": "prodA@beef.com",
    "givenName": "Aaron",
    "familyName": "Farmer"
  },
  "message": {
    "eventName": "Synchronisation",
    "item": {
      "itemType": "Animal",
      "animal": {
        "identifier": {
          "id": "982 123456789101",
          "scheme": "rfid"
        },
        "specie": "Cattle",
        "gender": "Female"
      }
    },
    "session": {
      "sessionID": "11871156",
      "totalInSession": 55
    },
    "event": {
      "insertionDate": "2021-10-01T07:30:23Z",
      "id": "Av897",
      "brand": "cue mate",
      "hormone": [
        {
          "name": "progesterone",
          "quantity": 11.5,
          "unit": "GRM"
        },
        {
          "name": "estrogen",
          "quantity": 7.5,
          "unit": "GRM"
        }
      ],
      "note": "used type a",
      "withdrawals": [
        {
          "productType": "Intercourse",
          "endDate": "2021-10-09T07:30:23Z"
        },
        {
          "productType": "Vaccination",
          "endDate": "2021-04-03T07:30:23Z"
        }
      ],
      "responsible": "Peter - Artificial Breeding Services CO."
    }
  }
}
\end{lstlisting}
\end{tcolorbox}

\begin{tcolorbox}[
breakable,
  arc=3mm,
  outer arc=3mm,
  boxrule=0.0mm,
  boxsep=0mm,
  left=3mm,
  right=3mm,
  top=3mm,
  bottom=3mm,
]
\begin{lstlisting}[language=json,caption= {Treatment event},label={treatment}]
{
  "eventDateTime": "2022-12-29T13:35:25Z",
  "source": {
    "id": "AM05 Automed Injector",
    "ip_address": "1.1.1.1",
    "manufacturer": {
      "id": "Automed"
    }
  },
  "owner": {
    "id": "A123ABCD",
    "name": "Producer A Beef",
    "email": "prodA@beef.com",
    "givenName": "Aaron",
    "familyName": "Farmer"
  },
  "message": {
    "eventName": "Treatment",
    "item": {
      "itemType": "Animal",
      "animal": {
        "identifier": {
          "id": "rfid",
          "scheme": "123"
        },
        "specie": "Cattle",
        "gender": "Male"
      }
    },
    "session": {
      "sessionID": "Dec-2022-Calves",
      "totalInSession": 20
    },
    "event": {
      "medicine": {
        "name": "DECTOMAX Injectable endectocide",
        "approved": "APVMA Approved",
        "registeredID": {
          "id": "46128",
          "scheme": "APVMA"
        }
      },
      "procedure": "injection",
      "batches": [
        {
          "id": "1122346T",
          "expiryDate": "2023-05-04T00:00:00Z"
        }
      ],
      "withdrawals": [
        {
          "productType": "Vaccination",
          "endDate": "2023-02-09T07:30:23Z"
        }
      ],
      "dose": {
        "doseQuantity": 8,
        "doseUnits": "MLT"
      },
      "site": "shoulder",
      "responsible": "Producer A Beef 0400 111 111"
    }
  }
}
\end{lstlisting}
\end{tcolorbox}

\begin{tcolorbox}[
breakable,
  arc=3mm,
  outer arc=3mm,
  boxrule=0.0mm,
  boxsep=0mm,
  left=3mm,
  right=3mm,
  top=3mm,
  bottom=3mm,
]
\begin{lstlisting}[language=json,caption= {Weaning event},label={weaning}]
{
  "eventDateTime": "2023-04-29T13:35:25Z",
  "source": {
    "id": "Aarons Computer",
    "ip_address": "1.1.1.1",
    "manufacturer": {
      "id": "Apple"
    }
  },
  "owner": {
    "id": "A123ABCD",
    "name": "Producer A Beef",
    "email": "prodA@beef.com",
    "givenName": "Aaron",
    "familyName": "Farmer"
  },
  "message": {
    "eventName": "Weaning",
    "item": {
      "itemType": "Animal",
      "animal": {
        "identifier": {
          "id": "900 000000000203",
          "scheme": "rfid"
        },
        "specie": "Cattle",
        "gender": "Male"
      }
    },
    "session": {
      "sessionID": "875421",
      "totalInSession": 5
    },
    "event": {
      "startdate": "2023-04-29T13:35:25Z",
      "age": 7,
      "reason": "Age",
      "weaningMethod": "Yard"
    }
  }
}
\end{lstlisting}
\end{tcolorbox}

\begin{tcolorbox}[
breakable,
  arc=3mm,
  outer arc=3mm,
  boxrule=0.0mm,
  boxsep=0mm,
  left=3mm,
  right=3mm,
  top=3mm,
  bottom=3mm,
]
\begin{lstlisting}[language=json,caption= {Weight event},label={weight}]
{
  "eventDateTime": "2021-02-08T16:13:23Z",
  "source": {
    "id": "Walkover Weigh",
    "serial": "2Vr93trD",
    "ip_address": "128.0.0.0",
    "manufacturer": {
      "id": "Agtech Company"
    }
  },
  "owner": {
    "id": "C123ABCD",
    "name": "Pro Cattle",
    "email": "proCattle@beef.com",
    "givenName": "Carol",
    "familyName": "Cattle"
  },
  "message": {
    "eventName": "Weight",
    "item": {
      "itemType": "Animal",
      "animal": {
        "identifier": {
          "id": "600 000000000198",
          "scheme": "rfid"
        },
        "specie": "Cattle",
        "gender": "Male"
      }
    },
    "session": {
      "sessionID": "1548",
      "totalInSession": 99
    },
    "event": {
      "weight": {
        "kind": "individual",
        "measurement": "KGM",
        "value": 352.5
      }
    }
  }
}
\end{lstlisting}
\end{tcolorbox}
\linenumbers

\break
\section{Figure}

\begin{figure}[ht!]
\centering
\begin{subfigure}{1\textwidth}
\centering
\includegraphics[scale=0.55]{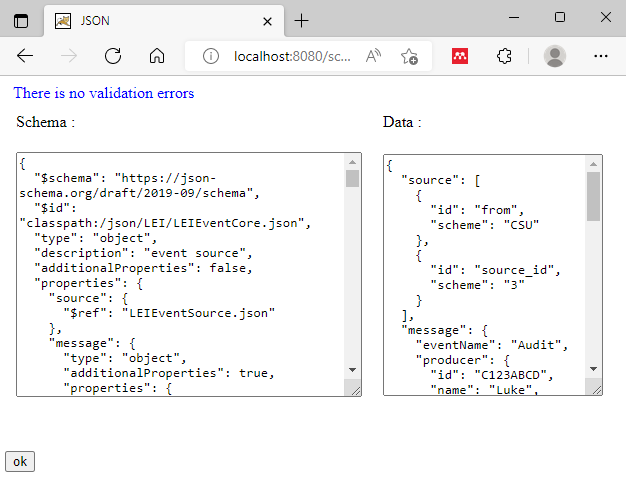}
\caption{}
\label{fig:validotionnoerror}
\end{subfigure}
\hfill
\begin{subfigure}{1\textwidth}
\centering
\includegraphics[scale=0.55]{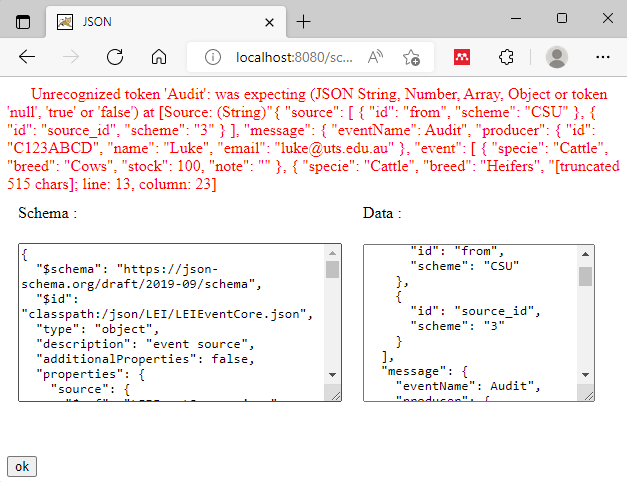} 
\caption{}
\label{fig:validotionerror}
\end{subfigure}
\caption{JSON validator}
\label{fig:validotionoerror}
\end{figure}
\end{document}